\documentclass[sigconf,nonacm]{acmart}%
\AtBeginDocument{%
  }

\newcommand{\topictask}{Topic Opinion Task}
\newcommand{\budgettask}{Budget Allocation Task}

\newcommand{\sigone}[1]{\textbf{#1}\textsuperscript{***}}
\newcommand{\sigfive}[1]{\textbf{#1}\textsuperscript{**}}
\newcommand{\sigten}[1]{#1\textsuperscript{*}}

\usepackage[T1]{fontenc}
\usepackage[utf8]{inputenc}
\usepackage{latexsym}
\usepackage{bm}
\usepackage{bbm}
\usepackage{graphicx}
\usepackage{breakcites}
\usepackage{mdframed}
\usepackage{lipsum}
\usepackage{algorithm}  
\usepackage[noend]{algpseudocode}
\usepackage[english]{babel}
\usepackage{delimset} %
\usepackage{tikz}
\usepackage{changepage}
\usepackage{url}            %
\usepackage{booktabs}       %
\usepackage{amsfonts}       %
\usepackage{microtype}      %
\usepackage{wrapfig}
\usepackage{caption}
\usepackage{graphicx}
\usepackage{subcaption}
\usepackage{booktabs, threeparttable, multirow, array}
\usepackage{makecell}
\usepackage{enumerate}
\usepackage{longtable}
\usepackage{booktabs}
\usepackage{enumitem}
\usepackage{threeparttable}
\usepackage{array}
\usepackage{xcolor}
\usepackage{pifont}

\newcommand{\Supported}{\textcolor{teal}{\ding{51}}}
\newcommand{\NotSupported}{\textcolor{red}{\ding{55}}}
\newcommand{\Partial}{\textcolor{orange}{$\sim$}}

\usepackage{hyperref}
\usepackage{cleveref}

\begin{document}

\title[How User Identity Shapes the Effectiveness of Model Disclaimers in Mitigating LLM Political Persuasion]{Whose Guardrail? User Identity Shapes the Effectiveness of Model Disclaimers in Mitigating LLM Political Persuasion}

\author{Jillian Fisher}
\email{fisherj@stanford.edu}
\orcid{0009-0005-9132-9374}
\correspondingauthor
\authornotemark[1]
\affiliation{%
  \institution{Stanford University}
  \department{Communications Department}
  \city{Stanford}
  \state{California}
  \country{USA}
}
  
\author{Daniel W. Fisher}
\affiliation{%
  \institution{University of Washington School of Medicine}
    \department{Department of Psychiatry and Behavioral Sciences}
  \city{Seattle}
  \state{Washington}
  \country{USA}}

\author{Jennifer Pan}
\affiliation{%
  \institution{Stanford University}
\department{Communications Department}
  \city{Stanford}
  \state{California}
  \country{USA}}

\author{Yulia Tsvetkov}
\affiliation{
\institution{University of Washington}
\department{Paul G. Allen School of Computer Science \& Engineering}
  \city{Seattle}
  \state{Washington}
  \country{USA}}

\author{Katharina Reinecke}
\affiliation{
\institution{University of Washington}
\department{Paul G. Allen School of Computer Science \& Engineering}
  \city{Seattle}
  \state{Washington}
  \country{USA}}

\renewcommand{\shortauthors}{Fisher et al.}

\begin{abstract}
Politically biased AI can shift users' opinions, yet it remains unclear whether model disclaimers protect against such persuasion. We report a preregistered experiment ($N=1,586$) examining how disclaimer design and user characteristics shape susceptibility to politically biased AI. We varied whether disclaimers identified political bias versus potential inaccuracies and whether they included an explanation. Participants completed two decision-making tasks with an LLM biased against their political views, with outcomes assessed immediately and one week later. Disclaimers identifying political bias and including an explanation reduced persuasion, but effects varied across contexts and did not persist. Individual differences were more consistent predictors: Republicans were generally more persuaded and reported higher trust than Democrats across tasks. Similarly, participants who used AI more frequently were also more persuaded, whereas those with greater self-reported AI knowledge were less persuaded. These findings suggest that generic disclaimers are not enough: effective safeguards must account for user heterogeneity and context.

\end{abstract}

\begin{teaserfigure}
 \includegraphics[width=\textwidth]{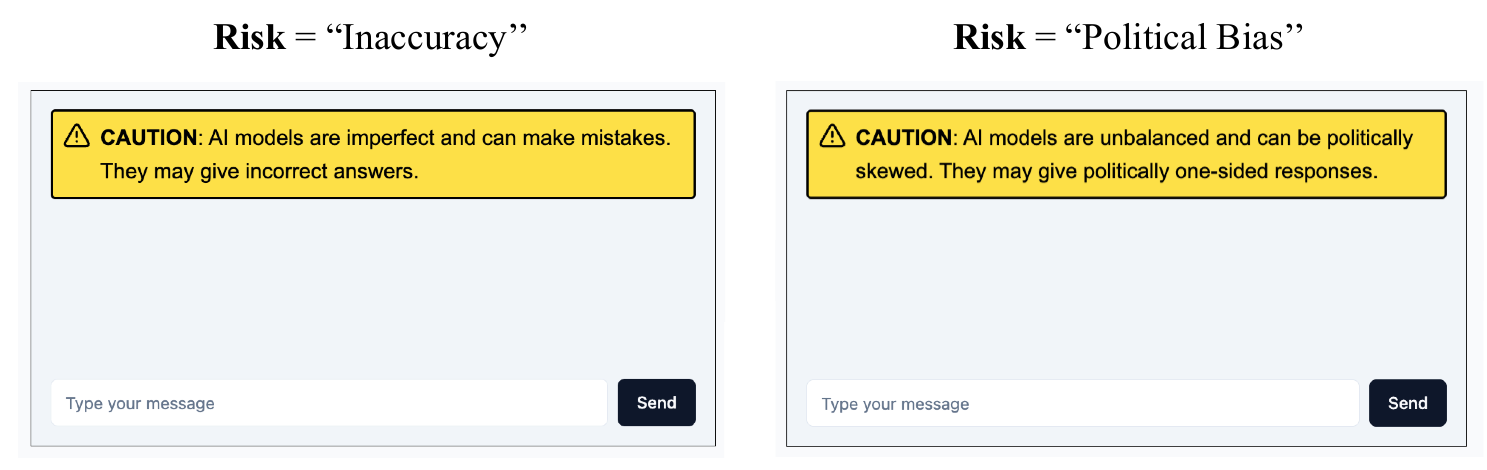}
  \caption{Examples of model disclaimers used in the study. All disclaimers followed the same format and differed only in wording. Disclaimers varied along two dimensions: \texttt{Risk} (shown here) and \texttt{Causal}.}
  \Description{Shown are two examples of model disclaimers used in the study. Both disclaimers share the same visual presentation: a bright yellow box at the top of the model chatbox, with black text and a warning symbol (a triangle containing an exclamation mark) on the left. The left disclaimer is labeled “Risk = Inaccuracy,” while the right is labeled “Risk = Political Bias.”}
  \label{fig:disclaimer_presentation_example}
\end{teaserfigure}

\maketitle
\section{Introduction}
As large language models (LLMs) are increasingly used to provide information and assist with decision-making on politically relevant issues, concerns have grown about their hidden biases and associated harms. Numerous studies have shown that LLMs exhibit measurable political biases in their responses \cite{rettenberger_assessing_2025,Rozado2024PoliticalPreferences, faulborn_only_2025}, and that LLM responses can influence users' opinions and decisions, even after relatively brief interactions \cite{fisher-etal-2025-biased, nature_persuasion, shu_how_2026, potter-etal-2024-hidden}. While prior work has made progress in \textit{measuring} LLMs' political bias and influence on users, little attention has been devoted to understanding \textit{how this influence can be mitigated}.

Existing mitigation approaches have primarily focused on technical interventions to reduce bias within the model's construction, such as post-training alignment, prompt engineering, and adversarial testing \cite{gallegos-etal-2024-bias, ganguli2022redteaminglanguagemodels, hort_bias_2024}. These approaches place the primary responsibility for mitigation on model developers and must be developed and maintained separately for every model. An alternative is to intervene at the point where users interact with AI models, such as the increasingly common approach of \textit{model disclaimers}: brief messages presented before or during interactions that communicate a model's limitations or risks directly to users \cite{cupac_regulate_2024, riesen_sociotechnological-system_2025, risk_management_gov, mitra_navigating_2024}. Prior work examined short-term effects of these disclaimers that communicate general model limitations or highlight specific concerns such as model sycophancy \cite{ibrahim2026, ye_observing_2026}. However, little is known about whether disclaimers can reduce the persuasive effects of politically biased AI and more importantly, existing research has not systematically examined which types of disclaimer information are most effective, either in reducing persuasion in the short term or in producing effects that persist beyond the immediate interaction. 

At the same time, research on persuasion and belief formation has shown that prior values, trust, knowledge, and other personal characteristics shape how people process and integrate new information, including the impact of disclaimers \cite{wegener1995flexible, koehler_explanation_1991, petty1986elaboration, friestad_persuasion_1994}. These individual differences may be particularly important when people
interact with AI systems, where users must decide how much to trust and rely on information generated by a system
whose limitations and biases may not be immediately apparent. We therefore examine how user characteristics shape susceptibility to politically-biased LLMs and how they impact the degree of bias mitigation achieved with different disclaimer designs. %
Specifically, we explore two research questions:
\begin{enumerate}[{}]
\item \textbf{RQ1 (Model Disclaimer Design):} To what extent does the information provided in model disclaimers affect users' susceptibility to political persuasion and their trust in the AI model? 
\item \textbf{RQ2 (User Characteristics):} How do user characteristics affect susceptibility to political persuasion and trust in AI systems, and how do these differences compare with the effects of model disclaimer design?
\end{enumerate}

To answer these questions, we conducted a preregistered\footnote{OSF Link: \url{https://osf.io/e5t8m/overview}} experiment recruiting 1,586 U.S.~participants who, after receiving one of several model disclaimers, interacted with an LLM politically biased in the direction opposite their own affiliation. Drawing on persuasion and bias correction theories, which suggest that warnings may be more effective when they both identify and explain risk \cite{wegener1995flexible, koehler_explanation_1991, petty1986elaboration, friestad_persuasion_1994}, we test two dimensions of disclaimer information: \textbf{\texttt{Risk}}, which varies whether the disclaimer communicates generic model inaccuracies or explicitly identifies political bias, and \textbf{\texttt{Causality}}, which varies whether the disclaimer also explains the source of that bias. We then examine how the effects of these disclaimer interventions on model persuasion compare with differences associated with users' inherent characteristics. We measure both immediate effects following the interaction and whether these effects persist at a one-week follow-up.

We find limited evidence that disclaimer design differentially influences political persuasion overall. Although some moderate effects emerged conditionally, depending on users' political affiliation and the type of task, these effects did not persist at the one-week follow-up. In contrast, individual differences among users emerged as more consistent predictors of susceptibility to persuasion, including political affiliation, trust in the AI system, and different forms of AI familiarity. Specifically, Republican participants were generally more susceptible to persuasion and reported higher post-task trust in the AI system than Democratic participants across tasks, regardless of disclaimer condition. AI use and self-reported AI knowledge also showed associations with susceptibility, with participants who reported using AI more frequently were more susceptible to persuasion, whereas those who reported above-average AI knowledge were less susceptible. Interestingly though, amount of formal education in AI was not associated with susceptibility to persuasion. Together, these findings suggest that \textbf{effective mitigation requires understanding both how model disclaimers are designed and who receives them.} We derive implications for the design and evaluation of AI transparency interventions, highlighting the importance of accounting for context and user heterogeneity and considering safeguards that extend beyond disclaimer wording alone.

\begin{comment}
This paper makes three primary contributions:
\begin{itemize}
\item We present a theory-driven, preregistered test of how model disclaimer content, specifically risk framing and causal explanation, affects resistance to political persuasion and trust in AI, showing that these effects are not uniform overall but emerge conditionally, varying by users' political affiliation and by task, and do not persist over time.

\item In exploratory analyses, we show that individual differences among users, including political affiliation, trust in the AI system, and distinct forms of AI familiarity, are more consistent predictors of susceptibility to political persuasion than disclaimer wording itself, suggesting that evaluations of AI transparency interventions should account for who receives them, not only how they are designed.

\item We derive implications for the design and evaluation of AI transparency interventions, highlighting the importance of accounting for context and user heterogeneity and considering safeguards that extend beyond disclaimer wording alone.

\end{itemize}
\end{comment}

 %
\section{Related Works}
\subsection{Political Bias and Persuasive Influence of LLMs}
Recent work has shown that political bias remains a persistent challenge for LLMs \cite{rettenberger_assessing_2025, feng2023pretrainingdatalanguagemodels, Rozado2024PoliticalPreferences, santurkar_whose_2023, faulborn_only_2025}. This has led to a growing body of work developing benchmarks and methodologies to measure political bias across models \cite{rottger_issuebench_2025, stray_political_2026, rottger_political_2024, measure_slant}. Even more, the issue has become sufficiently important that leading AI companies, including Anthropic and X.AI, now explicitly report measures of political bias in their model documentation and system cards \cite{xai2025grok4fast, anthropic2026mythos}.

While considerable effort has focused on identifying and quantifying political bias, a more fundamental question is whether such biases meaningfully influence users. Recent studies have shown that politically biased LLMs can shape users' opinions, decisions, and judgments across a range of settings \cite{fisher-etal-2025-biased, nature_persuasion, shu_how_2026, potter-etal-2024-hidden}. Most relevant to the present work, \citet{fisher-etal-2025-biased} demonstrated that interactions with politically biased LLMs shifted participants' political opinions and budget allocation decisions, regardless of whether participants initially agreed with the model's position. Notably, these effects emerged after only brief interactions (an average of approximately five conversational turns) and on politically polarized topics, highlighting both the speed and robustness with which biased LLMs can influence users. These findings suggest that mitigating the persuasive effects of political bias is an important challenge for human-AI interaction. 

\subsection{AI Transparency and Model Disclaimers}
Emerging bias mitigation strategies for LLMs include user-facing transparency mechanisms, particularly model disclaimers. Most commercial foundation models now present users with brief disclaimers before or during interaction, typically warning that the model may make mistakes or generate inaccurate information (see current examples in \Cref{appx:current_disclaimers}). These disclaimers are appealing because they are simple to deploy, model-agnostic, and preserve user autonomy by informing rather than restricting model use. However, existing disclaimers are generally brief, generic, and focused almost exclusively on factual inaccuracies rather than other risks such as bias.

Existing evaluations of model disclaimers remain limited. Generic disclaimers appear to have little effect on trust calibration or reliance \cite{metzger2024empowering}. Likewise, \citet{henestrosa2025} found that varying disclaimer wording produced inconsistent effects on perceptions of AI-generated content, although further downstream effects were not measured. However, more targeted disclaimer designs have shown some promise. For example, multiple works have found that direct warnings about LLMs' sycophantic behaviors has improved user recognition of these behaviors \cite{ye_observing_2026, ibrahim2026}, while \citet{gamage2025} demonstrated that AI-generated content labels influence perceptions depending on their design. Collectively, these studies suggest that disclaimer wording might matter, but provide little guidance on which information should be communicated to maximize effectiveness or whether that effectiveness holds uniformly across users with different inherent characteristics or \textit{a priori } beliefs.

Most directly relevant to the present work, \citet{digiuseppe_perceived_2026} found that informing users a model held a directional, out-party political bias reduced its persuasive influence on economics misconceptions by 28\% in a live conversational setting; however, their strongest manipulation conflated the bias disclosure itself with source-credibility information (the model developers' political donations and affiliations), leaving open whether disclosing bias \textit{content} alone, independent of attacks on the source, is sufficient to attenuate persuasion, a question the present \texttt{Risk} and \texttt{Causal} factors are designed to isolate. Additionally, this work focused specifically on misconceptions, instead of the more general bias framing we explore in our work.

\subsection{Warning Labels and Resistance to Persuasion}
Across a variety of domains, the design of warning labels has been shown to substantially influence their effectiveness. In particular, research has examined how to effectively label misinformation on social media platforms, finding that the wording of warning labels, specifically their specificity and strength, influences users' responses to flagged content \cite{clayton2020, martel_misinformation_2023} or reducing reliance on outdated or incorrect information \cite{ecker_explicit_2010}. Other work has shown that both the timing and placement of warnings are important, with labels presented before false content proving more effective than those shown afterward \cite{grady_nevertheless_2021}. However, these effects diminished over time,  highlighting the importance of considering not only the immediate effects of warning labels but also their long-term effectiveness. Additionally, this literature points to a potential downside of using labels: the ``implied truth effect,'' whereby labeling some content can lead users to perceive unlabeled content as implicitly accurate \cite{pennycook_implied_2020}. Although these design principles and potential challenges have been studied extensively in misinformation and warning-label research, they have not been systematically applied to AI model disclaimers.

\subsection{Theories of Persuasion and Bias Correction}
Persuasion research provides several theoretical frameworks for understanding how people recognize and resist biased information. One of the initial models, the Flexible Correction Model, proposes that people correct their judgments only when they are both motivated to do so and able to identify the source of the bias \cite{wegener1995flexible}. Subsequent work suggests that effective correction requires sufficiently precise knowledge about both the nature and magnitude of the bias \cite{wegener1997flexible}. Similarly, the Elaboration Likelihood Model (ELM) argues that persuasion depends not only on an individual's motivation but also on their ability to process available information, including prior knowledge about the persuasive source \cite{petty1986elaboration}. The Persuasion Knowledge Model likewise suggests that people rely on their beliefs about a persuasive agent's motives and characteristics to recognize and discount persuasion attempts \cite{friestad_persuasion_1994}. Together, these theories suggest that warnings that explicitly identify a specific source of bias better support resistance to persuasion than generic warnings and thus motivate our \texttt{Risk} dimension.

A complementary line of work emphasizes the importance of explanations. Psychology research has shown that explanations increase the perceived plausibility of claims and shape subsequent judgments \cite{koehler_explanation_1991}. Similar effects have been observed across a range of domains, including medical decision-making and implicit bias \cite{FitzGerald2017, Vela2022, Williams2019}, research evaluation \cite{Pannucci2010}, and the consumption of biased media \cite{10.1093/hcr/hqac004, Spinde2022}. Collectively, this literature suggests that warnings may be more effective when they explain the mechanisms of risks and therefore we include the \texttt{Causal} dimension in our study.

\subsection{User Characteristics and Susceptibility to Persuasion}
Individual differences in how people process and respond to persuasive information may shape susceptibility to influence. One important source of individual variation is stable personality differences, as the Big Five personality traits predict heterogeneous responses to persuasion. Specifically, \citet{gerber_big_2013} find that Openness is associated with broader responsiveness to persuasive appeals in voter-turnout experiments, while other traits predict responses to particular types of messages. Self-esteem has similarly been linked to suggestibility, including evidence of a curvilinear relationship in which individuals with moderate self-esteem may be more suggestible than those with either low or high self-esteem \cite{rhodes1992self}.

Political identities also shape how individuals seek out, interpret, remember, and evaluate information, with partisans often favoring information consistent with their political identities \cite{taber_lodge_2006, stroud2010polarization, ditto_partisan_2025}. These processes can shape how individuals respond to political messages depending on both their prior beliefs and the political content of the message \cite{leeper_political_2014, cruz_roles_2024}. For example, prior research has found that conservatives tend to exhibit a more ``intuitive,'' rather than ``analytic,'' thinking style \cite{jost_left_2021}, greater avoidance of politically dissonant information \cite{https://doi.org/10.1111/ssqu.12138}, and greater reluctance to update their beliefs when confronted with contrary information \cite{sinclair_closed-minded_2020}. These and other differences may, in turn, shape how political partisanship affects users response to politically biased AI-generated information.

Finally, AI literacy is related to how users understand, evaluate, and trust AI-generated information \cite{zhang_college_2026, KIM2026102992}. More broadly, users with greater experience using AI may have developed more reliance and more cognitive off-loading \cite{kosmyna2025brainchatgptaccumulationcognitive}. We therefore examine three aspects of AI familiarity: self-reported AI knowledge, prior AI-related education, and the current amount of AI use. 

\section{Methods}
To answer our research questions, we conducted a preregistered\footnote{OSF Link: \url{https://osf.io/e5t8m/overview}} online experiment varying two disclaimer dimensions, \texttt{Risk} and \texttt{Causal}, followed by a one-week survey to assess whether their effects on persuasion persisted. 

\begin{figure}[ht]
    \centering
    \includegraphics[width=1\linewidth]{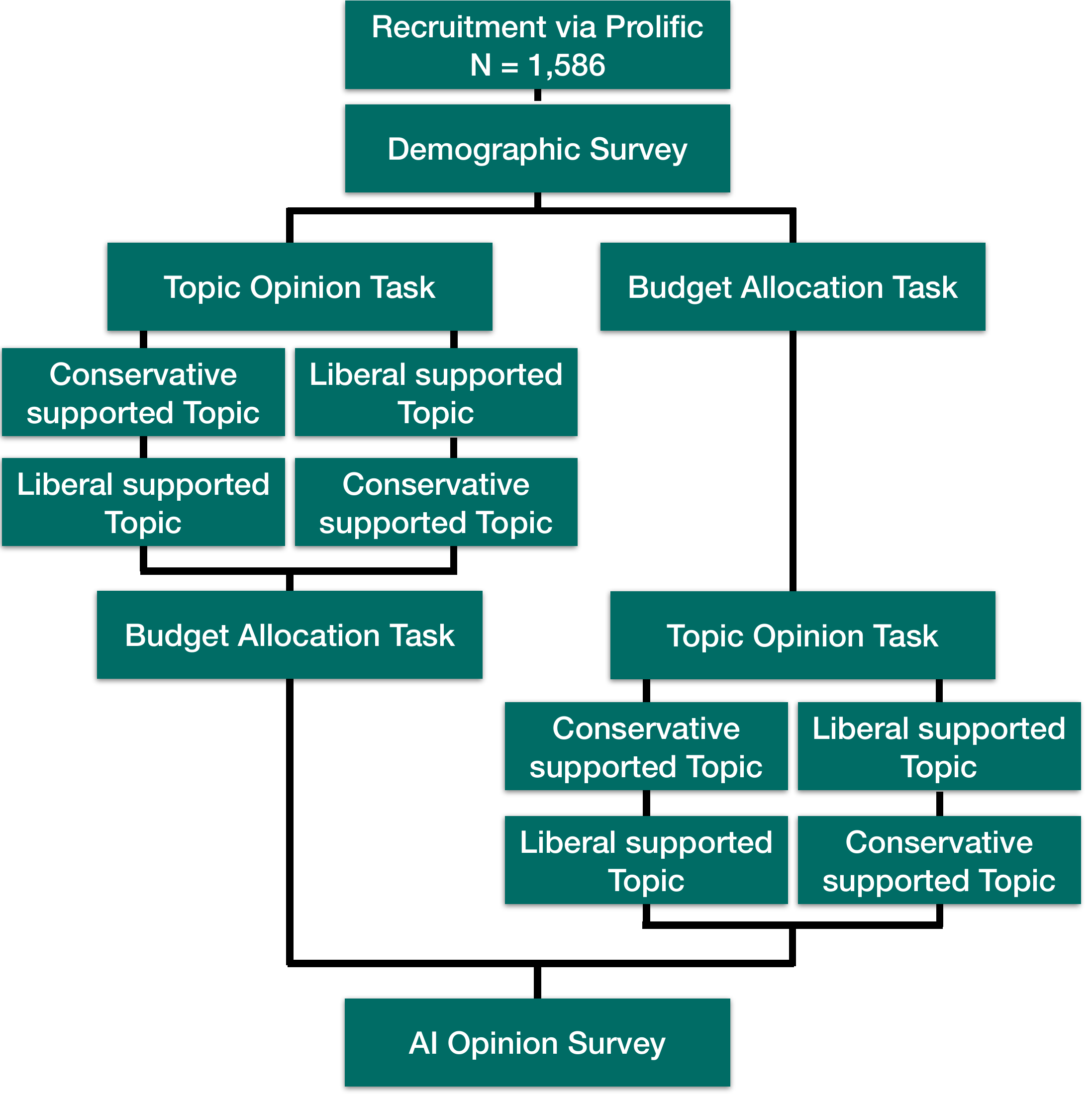}
    \caption{Experimental design and randomization overview.}
    \label{fig:experiemental_flow_diagram}
    \Description{Flowchart showing the study design. Participants were recruited via Prolific (N = 1,586) and completed a demographic survey before being assigned to either the Topic Opinion Task or Budget Allocation Task. Participants in the Topic Opinion Task were assigned one conservative-supported topic and one liberal-supported topic, while participants in the Budget Allocation Task subsequently completed the Topic Opinion Task with the same two topic orientations. Both task sequences concluded with an AI Opinion Survey.}
\end{figure}

\subsection{Hypotheses}
Our experiment formally tested the following five hypotheses: 
\begin{enumerate}[{}]
    \item \textbf{H1 (Risk).} Disclaimers that explicitly identify political bias as a potential risk will reduce the persuasive influence of politically biased LLMs more than disclaimers that warn of general inaccuracies.

    \item \textbf{H2 (Causal).} Disclaimers that explain the cause of potential risks will reduce the persuasive influence of politically biased LLMs more than disclaimers that merely identify the risk.

    \item \textbf{H3 (Trust).} Disclaimers that explicitly identify political bias or its causes will reduce users' trust in the AI system relative to generic accuracy disclaimers.

    \item \textbf{H4 (Persistence).} The mitigating effects of specific and causal disclaimers will persist one week after the interaction.

    \item \textbf{H5 (User Differences).} Individual differences among users, including political affiliation, AI familiarity, and post-task trust in the AI system, will be associated with susceptibility to political persuasion and trust in the AI system.
\end{enumerate}
\begin{table*}[h]
\centering
\begin{tabular}{cc p{12cm}}
\toprule
\textbf{Risk} & \textbf{Causal} & \textbf{Disclaimer Wording} \\
\midrule
Generic & Absent & ``AI models are imperfect and can make mistakes. They may give incorrect answers.'' \\
Specific & Absent & ``AI models are unbalanced and can be politically skewed. They may give politically one-sided responses.'' \\
Generic & Present & ``This AI was trained on imperfect internet data that can contain faulty information. Therefore, it can produce incorrect responses.'' \\
Specific & Present & ``This AI was trained on unbalanced internet data that can contain politically skewed information. Therefore, it can produce politically one-sided responses.'' \\
\bottomrule
\end{tabular}
\caption{Final disclaimer text by condition.}
\label{tab:disclaimers}
\end{table*} 
\subsection{Participants}
We recruited a balanced sample of 800 Democrats and 800 Republicans through Prolific\footnote{\url{https://www.prolific.com}}. Eligibility criteria required participants to be at least 18 years old, current U.S. residents and nationals, proficient in English, and self-identified as either Republican or Democrat. Following Prolific's recommendations, we limited participation to users who had completed at least 30 prior studies with an average approval rating of at least $97\%$. We excluded 4 Democrats and 10 Republicans who failed bot checks (see \Cref{appx:data} for data collection details; see \Cref{appx:power_analysis} for power analysis details).

This yielded a final analytic sample of $N = 1,586$ participants ($64\%$ female, $36\%$ male; $M_{\text{age}} = 43.57$, $SD = 14.45$), compensated at a rate of \$15/hr. Republicans and Democrats differed significantly in age, education, race, and income (balance checks in \Cref{appx:balance_checks_all}); additional pooled analyses included these demographic variables as covariates are included in the appendix (\Cref{appx:pooled_full_models}). This study was deemed exempt by an Internal Review Board, though all participants provided informed consent prior to the study; see \Cref{appx:irb} for a full discussion of ethical considerations. 

Of the 1,586 participants in the main study, 1,242 (78.3\%) completed the one-week follow-up. Attrition did not significantly differ by disclaimer condition or partisanship. Full attrition analyses and balance checks for the follow-up are reported in \Cref{appx:balance_checks_FU_all}.

\subsection{Experimental Setup}
The study used a $2 \times 2$ design crossing two experimental factors: \texttt{Risk} (general vs. specific  warning) and \texttt{Causal} (absent vs. present), with political affiliation serving as a participant blocking factor. Participants were randomly assigned to one of four model disclaimer conditions: generic/no-causal, specific/no-causal, generic/causal, or specific/causal. The disclaimer presentation format was identical across conditions, with only its wording varied. Critically, all participants interacted with a model biased in the direction opposite their own political views (i.e. Democrats interacted with a conservative-biased model, and Republicans with a liberal-biased model), and participants were not informed that the model was biased in any way.

\paragraph{Main Study} Participants were told that they would be interacting with AI language models to complete a series of tasks, with no mention of the study's focus on AI bias. Participants first completed a demographic questionnaire (age, gender, race/ethnicity, education, and income), followed by the single-item self-esteem scale \cite{robins2001measuring} and the Ten-Item Personality Inventory (TIPI) \cite{gosling2003very}, which measures the Big Five personality traits \cite{john1999bigfive}. 

Participants then completed two tasks, identical to those used in \cite{fisher-etal-2025-biased}: the \topictask{} and \budgettask{} (in randomized order). In the \topictask, participants reported baseline opinions on two politically divisive but relatively obscure topics, selected to minimize strong prior beliefs while reflecting documented ideological divides in the U.S. population. Participants then freely interacted with an AI chatbot for 3–20 turns per topic to learn more about the topic before reassessing their opinions. This allowed us to examine opinion change on topics with limited prior knowledge. For a balanced design, each participant saw one topic U.S. conservatives generally support (Conserv+ - Covenant Marriage or Unilateralism) and one which U.S. liberals generally support (Liberal+ - Postal Banking or Nonprofit Tax Uniformity). Topics were selected with expert guidance from a political scientist researcher, with full rationale detailed in \Cref{appx:chosen_topics}. 

In the \budgettask, participants role-played as a mayor allocating government funds across four categories representing politically divergent funding priorities: Public Safety, Education, Veteran Services, and Welfare. Public Safety and Veteran Services represent priorities more strongly associated with U.S. conservative policy preferences, whereas Education and Welfare represent priorities more strongly associated with U.S. liberal policy preferences. After submitting an initial allocation, participants received feedback from and freely discussed their allocation with an AI chatbot for 3–20 turns before submitting a final allocation. Again, we measured the change in allocation from pre- to post- interaction with the  model.

After completing both tasks, participants answered follow-up questions about their experience interacting with this specific AI language model and their general AI familiarity. Finally, all participants were debriefed on the true bias of the model and given the option to withdraw their data. No participant chose to opt out.

\paragraph{Follow Up}
In the one-week follow-up, participants completed only the post-interaction measures of the \topictask{} which asked their current opinion on each topic, their self-reported topic knowledge, and whether (and how) they had engaged with the topic since the initial session. Similarly, for the \budgettask{}, the one-week follow-up asked participants to provide a current allocation for the pretend budget and to report whether and how they had engaged with the task in the interim. Full follow-up procedures are described in \Cref{appx:exp_details}. No additional interaction with a chatbot was completed in the follow-up.

\subsection{Model Biasing}
Each participant interacted with a model whose political views were opposite their self-reported affiliation. We implemented this bias using the same method as \citet{fisher-etal-2025-biased}, prepending a bias-inducing prefix to each participant input on the backend without their knowledge. We used prefixing rather than fine-tuning because it does not require extensive training data while preserving the fluency of the pretrained model. All interactions used GPT-4o-mini \citep{gpt4omini}. See \Cref{appx:model_prefix} for the full set of prefixes used across all tasks.

\subsection{Disclaimers} 
We constructed four model disclaimers that systematically varied \texttt{Risk} (generic inaccuracy vs. specific political bias language) and \texttt{Causal} (explanation absent vs. present), while holding their format and lexical characteristics as constant as possible. We used a template-based procedure and pretested the resulting disclaimers for readability and stylistic comparability. The final disclaimers are shown in Table~\ref{tab:disclaimers}; full construction and validation procedures are reported in \Cref{appx:disclaimer_selection_details}.

We note that the control condition was itself a generic disclaimer rather than the absence of one. This was intentional, as generic disclaimers resemble those currently used by foundation model products (see \Cref{appx:current_disclaimers}), making them an ecologically valid baseline. Moreover, prior work found that a single generic disclaimer does not meaningfully affect trust calibration relative to no disclaimer \cite{metzger2024empowering,henestrosa2025}, suggesting that our control provides a reasonable baseline for comparison. 

\paragraph{Disclaimers Presentation} Disclaimers were displayed at the top of the interface throughout the interaction, using a high-contrast, safety-yellow disclaimer box with a standard caution icon, following established human factors guidelines for disclaimer salience and interpretability \cite{WogalterDeJoyLaughery1999Warnings, ANSI1998SafetyColorCode}. See \Cref{fig:disclaimer_presentation_example} for an example. Full details of the visual design rationale are provided in \Cref{appx:warning_presentation_details}. 
\begin{table*}[t]
\centering
\caption{Summary of hypotheses and results.}
\label{tab:hypothesis-summary}
\begin{threeparttable}
\begin{tabular}{@{}
    p{0.14\textwidth}
    p{0.28\textwidth}
    cc
    p{0.41\textwidth}
@{}}
\toprule
\textbf{Hyp.} & \textbf{Prediction} & \textbf{Dem.} & \textbf{Rep.} & \textbf{Key Findings} \\
\midrule
H1 (Risk) &
Risk-framed disclaimers reduce persuasive influence more than generic disclaimers. &
\Partial &
\Supported &
Budget bias reduced among Republicans; marginal Conserv+ opinion reduced among Democrats \\

H2 (Causal) &
Causal disclaimers reduce persuasive influence more than non-causal disclaimers. &
\Supported &
\NotSupported &
Conserv+ topic opinion reduced among Democrats; no effect among Republicans \\

H3 (Trust) &
Politically-specific / causal disclaimers reduce trust relative to generic / non-causal disclaimers. &
\NotSupported &
\Partial &
No main effects for either party; Significant interaction reduction among Republicans\\

H4 (Persistence) &
Effects from H1/H2 persist at one-week follow-up. &
\NotSupported &
\NotSupported &
All follow-up coefficients were non-significant; estimates were directionally consistent with attenuation rather than reversal. \\

\bottomrule
\end{tabular}

\begin{tablenotes}
\small
\item \Supported\ = supported (at least one task), 
\NotSupported\ = not supported, 
\Partial\ = partial/marginal support ($p<.10$ or interaction only).
\end{tablenotes}
\end{threeparttable}
\end{table*} 
\subsection{Analysis}
We first examined whether model disclaimer design affected political persuasion and post-task AI trust and whether any effects persisted at the one-week follow-up (H1--H4). We estimated linear regression models with \texttt{Risk} (general vs. specific) and \texttt{Causal} (absent vs. present) as fixed effects, their interaction (\texttt{Risk} $\times$ \texttt{Causal}), and HC3 robust standard errors. We used one-tailed tests at $\alpha=.05$, consistent with our directional preregistered hypotheses. Analyses were conducted separately within political affiliation groups to account for potential differences in outcome and covariate distributions. As a sensitivity check, we also conducted pooled analyses across partisan groups. We present minimal models without covariates in the main text because we did not preregister any covariates. However, in \Cref{appx:pooled_full_models}, we report the same pooled analyses including demographic covariates that could potentially be considered confounders. The results are substantively unchanged, with no substantial differences between the minimal and covariate-adjusted models.

For H1 and H2, we modeled post-interaction opinion in the Topic Task (\texttt{POST\_OPINION}) and budget bias in the Budget Task (\texttt{POST\_BUDGET\_BIAS}) using \texttt{Risk}, \texttt{Causal}, their interaction, and the corresponding mean-centered pre-interaction outcome as fixed effects. For all outcomes, \textit{positive values indicate movement toward the model's biased position}. For H4, we re-estimated these models using the corresponding one-week follow-up outcomes (\texttt{POST\_1WK}). For H3, we used the same fixed-effects specification with post-task AI trust as the outcome, without a pre-treatment covariate because trust was measured only after the interaction.

For H5, we extended the persuasion models to include user characteristics as fixed effects. We examined political affiliation, self-esteem and the Big Five personality traits, and AI familiarity, including self-reported AI knowledge, AI-related education, and prior AI use. These characteristics were included alongside \texttt{Risk}, \texttt{Causal}, their interaction, and the mean-centered pre-interaction outcome. We report the associations between each user characteristic and post-interaction persuasion and compare these with the effects of disclaimer design. In a separate exploratory analysis, we examined post-task AI trust as a correlate of persuasion. We applied a Holm correction within each group of tests. As this was exploratory, we use two-side tests with $\alpha=0.05$.
 
All analyses were conducted in R using the \texttt{sandwich} and \texttt{lmtest} packages, with \texttt{emmeans} used for estimated marginal means and simple-effects comparisons. Full model specifications are provided in \Cref{appx:analysis}.

\paragraph{Deviation from Preregistration} We made two deviations from our preregistered analysis plan to avoid conditioning our primary analyses on post-treatment variables. First, rather than restricting the main analyses to participants who passed the post-experiment attention check, we analyze the full sample and report the preregistered restricted-sample analyses, as well as Lee bounds \cite{Lee2005TrainingWA} in \Cref{appx:results_excluded}. Second, for H4, we use the pre-task rather than post-task outcome as the baseline covariate, as the post-task measure may itself be affected by the disclaimer; the preregistered analysis and a full discussion of this deviation are reported in \Cref{appx:fu_post_analysis}.
\section{RQ1: Effect of Model Disclaimers on Persuasion and Trust(\textbf{H1, H2, H3, H4})}
For a summary of results see \Cref{tab:hypothesis-summary}.

\subsection{Effect of Disclaimer Type on Persuasion of Partisan Biased AI (H1/H2)}\label{sect:H1_H2_results}

\begin{figure*}
    \centering
    \caption{Estimated Marginal Mean Change in Persuasion and Trust Outcomes Per Disclaimer Condition and Partisanship}
    \includegraphics[width=0.8\linewidth]{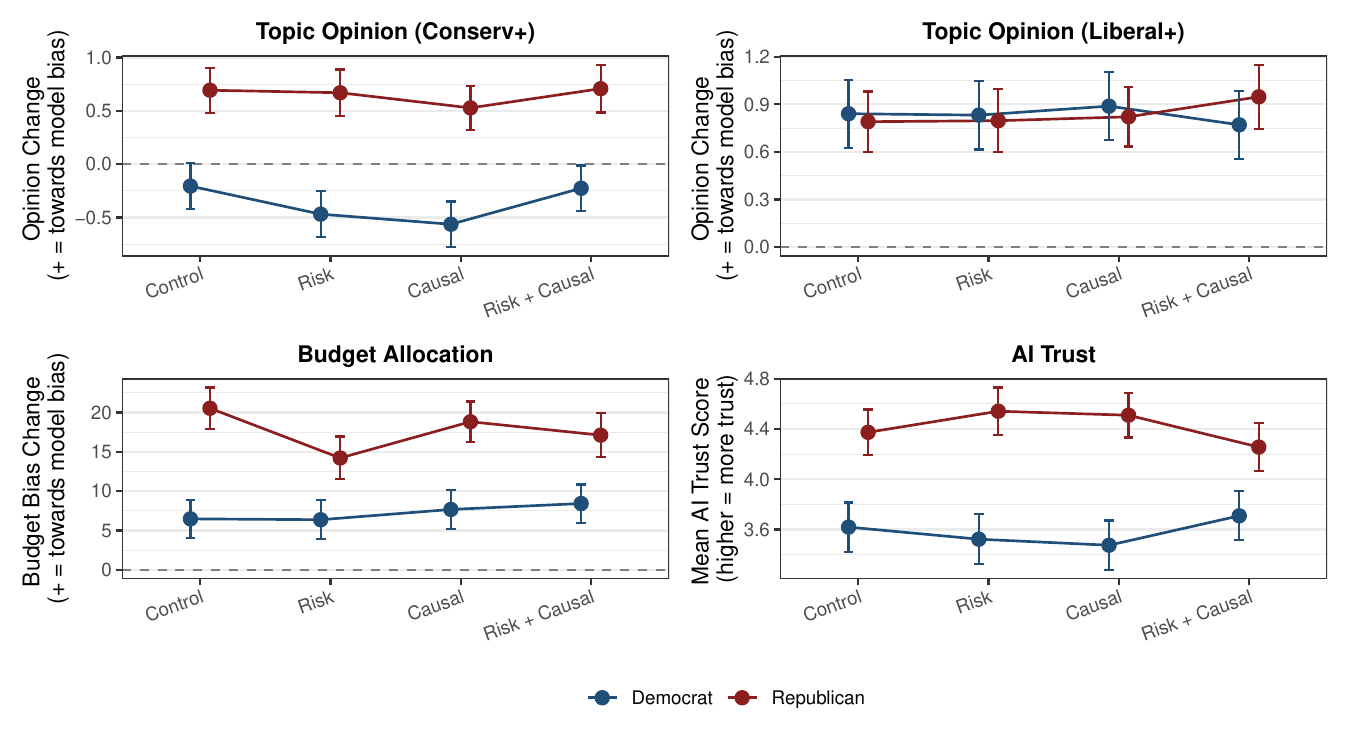}
    \begin{minipage}{0.8\textwidth}
    \footnotesize %
    \textit{Note:} Estimated marginal means (± 90\% CI) for each disclaimer condition (Control, Risk, Causal, Risk + Causal), by party. For the three persuasion outcomes (Topic Opinion: Conserv+/Liberal+, Budget Allocation), values are centered on the sample grand mean of the corresponding pre-task measure, so the y-axis reflects change from average baseline in the direction of the model's bias (positive = towards model bias). The AI Trust panel (H3) shows the absolute mean trust score, as no pre-task measure was collected for this outcome.
  \end{minipage}
    \label{fig:main_mean_withsubstract_all}
    \Description{Four-panel figure showing estimated mean outcomes for Democrats and Republicans across four experimental conditions: Control, \texttt{Risk}, \texttt{Causal}, and \texttt{Risk} + \texttt{Causal}. The panels show opinion change for Conservative-supported topics, opinion change for Liberal-supported topics, budget allocation change, and mean AI trust score. Democrats are shown in blue and Republicans in red. The opinion and budget panels indicate change in the direction of model bias, while the AI trust panel indicates that higher scores represent greater trust.}
\end{figure*}

To test H1 and H2, we examined whether \texttt{Risk}, \texttt{Causal}, and their interaction affected persuasion across the Topic and Budget Tasks. We estimated these models separately for Democrats and Republicans (\Cref{tab:main_all}); a pooled sensitivity analysis is reported in \Cref{tab:main_minimal_all}. Figure~\ref{fig:main_mean_withsubstract_all} presents the corresponding model-adjusted estimated means for each disclaimer condition and partisan group, expressed as change from the sample's average pre-task baseline.

\textit{Topic Task Results.} Among Democrats, \Cref{fig:main_mean_withsubstract_all} shows a downward step in adjusted opinion change on Conserv+ topics for the \texttt{Causal} disclaimers ($\beta = -0.36, p<0.05$) producing the largest, statistically significant reduction compared to the generic/non-causal control disclaimer. However, neither disclaimer reliably shifted opinions on the Liberal+ topics, whose estimated means remain similar across all four conditions. Among Republicans, estimated means for both Conserv+ and Liberal+ topics were similar to their control-condition level across all disclaimer conditions (all $\vert \beta\vert <0.20, p>0.10$), and the figure shows comparatively flat or even increasing values for both outcomes. We found no evidence of a synergistic effect between \texttt{Risk} and \texttt{Causal}, as the interaction term was non-significant

\textit{Budget allocation.} The partisan pattern reversed for budget allocation bias. Republicans' estimated means drop most sharply under the \texttt{Risk} condition, corresponding to a significant reduction in pro-model budget bias relative to control ($\beta=-6.32, p < 0.01$) but their estimated means remain high under \texttt{Causal} and \texttt{Risk} $\times$ \texttt{Causal}, consistent with the non-significant coefficients for those conditions. Democrats showed no significant effect of the main or the interaction term.

\textit{Pooled model.} In the full-sample sensitivity model shown in \Cref{tab:main_minimal_all}, both the \texttt{Causal} disclaimer's effect on Conserv+ topic remained significant ($\beta= -0.27, p<0.05$) and the \texttt{Risk} disclaimer's effect on budget allocation remained significant ($\beta= -3.20, p<0.05$). However, we do see that participant partisanship itself remained a strong, significant covariate across all four outcomes, with Republicans generally having statistically higher values meaning more susceptible to persuasion.

\begin{table*}[t]
\centering
\begin{threeparttable}
\caption{Effect of Disclaimer Conditions on Opinion, Budget Allocation, and AI Trust}
\label{tab:main_all}
\small
\begin{tabular}{ll r@{} l r@{} l r@{} l r@{} l}
\toprule
& & \multicolumn{2}{c}{Conserv+ Opinion} & \multicolumn{2}{c}{Liberal+ Opinion} & \multicolumn{2}{c}{Budget Allocation} & \multicolumn{2}{c}{AI Trust} \\
\cmidrule(lr){3-4} \cmidrule(lr){5-6} \cmidrule(lr){7-8} \cmidrule(lr){9-10}
Party & Predictor & \multicolumn{2}{c}{$b$ (SE)} & \multicolumn{2}{c}{$b$ (SE)} & \multicolumn{2}{c}{$b$ (SE)} & \multicolumn{2}{c}{$b$ (SE)} \\
\midrule
\multirow{3}{*}{Democrat}
 & Risk       & \sigten{$-0.26$} & $(0.19)$  & $-0.01$ & $(0.19)$  & $-0.11$ & $(1.93)$  & $-0.10$ & $(0.17)$  \\
 & Causal     & $\sigfive{-0.36}$ & $(0.19)$  & $0.05$ & $(0.18)$  & $1.19$ & $(2.21)$  & $-0.14$ & $(0.17)$  \\
 & Risk$\times$Causal & $0.60$ & $(0.26)$   & $-0.11$ & $(0.26)$   & $0.86$ & $(3.00)$   & $0.33$ & $(0.24)$   \\[4pt]
\multirow{3}{*}{Republican}
 & Risk       & $-0.02$ & $(0.18)$   & $0.01$ & $(0.17)$  & $\sigfive{-6.32}$ & $(2.28)$  & $0.17$ & $(0.15)$   \\
 & Causal     & $-0.17$ & $(0.18)$   & $0.03$ & $(0.16)$   & $-1.71$ & $(2.23)$  & $0.14$ & $(0.16)$   \\
 & Risk$\times$Causal & $0.20$ & $(0.26)$   & $0.12$ & $(0.24)$   & $4.61$ & $(3.25)$   & $\sigfive{-0.42}$ & $(0.22)$   \\
\bottomrule
\end{tabular}
\begin{tablenotes}
\small
\item \textit{Note.} Entries are OLS regression coefficients with HC3 robust standard errors in parentheses. All hypothesis tests are one-tailed ($\alpha = .05$), directional negative. Models for opinion and budget allocation include the corresponding pre-measure as a covariate (mean-centered); the trust model does not. Values \sigfive{} = $p < .05$; \sigten{} = $p < .10$.
\end{tablenotes}
\end{threeparttable}
\end{table*} %
\begin{table*}[ht]
\centering
\begin{threeparttable}
\caption{Effect of Disclaimer Conditions on Opinion, Budget Allocation, and AI Trust (Pooled Across Partisans)}
\label{tab:main_minimal_all}
\small
\begin{tabular}{l r@{} l r@{} l r@{} l r@{} l}
\toprule
& \multicolumn{2}{c}{Conserv+ Opinion} & \multicolumn{2}{c}{Liberal+ Opinion} & \multicolumn{2}{c}{Budget Allocation} & \multicolumn{2}{c}{AI Trust} \\
\cmidrule(lr){2-3} \cmidrule(lr){4-5} \cmidrule(lr){6-7} \cmidrule(lr){8-9}
Predictor & \multicolumn{2}{c}{$b$ (SE)} & \multicolumn{2}{c}{$b$ (SE)} & \multicolumn{2}{c}{$b$ (SE)} & \multicolumn{2}{c}{$b$ (SE)} \\
\midrule
Risk
& $\sigten{-0.14}$ & $(0.13)$
& $-0.00$ & $(0.13)$
& $\sigfive{-3.20}$ & $(1.49)$
& $0.03$ & $(0.12)$ \\

Causal
& $\sigfive{-0.27}$ & $(0.13)$
& $0.04$ & $(0.12)$
& $-0.24$ & $(1.56)$
& $-0.00$ & $(0.12)$ \\

Risk$\times$Causal
& $0.40$ & $(0.19)$
& $0.00$ & $(0.18)$
& $2.73$ & $(2.20)$
& $-0.04$ & $(0.16)$ \\

Republican
& $\sigfive{1.04}$ & $(0.09)$
& $\sigfive{-0.54}$ & $(0.10)$
& $\sigfive{5.21}$ & $(1.15)$
& $\sigfive{0.84}$ & $(0.08)$ \\
\bottomrule
\end{tabular}

\begin{tablenotes}
\small
\item \textit{Note.} Entries are OLS regression coefficients with HC3 robust standard errors in parentheses. Models are estimated on the full sample; Republican (vs.\ Democrat) is included as a main-effect covariate only. Hypothesis tests for disclaimer main effects (Risk, Causal, Risk$\times$Causal) are one-tailed, directional negative; Republican is tested two-tailed. Models for opinion and budget allocation include the corresponding pre-measure as a covariate (mean-centered); the trust model does not. \sigfive{} = $p < .05$; \sigten{} = $p < .10$.
\end{tablenotes}

\end{threeparttable}
\end{table*}

\subsection{Effect of Disclaimer Type on Post-Task AI Trust (H3)}\label{sect:H3_results}
Similarly to test H3, we examined whether \texttt{Risk}, \texttt{Causal}, and their interaction affected post-task AI trust, estimating the models separately for Democrats and Republicans (\Cref{tab:main_all}). A pooled sensitivity analysis is reported in \Cref{tab:main_minimal_all}. Figure~\ref{fig:main_mean_withsubstract_all} (bottom right) presents the corresponding model-adjusted estimated means by condition and partisan group.

Contrary to our H3 hypothesis, we found no evidence that the \texttt{Risk} or \texttt{Causal} disclaimer reduced AI trust among either party. Among Democrats, the coefficients were small and non-significant ($\beta= -0.10/-0.14$, ns) and among Republicans, there was directionally positive and non-significant ($\beta=0.17/0.14$, ns). However, for Republicans we observed a significant Risk × Causal interaction ($\beta = -0.42$, $p<0.05$), such that combining both disclaimer types reduced trust more than either alone, even though neither component individually reached significance. 

\textit{Pooled model.} The pooled minimal model confirms this null main effect ($\beta=0.03/0.00$, ns) and a null interaction ($\beta=-0.04$, ns) when pooled across parties. However, as with H1 and H2, we found a significant difference between the two partisan parties. Republican identity was a strong, significant predictor of trust ($\beta=0.84, p<0.001$), reflecting substantially higher post-task trust in the AI among Republican participants than among Democrats regardless of condition.  

\subsection{Effect of Disclaimer Type on Long-Term Persuasion of Partisan Biased AI (H4)}\label{sect:H4_results}
To test H4, we regressed each follow-up persuasion outcome on disclaimer condition (\texttt{Risk}, \texttt{Causal}, and their interaction), controlling for the original pre-task score (mean-centered) as the baseline covariate. 

\Cref{tab:followup_pre_all} shows that neither the \texttt{Risk} nor \texttt{Causal} disclaimers significantly reduced persuasion at the one-week follow-up for any outcome or political party (all $p>.05$). However, some estimated effects were directionally consistent with the immediate post-task effects, suggesting attenuation rather than reversal over time. For example, among Democrats, the \texttt{Risk} and \texttt{Causal} effects on the Conserv+ topic outcome remained negative at follow-up ($\beta=-0.28$ and $\beta=-0.30$, respectively), compared with immediate effects of $\beta=-0.26$ and $\beta=-0.36$. No \texttt{Risk} $\times$ \texttt{Causal} interactions reached statistical significance. Thus, the results are consistent with disclaimer effects fading over time; however, we caution that the reduced follow-up sample size may limit our ability to detect a lasting but attenuated effect of the disclaimer over this interval. Additionally, we note that users were informed that the model they interacted with had a bias at the end of the main study (due to ethical concerns), which could have also attenuated the results. Lastly, we note that our sensitivity analysis did show significant results for the Causal condition \Cref{tab:followup_pre_minimal_all}, further indicating possible reversal rather than attenuation. 

\begin{table*}[t]
\centering
\begin{threeparttable}
\caption{Effect of Disclaimer Conditions on Follow-Up Opinion and Budget Allocation (Pre-Task Baseline)}
\label{tab:followup_pre_all}
\small
\begin{tabular}{ll r@{} l r@{} l r@{} l}
\toprule
& & \multicolumn{2}{c}{Conserv+ Opinion} & \multicolumn{2}{c}{Liberal+ Opinion} & \multicolumn{2}{c}{Budget Allocation} \\
\cmidrule(lr){3-4} \cmidrule(lr){5-6} \cmidrule(lr){7-8}
Party & Predictor & \multicolumn{2}{c}{$b$ (SE)} & \multicolumn{2}{c}{$b$ (SE)} & \multicolumn{2}{c}{$b$ (SE)} \\
\midrule
\multirow{3}{*}{Democrat}
 & Risk               & $\sigten{-0.28}$ & $(0.19)$ & $0.00$ & $(0.22)$ & $-0.08$ & $(2.25)$ \\
 & Causal             & $\sigten{-0.30}$ & $(0.20)$ & $-0.14$ & $(0.21)$ & $-0.62$ & $(2.67)$ \\
 & Risk$\times$Causal & $0.04$  & $(0.28)$ & $-0.11$ & $(0.30)$ & $3.44$  & $(3.49)$ \\[4pt]
\multirow{3}{*}{Republican}
 & Risk               & $0.06$  & $(0.20)$ & $0.11$  & $(0.19)$ & $-1.57$ & $(2.92)$ \\
 & Causal             & $-0.19$ & $(0.20)$ & $0.06$  & $(0.19)$ & $3.33$  & $(2.74)$ \\
 & Risk$\times$Causal & $0.19$  & $(0.29)$ & $0.09$  & $(0.27)$ & $0.51$  & $(4.33)$ \\
\bottomrule
\end{tabular}
\begin{tablenotes}
\small
\item \textit{Note.} Entries are OLS regression coefficients with HC3 robust standard errors in parentheses. Dependent variables are one-week follow-up measures of opinion and budget allocation; the original pre-task value of the corresponding outcome (mean-centered) is included as a covariate. All hypothesis tests are one-tailed, directional negative. \sigfive{} = $p < .05$; \sigten{} = $p < .10$.
\end{tablenotes}
\end{threeparttable}
\end{table*} 

\begin{table*}[ht]
\centering
\begin{threeparttable}
\caption{Effect of Disclaimer Conditions and Accurate Disclaimer Recall on Persuasion and Trust}
\label{tab:manip_check}
\small
\begin{tabular}{ll r@{} l r@{} l r@{} l r@{} l}
\toprule
& & \multicolumn{2}{c}{Conserv+ Opinion} & \multicolumn{2}{c}{Liberal+ Opinion} & \multicolumn{2}{c}{Budget Allocation} & \multicolumn{2}{c}{AI Trust} \\
\cmidrule(lr){3-4} \cmidrule(lr){5-6} \cmidrule(lr){7-8} \cmidrule(lr){9-10}
Party & Predictor & \multicolumn{2}{c}{$b$ (SE)} & \multicolumn{2}{c}{$b$ (SE)} & \multicolumn{2}{c}{$b$ (SE)} & \multicolumn{2}{c}{$b$ (SE)} \\
\midrule
\multirow{4}{*}{Democrat}
 & Risk                       & $-0.23$ & $(0.19)$ & $0.02$  & $(0.19)$& $0.33$  &$(2.00)$ & $-0.05$ &$(0.17)$ \\
 & Causal                     & $\sigfive{-0.34}$ &$(0.19)$ & $0.06$  &$(0.18)$ & $1.40$  &$(2.24)$ & $-0.12$ & $(0.17)$ \\
 & Risk$\times$Causal         & $0.57$  &$(0.26)$ & $-0.13$ &$(0.26)$ & $0.51$  &$(3.02)$ & $0.29$  & $(0.23)$ \\
 & \textbf{Accurate Disclaimer Recall} & \sigten{$-0.27$} & $(0.15)$& $-0.18$ &$(0.15)$ & \sigten{$-3.33$} & $(1.90)$& $\sigfive{-0.35}$ & $(0.13)$\\[4pt]
\multirow{4}{*}{Republican}
 & Risk                       & $-0.03$ &$(0.18)$ & $0.01$  & $(0.17)$& $\sigfive{-6.34}$ &$(2.30)$ & $0.18$  & $(0.15)$\\
 & Causal                     & $-0.17$ &$(0.19)$ & $0.04$  & $(0.16)$& $-1.79$ & $(2.25)$& $0.16$  &$(0.16)$ \\
 & Risk$\times$Causal         & $0.20$  & $(0.26)$& $0.12$  & $(0.24)$& $4.61$  &$(3.26)$ & $\sigfive{-0.42}$ & $(0.22)$\\
 & \textbf{Accurate Disclaimer Recall} & $0.09$  &$(0.15)$ & $-0.10$ & $(0.13)$& $1.02$  &$(1.68)$ & $\sigfive{-0.39}$ & $(0.12)$\\
\bottomrule
\end{tabular}
\begin{tablenotes}
\small
\item \textit{Note.} Entries are OLS regression coefficients with HC3 robust standard errors in parentheses. Hypothesis tests for disclaimer conditions (Risk, Causal, Risk$\times$Causal) are one-tailed, directional negative; Accurate Disclaimer Recall is tested two-tailed. Models for opinion and budget allocation include the corresponding pre-measure as a covariate (mean-centered); the trust model does not. \sigfive{} = $p < .05$; $\sigten{} = p < .10$.
\end{tablenotes}
\end{threeparttable}
\end{table*} 
\section{RQ2: User Characteristics Effect on Persuasion and Trust (\textbf{H5})}
We next present a set of exploratory analyses to better understand how inherent user characteristics, specifically political partisanship, personality traits, and AI familiarity correlate with susceptibility to persuasion and AI trust (H5). However, we note that because these analyses were not part of our preregistered hypotheses, they should be interpreted as hypothesis-generating rather than confirmatory.

\subsection{Partisan Difference in Susceptibility and Post AI Trust}\label{sect:partisan_trust_susceptibility_results}

Across the H1, H2 and H3 models, political partisanship emerged as a strong, consistent predictor of both persuasion and trust outcomes, independent of disclaimer condition (Table~\ref{tab:main_minimal_all}). Specifically, Republicans showed significantly greater pro-model shift on the Conserv+ topics ($\beta=1.04, p<.001$) and significantly less pro-model shift on the Liberal+ topics ($\beta=-0.54, p<.001$), relative to Democrats, holding condition and pre-task opinion constant. Republicans also showed a significantly larger pro-model shift in budget allocation ($\beta=5.21, p<.001$) and reported significantly higher post-task trust in the AI ($\beta=0.84, p<.001$). These results indicate a consistent partisan asymmetry in susceptibility to the model's biased framing that holds across every outcome measured, independent of which disclaimer participants received. Notably, this partisan main effect was substantially larger in magnitude and more consistently significant than any individual disclaimer effect reported in H1, H2 or H3. 

This partisan asymmetry extended to participants' own awareness of the model's bias. As part of the post-task AI Opinion Survey, participants were asked whether they believed the models' responses reflected a political lean. We found that Republicans were significantly less likely than Democrats to report noticing the model's bias (48.36\% vs. 63.95\%, $\chi^2 = 38.52$, $p < .001$). This gap in awareness is consistent with the suggestibility differences reported above and suggests that Republican participants were, on average, not only more persuaded by the model but also less likely to recognize that persuasion as arising from a politically biased source.

\subsection{Partisan Difference in How Disclaimer Attention Affects Persuasion}\label{sect:partisan_attention}
To assess whether accurate recall of the disclaimer itself, beyond mere assignment to condition, related to persuasion or trust, we re-estimated the H1, H2, and H3 models including an added binary ``Accurate Disclaimer Recall '' predictor alongside the assigned conditions. However, because manipulation check accuracy was not randomly assigned, unlike disclaimer condition, these analyses are correlational rather than causal. 

In \Cref{tab:manip_check} we again observed partisan differences, consistent with our primary analyses. Among Democrats, correctly identifying the disclaimer was associated with marginally less pro-model persuasion on the Conserv+ opinion task ($\beta=-0.27$, $p=.07$) and budget allocation task ($\beta=-3.33$, $p=.07$), as well as significantly lower post-task trust ($\beta=-0.35$, $p<.05$) independent of assigned condition. Among Republicans, accurate disclaimer recall was not significantly associated with either persuasion outcome, but was significantly associated with lower post-task trust ($\beta=-0.39$, $p<.01$), mirroring the Democrat trust result. Additionally, the key condition effects for H1, H2 and H3 remained robust.

These results give some indication that accurately recalling or identifying the disclaimer received might be associated with reduced trust (for both parties) and, among Democrats specifically, reduced persuasion.

\subsection{Personality Traits as Predictors of Persuasion and AI Trust}\label{sect:personality_traits}
\begin{table*}[ht]
\centering
\begin{threeparttable}
\caption{Personality Traits as Predictors of Persuasion and AI Trust (Pooled Model)}
\label{tab:personality_allresults_all}
\small
\begin{tabular}{l r@{} l r@{} l r@{} l r@{} l}
\toprule
& \multicolumn{2}{c}{Conserv+ Opinion} & \multicolumn{2}{c}{Liberal+ Opinion} & \multicolumn{2}{c}{Budget Allocation} & \multicolumn{2}{c}{AI Trust} \\
\cmidrule(lr){2-3} \cmidrule(lr){4-5} \cmidrule(lr){6-7} \cmidrule(lr){8-9}
Trait & \multicolumn{2}{c}{$b$ (SE)} & \multicolumn{2}{c}{$b$ (SE)} & \multicolumn{2}{c}{$b$ (SE)} & \multicolumn{2}{c}{$b$ (SE)} \\
\midrule
Risk
& $-0.13$ & $(0.13)$ & $0.02$ & $(0.13)$ & $\sigfive{-3.17}$ & $(1.49)$ & $0.04$ & $(0.11)$ \\

Causal
& $\sigfive{-0.28}$ & $(0.13)$ & $0.01$ & $(0.12)$ & $-0.38$ & $(1.57)$ & $-0.05$ & $(0.11)$ \\

Risk$\times$Causal
& $\sigfive{0.40}$ & $(0.19)$ & $0.01$ & $(0.18)$ & $2.85$ & $(2.18)$ & $0.03$ & $(0.16)$ \\

Republican
& $\sigfive{0.99}$ & $(0.10)$ & $\sigfive{-0.51}$ & $(0.10)$ & $\sigfive{5.27}$ & $(1.25)$ & $\sigfive{0.75}$ & $(0.08)$ \\
\midrule
Extraversion
& $0.05$ & $(0.03)$ & $0.07^*$ & $(0.03)$ & $0.18$ & $(0.38)$ & $\sigfive{0.10}$ & $(0.03)$ \\

Openness
& $-0.04$ & $(0.04)$ & $0.05$ & $(0.04)$ & $0.11$ & $(0.55)$ & $0.01$ & $(0.04)$ \\

Conscientiousness
& $0.09^*$ & $(0.04)$ & $-0.02$ & $(0.04)$ & $-0.12$ & $(0.54)$ & $0.01$ & $(0.04)$ \\

Emotional Stability
& $0.00$ & $(0.04)$ & $\sigfive{-0.12}$ & $(0.04)$ & $-0.38$ & $(0.47)$ & $0.04$ & $(0.03)$ \\

Agreeableness
& $0.01$ & $(0.04)$ & $\sigfive{0.14}$ & $(0.04)$ & $1.16^*$ & $(0.49)$ & $\sigfive{0.16}$ & $(0.03)$ \\

Self-Esteem
& $-0.04$ & $(0.05)$ & $0.04$ & $(0.05)$ & $-0.00$ & $(0.60)$ & $0.11^*$ & $(0.05)$ \\
\bottomrule
\end{tabular}

\begin{tablenotes}
\small
\item \textit{Note.} Entries are OLS regression coefficients with robust standard errors in parentheses. Models are estimated on the pooled sample, controlling for party, disclaimer condition (Risk, Causal, Risk$\times$Causal), and the corresponding pre-measure (opinion and budget models only). \sigfive{} = $p_{\text{Holm}}<.05$ or $p<0.05$ for disclaimer condition or Republican; $^*$ = $p_{\text{raw}}<.05$.
\end{tablenotes}

\end{threeparttable}
\end{table*}
 To explore whether individual differences in personality relate to susceptibility to partisan-biased AI persuasion or post-task trust, we added the Big Five personality traits (Extraversion, Openness, Conscientiousness, Emotional Stability, Agreeableness) and self-esteem to the H1, H2, and H3 models as simultaneous predictors, pooled across party, and controlling for disclaimer condition. Because six personality traits were tested simultaneously for each outcome, we applied a Holm correction within each outcome (6 tests per outcome). The six measures were moderately intercorrelated ($r=.17 - .42$), but variance inflation factors remained low (all $\leq 2.02$), indicating multicollinearity did not meaningfully distort these estimates. 

Results in \Cref{tab:personality_allresults_all} showed personality traits did not significantly predict Conserv+ opinion, or budget allocation change after correction for multiple comparisons (all $p_{Holm}>.14$), however both Emotional Stability ($\beta=-0.12, p_{Holm} < 0.01$) and Agreeableness ($\beta=0.14, p_{Holm} < 0.01$) had significant associations for the Liberal+ topic opinion. Additionally, Extraversion ($\beta=0.10, p_{Holm}<0.01$) and Agreeableness ($\beta=0.16, p_{Holm} < 0.05$) remained significant predictors of trust after correction, with more agreeable and more extraverted participants reporting higher post-task trust in the AI. Self-esteem was nominally significant as a predictor for post-task AI Trust but did not survive correction ($\beta=0.11, p=.034, p_{Holm}=0.17$). Condition and party coefficients remained substantively consistent with our primary models. 

\subsection{AI Familiarity (Knowledge, Education, Use) as Predictors of Persuasion and AI Trust}\label{sect:ai_familiarity}
\begin{table*}[ht]
\centering
\begin{threeparttable}
\caption{AI Familiarity Measures as Predictors of Persuasion and AI Trust (Pooled Model)}
\label{tab:ai_familiarity_allresults_all}
\small
\begin{tabular}{l r@{} l r@{} l r@{} l r@{} l}
\toprule
& \multicolumn{2}{c}{Conserv+ Opinion} & \multicolumn{2}{c}{Liberal+ Opinion} & \multicolumn{2}{c}{Budget Allocation} & \multicolumn{2}{c}{AI Trust} \\
\cmidrule(lr){2-3} \cmidrule(lr){4-5} \cmidrule(lr){6-7} \cmidrule(lr){8-9}
Predictor & \multicolumn{2}{c}{$b$ (SE)} & \multicolumn{2}{c}{$b$ (SE)} & \multicolumn{2}{c}{$b$ (SE)} & \multicolumn{2}{c}{$b$ (SE)} \\
\midrule
Risk
& $-0.13$ & $(0.13)$ & $0.01$ & $(0.13)$ & $\sigfive{-3.04}$ & $(1.49)$ & $0.05$ & $(0.11)$ \\

Causal
& $\sigfive{-0.25}$ & $(0.13)$ & $0.05$ & $(0.12)$ & $0.00$ & $(1.54)$ & $0.02$ & $(0.11)$ \\

Risk $\times$ Causal
& $\sigfive{0.38}$ & $(0.18)$ & $-0.03$ & $(0.18)$ & $2.24$ & $(2.18)$ & $-0.04$ & $(0.16)$ \\

Republican
& \sigfive{1.01} & $(0.09)$ & \sigfive{-0.56} & $(0.10)$ & \sigfive{4.94} & $(1.14)$ & \sigfive{0.81} & $(0.08)$ \\

\midrule
AI Knowledge
& \sigfive{-0.18} & $(0.07)$ & \sigfive{-0.23} & $(0.07)$ & \sigfive{-3.51} & $(0.90)$ & \sigten{-0.14} & $(0.06)$ \\

AI Education
& $0.02$ & $(0.08)$ & $0.06$ & $(0.07)$ & $0.16$ & $(1.06)$ & \sigfive{0.20} & $(0.07)$\ \\

AI Use
& \sigfive{0.23} & $(0.06)$ & \sigfive{0.18} & $(0.05)$ & \sigfive{2.88} & $(0.69)$ & \sigfive{0.37} & $(0.05)$ \\
\bottomrule
\end{tabular}

\begin{tablenotes}
\small
\item \textit{Note.} Entries are OLS regression coefficients with HC3 robust standard errors in parentheses. Models are estimated on the pooled sample, controlling for party, disclaimer condition (Risk, Causal, Risk$\times$Causal), and the corresponding pre-measure (opinion and budget models only). \sigfive{} = $p_{\text{Holm}}<.05$ or $p<0.05$ for disclaimer condition or Republican; $^*$ = $p_{\text{raw}}<.05$.
\end{tablenotes}
\end{threeparttable}
\end{table*}
 To explore whether AI familiarity relates to susceptibility to partisan-biased persuasion or post-task trust, we added three ordinal self-report measures, self-rated AI Knowledge, AI Education, and frequency of AI Use, to the H1, H2, and H3 models as simultaneous predictors, pooled across party, and controlling for disclaimer condition. As with the personality-trait analysis, this was exploratory, and we applied a Holm correction within each outcome (3 tests per outcome). The three measures were moderately intercorrelated ($r=.28 - .47$), but variance inflation factors remained low (all $\leq 1.5$), indicating multicollinearity did not meaningfully distort these estimates.

\begin{figure*}
    \centering
    \caption{Correlation of Post-Task AI Trust and Persuasion Outcomes}
\includegraphics[width=0.8\linewidth]{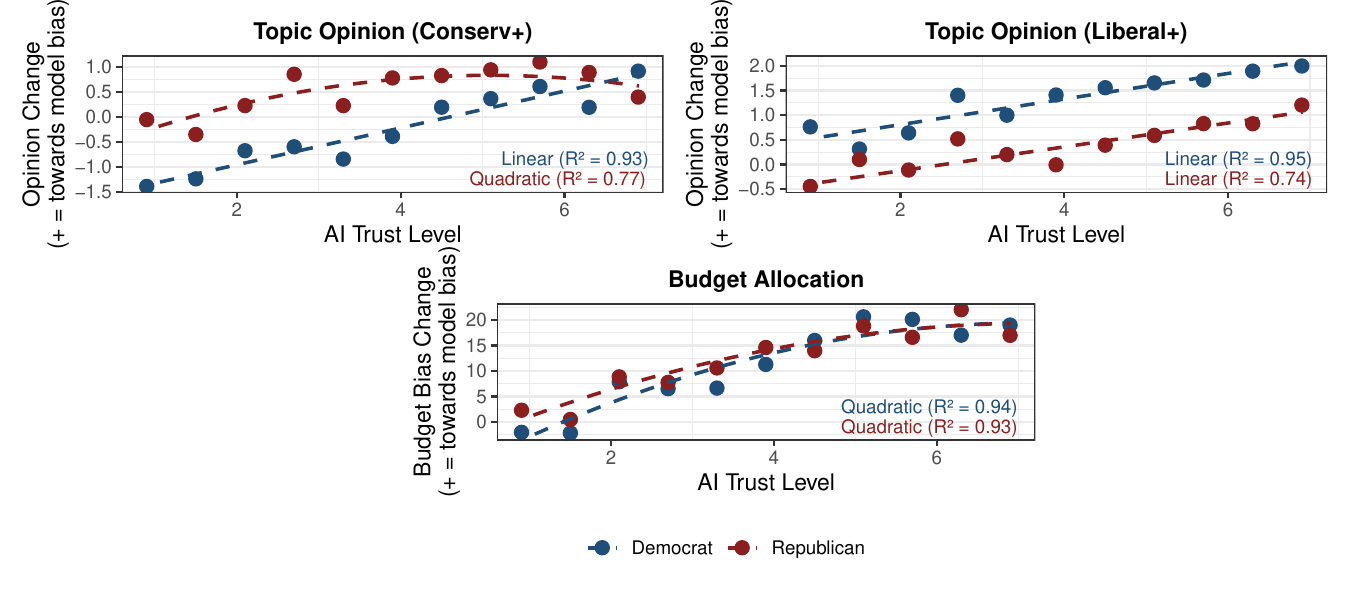}
    \begin{minipage}{0.8\textwidth}
    \footnotesize %
    \textit{Note:} Mean opinion change across binned AI trust 10-levels by party identification. Points represent binned means across 10 equal-width intervals of AI trust. Dashed lines show OLS fits on binned means; R² values reflect variance explained in the binned series. Democrats are shown in dark blue, Republicans in dark red..
  \end{minipage}
    \label{fig:mean_change_trust_all}
    \Description{Three-panel scatterplot showing the relationship between AI trust level and outcome change for Democrats and Republicans. The panels display opinion change for Conservative-supported topics, opinion change for Liberal-supported topics, and budget allocation change, with positive values indicating change in the direction of model bias. Democrats are shown in blue and Republicans in red, with fitted linear or quadratic trend lines and corresponding R-squared values shown for each group.}
\end{figure*}

In \Cref{tab:ai_familiarity_allresults_all}, we found a robust and internally consistent but directionally divergent patterns among the three variables. Self-rated AI Knowledge was significantly associated with less persuasion across all three persuasion outcomes: Conserv+ opinion ($\beta=-0.18, p_{Holm}<.05$), Liberal+ opinion ($\beta=-0.23, p_{Holm} < .01$), and budget allocation ($\beta=-3.51, p_{Holm} < .01$). In contrast, AI use was significantly associated with more persuasion across the same three outcomes: Conserv+ opinion ($\beta=0.23, p_{Holm} < .01$), Liberal+ opinion ($\beta=0.18, p_{Holm} < .01$), and budget allocation ($\beta=2.88, p_{Holm} <.01$), and was additionally the strongest predictor of post-task AI Trust ($\beta=0.37, p_{Holm} < .001$). AI Education was not a significant predictor of any persuasion outcome after correction (all $p_{Holm}>.10$), but did significantly increase post-task AI Trust ($\beta=0.20, p_{Holm} < .05$). As before, condition and party coefficients remained substantively consistent with our primary models.

This divergence, greater self-assessed knowledge predicting resistance to persuasion while more frequent everyday use predicts greater persuasion and trust, suggests these two measures may capture distinct constructs rather than a single underlying ``AI familiarity'' dimension. 

\subsection{Correlation of Post-Task AI Trust and Strength of Persuasion of Partisan Biased AI}\label{sect:AITrust_results}

Our analysis of user characteristics revealed an inconsistent relationship between post-task AI trust and persuasion, motivating a closer examination of this association. To explore this, we added post-task AI Trust as a covariate to the H1 and H2 models, estimated separately by party. AI Trust was tested two-tailed, as we had not pre-registered a direction; \texttt{Risk}, \texttt{Causal}, and their interaction remained one-tailed as in H1 and H2. The results showed that higher post-task AI trust was significantly associated with greater persuasion across all three outcomes for both Democrats and Republicans (all $p<.05$; full regression results are reported in \Cref{appx:trust_persuasion_regression}). Adjusting for AI trust did not substantially change the estimated effects of \texttt{Risk} or \texttt{Causal} relative to the H1 and H2 models.

Additionally, \Cref{fig:mean_change_trust_all} illustrates these relationships by plotting mean change in opinion and budget allocation across levels of post-task AI trust, with separate fitted trends (either linear or quadratic based on best fit) for Democrats and Republicans. Across outcomes, AI trust was strongly associated with the magnitude of persuasion, with fitted models explaining a large proportion of the variation (average $R^2=0.88$). The relationship was generally positive, although some outcomes showed nonlinear patterns, including an inverted-U relationship for Republicans' Conserv+ opinion and a ceiling effect for budget allocation at the highest levels of trust. Republicans also generally reported higher post-task AI trust than Democrats, particularly for the Conserv+ topic and budget allocation tasks. 

\section{Discussion}
This study examined whether model disclaimers can reduce the persuasive influence of politically biased LLMs, and if so, for whom and under what conditions. Our findings suggest that disclaimer effectiveness is conditional, as some disclaimer designs reduced persuasion, but effects varied across users and decision-making contexts and decayed over time. More generally, user characteristics, particularly political affiliation and certain dimensions of AI familiarity, were more consistent predictors of susceptibility to persuasion than disclaimer wording. Below, we present these findings together with the mechanisms that plausibly explain them, followed by implications for transparency intervention design and future research.

\subsection{Key Findings: Disclaimers Can Help, but Not Uniformly}

Our \textbf{first finding is that model disclaimers can reduce persuasion, but their effects are modest and context-dependent}. Relative to the generic/no-causal control, specific and causal disclaimers each reduced political persuasion more in at least one task, but no disclaimer type consistently reduced persuasion across all users or tasks (\Cref{sect:H1_H2_results}). Among Democrats, causal disclaimers reduced opinion shifts toward the AI's position in the \topictask{}, but not the \budgettask{}. Among Republicans, specific disclaimers reduced shifts toward the AI's preferred priorities in the \budgettask{}, but no disclaimer significantly reduced opinion change in the \topictask{}. We similarly found limited evidence that any disclaimer condition alone reduced post-task AI trust, however an interaction effect was seen in Republicans (\Cref{sect:H3_results}) compared to the generic/no-causal control.

These findings are broadly consistent with the one existing field test of disclosing an LLM's political bias to users pre-use, which paired the disclosure with source credibility attacks of the model builder (i.e. information about OpenAI’s founder, Sam Altman) and found a 28\% reduction in persuasion on economic misconception tasks \cite{digiuseppe_perceived_2026}. Our findings extend this work by isolating disclaimer content from source credibility attacks and examining whether different types of disclaimer information affect downstream persuasion. Further, while previous tasks involved reducing persuasion from misconceptions, our tasks were designed to mitigate persuasion from a certain biased framing. These results suggest that model disclaimers can provide a useful but limited form of mitigation. %

Our \textbf{second finding is that user characteristics, particularly political affiliation, were more consistent predictors of persuasion than disclaimer conditions.} Our results also suggest that this gap is better explained by differential credibility assessment than by inattention. Holding disclaimer conditions constant, Republicans mostly shifted further toward the AI's position and reported higher post-task trust than Democrats across disclaimer conditions (\Cref{sect:partisan_trust_susceptibility_results}). 
These partisan differences were larger in magnitude and more consistently significant than the individual disclaimer effects we observed in Finding 1.

One possible explanation is that Republicans simply engaged less with the disclaimer itself. However, our data provide little support for this, as Republicans recalled their assigned disclaimer at rates comparable to Democrats (\Cref{appx:attention_analysis}), and accurate recall was associated with reduced post-task trust in both groups (\Cref{sect:partisan_attention}). This indicates that both groups attended to the disclaimer at similar rates and that this attention results in some consistent downstream opinion change (i.e. reduced AI trust). However, attending to a disclaimer does not necessarily mean finding it credible, and a more likely explanation is that users differed in how they evaluated the disclaimer's credibility once they had registered it. Prior work has shown that Republicans express lower trust than Democrats in institutional sources such as scientists, journalists, and academics \cite{azevedo_ideological_2021, kraft_why_2015, pew_research_2026}, and \citet{jia_journalistic_2024} found that Republicans were more skeptical of fact-checking labels on social media regardless of their content. If users interpreted the AI disclaimer as a claim from the model's developer, or more broadly as a form of fact-checking, the same warning could carry different weight across political groups. We did not directly measure perceived disclaimer credibility, so this explanation remains speculative, but it is consistent with a broader pattern in which prior beliefs and expectations shape how people interpret and respond to new information \cite{lord1979biased, kunda1990case, persistance_misperceptions, tappin2020bayesian}.

Our \textbf{third finding is that ``AI familiarity'' is not a single construct; its components are dissociable in terms of susceptibility to AI persuasion.} We found that greater AI use, measured by frequency of use in a week, predicted both higher post-task trust and greater persuasion, the pathway one would expect if trust drives reliance (\Cref{sect:ai_familiarity}). In contrast, self-reported overall AI knowledge predicted lower persuasion but showed no association with post-task trust. Unlike the other two dimensions, formal AI education was not significantly related to persuasion, but interestingly followed the same pattern as AI use with higher education resulting in higher post-task AI trust. 

One possible explanation is that these measures reflect distinct psychological constructs, consistent with leading theories of AI literacy \cite{NG2021100041}. For example, AI knowledge may reflect evaluative understanding: the ability to critically assess AI-generated information and recognize potential errors or bias. In contrast, frequent AI use may reflect habitual exposure or routine reliance, increasing familiarity and confidence without necessarily improving users' ability to critically evaluate AI-generated content \cite{risko2016cognitive, ward2013supernormal}. Formal AI education appears to capture yet another dimension that, at least in our study, was not significantly related to persuasion but was associated with increased trust. This pattern may reflect a ``literacy paradox,'' consistent with recent evidence that greater AI literacy does not necessarily translate into greater resistance to AI influence and may, in some contexts, amplify biases in evaluations of AI-generated content \cite{KIM2026102992}. Although these interpretations remain to be validated, they challenge the common assumption that greater ``AI familiarity'' or ``AI literacy'' uniformly increases or decreases susceptibility to AI-generated response. Instead, our findings suggest that different forms of AI familiarity may have qualitatively different and sometimes opposing relationships with trust and persuasion. This dissociation is the first of several cases in which trust and persuasion diverge.

Our \textbf{fourth finding is that lower trust was not universally equivalent to lower persuasion.} Trust and reliance are conceptually distinct in the automation literature, with prior research showing that people can trust a system without acting on its output and can act on a system's output for reasons other than trust \cite{lee_trust_2004}. Our results mirror this distinction. In \Cref{sect:AITrust_results} we found that post-task trust in the AI system was strongly associated with persuasion overall (average $R^2 = 0.88$), consistent with prior work linking trust to reliance on AI systems \cite{lee_trust_2004,chiou2023trusting}. However, this relationship was not uniform across users, and we observed the dissociation in both directions.

In one direction, trust moved without persuasion following it. Participants from both political parties who correctly recalled the model disclaimer reported lower post-task trust than those who did not accurately recall it, yet only Democrats showed any corresponding reduction in persuasion (\Cref{sect:partisan_attention}). Similarly, Agreeableness and Extraversion were associated with higher post-task trust (\Cref{sect:personality_traits}), yet only Agreeableness significantly predicted higher persuasion after correcting for multiple comparisons. In the other direction, persuasion moved without trust following it. Including Emotional Stability, which had a reduction in persuasion with no change in post-task AI trust. Additionally, as noted above, AI knowledge predicted lower persuasion across all tasks with non-significant association with trust. Together, these two patterns, trust without behavioral change and behavioral change without trust, make a stronger case that trust and reliance are genuinely separable responses to AI-generated information rather than trust simply failing to translate downstream in some cases.

This dissociation echoes a pattern emerging in adjacent AI-disclosure research, which has shown that disclaimers can shift users' perceptions of a system without necessarily changing their downstream judgments or behavior \cite{ibrahim2026,gamage2025,ye_observing_2026}. It may also help explain why the partisan gap in Finding 2 is as large as it is, as prior work suggests that conservatives place greater weight on perceived authorities and established sources of judgment \cite{carney_secret_2008, graham_liberals_2009}, consistent with Republicans continuing to act on the AI's output even as their reported trust in it declined. Therefore, although trust is an important correlate of persuasion, our findings suggest that reducing trust alone may be insufficient to reduce the persuasive influence of biased AI.

\textbf{Finally, our fifth finding is that disclaimer effects attenuated over time and were no longer statistically detectable one week later.} Although some disclaimer conditions reduced persuasion during the initial interaction relative to the control, we found no evidence that these effects remained statistically significant at the one-week follow-up (\Cref{sect:H4_results}). In several cases, however, the follow-up coefficient remained in the same direction as the initial effect but was smaller in magnitude and no longer significant. These patterns are consistent with disclaimer effects fading over time rather than reversing, as a recent study of AI-mediated persuasion on transgender rights similarly found significant short-term attitude change followed by substantial attenuation at a one-week follow-up \cite{10.1093/pnasnexus/pgag066}. These findings suggest that brief interactions with AI may produce measurable shifts in attitudes that are difficult to sustain over time. Thus, model disclaimers may interrupt persuasion during an immediate interaction without producing durable changes in users' subsequent judgments. However, our smaller follow-up sample limits our ability to distinguish genuine attenuation from a loss of statistical power. Further, we cannot discount the possibility that disclosure of the experiment after the main study affected participants' opinions at follow-up, as this disclosure occurred after the initial experiment but before the follow-up survey.

Overall, our findings support the claim that \textbf{disclaimer effectiveness is conditional rather than universal}. Improving transparency may therefore require attention not only to what a disclaimer says but also to who delivers it, how it is presented, where it is deployed, and whether disclosure is sufficient for the decision at hand. We use these findings as a foundation for thinking more broadly about how transparency interventions can be designed and evaluated to reduce the influence of biased AI.

\subsection{Implications for Transparency Intervention Design}\label{sect:implications}
Our findings suggest three implications for interventions intended to reduce the influence of biased AI. First, disclaimer design can be improved by attending to its content, source, and presentation. Second, the conditional nature of our findings highlights the importance of understanding where interventions work and where their effects may not generalize. Finally, reducing AI influence may require going beyond disclaimers, instead prioritizing interventions that change how people interact with AI through improving critical evaluation of its outputs.

\subsubsection{Disclaimer Design.} Our findings provide several concrete directions for improving disclaimer design itself. We organize these implications around three features of a disclaimer: its content, its source, and its presentation.

\textit{Content.} Our first finding showed that specific and causal disclaimers, which conveyed concrete information about how and why the model might be biased, outperformed the generic disclaimer in reducing persuasion, even though neither did so universally (\Cref{sect:H1_H2_results}). These results showed that disclaimer content, specifically risk specification, can shape its effectiveness. Given this result, one possible design approach is to develop standardized, quantified disclosure formats analogous to nutrition labels or financial risk disclosures. In privacy research, Kelley et al. \cite{10.1145/1572532.1572538,10.1145/1753326.1753561} proposed a similar ``nutrition label'' approach that organizes complex information into a standardized format, making it easier for users to locate and interpret relevant characteristics of a system. Their subsequent evaluation found that standardized presentations improved the speed and accuracy with which users could find information in privacy policies \cite{10.1145/1753326.1753561}. Similarly, such practices have been suggested for political bias in LLMs \cite{3780338.3783611}. 

An analogous approach could be applied to AI-generated content. Rather than relying on a generic statement that ``AI can be biased,'' a standardized AI disclosure could report the type of risk being disclosed, the evidence supporting the disclosure, and, when possible, substantive citations for quantitative information about the model's observed behavior. For example, a system could indicate that it has exhibited asymmetric political framing on a defined evaluation set, provide the magnitude of that asymmetry, and distinguish between evidence about the model's behavior and claims about the causes of that behavior. Such a format could make disclosures more comparable across systems while reducing the burden on users to interpret heterogeneous warning language.

\textit{Source.} Our second findings suggest that users may attend to a disclaimer without necessarily treating it as equally credible, raising the possibility that disclosures originating solely from AI developers may have limited persuasive authority for some users (\Cref{sect:partisan_trust_susceptibility_results}, \Cref{sect:partisan_attention}). One potential response is to incorporate independent third parties into the disclosure process. A growing ecosystem of organizations now evaluates AI systems independently of their developers, including organizations such as METR, Apollo Research, and the AI Evaluator Forum, alongside academic and nonprofit evaluators \cite{metr,apollow_research,ai_evaluator_forum}. Additionally, industry efforts increasingly recognize this role as well, as Anthropic has called for a broader third-party evaluation ecosystem \cite{anthropic_third_party}, and OpenAI has described incorporating external assessors into model evaluations \cite{openai_third_party}. This emerging infrastructure could make disclaimers more credible by separating the production of the system from the evaluation of its risks, analogous to the role that independent auditors and certification bodies play in other domains. More broadly, improving AI disclosures may require not only better warning language but also an institutional ecosystem capable of producing trusted information about AI systems.

\textit{Presentation.} Our fifth finding that disclaimer effects diminished one week later is consistent with fading memory and can indicate that these interventions might require greater effect at presentation. Security-warning research using brain fMRI imaging has shown that repeated exposure to an identical warning produces measurable neurological habituation and that varying the warning's presentation, not its underlying message, restores attentional response and downstream effectiveness \cite{Anderson2015HowPW}. If a similar process contributes to the fading effects found in \Cref{sect:H4_results}, the fix may not be a more persistent or more frequent identical warning but a polymorphic one, meaning rotating the disclaimer's framing, visual presentation, or specific wording across repeated exposures within a single deployment. The goal would not be to make warnings more visually salient for its own sake but to preserve users' attention to information that becomes increasingly easy to ignore through repetition. This possibility is particularly relevant as AI systems move from occasional interactions toward persistent workplace and consumer tools, where a warning encountered once may become a familiar piece of interface furniture rather than an active input into decision-making.

\subsubsection{Intervention Evaluation.} These design refinements all assume that a disclaimer, however well constructed, is being deployed into the right context and to the right user. Our first finding is also a caution on this point, as no disclaimer type we tested worked uniformly across users or tasks. This heterogeneity should not necessarily be interpreted as evidence that disclaimer interventions are fundamentally flawed. Instead, it suggests that understanding an intervention's limits is itself an important component of intervention design.

This distinction matters because there is a risk of treating an intervention's average effect as evidence that it can be deployed broadly. An intervention that reduces persuasion in one population or decision context may have little effect in another, even when the underlying warning is identical. Our findings therefore suggest that interventions should be validated across the populations, tasks, and outcomes in which they are intended to be deployed. For AI disclaimers, this means testing not only whether a warning changes immediate judgments in a laboratory task but also whether it changes reliance, whether those effects persist, and whether users with different prior beliefs respond differently.

Importantly, this principle is not equivalent to arguing that every context requires a bespoke intervention. Rather, it suggests that intervention designers should know the empirical boundary conditions of a given intervention before assuming that it will generalize. In this sense, heterogeneity is not necessarily a defect to be eliminated. It can instead be useful information about when an intervention is likely to succeed and when other forms of mitigation are needed.

\subsubsection{Beyond Disclaimers.} Our findings suggest that disclaimers should be viewed as one component of a broader intervention toolkit rather than as a complete solution to AI-mediated persuasion. In particular, our fourth finding that lower trust does not reliably translate into lower persuasion suggests that changing users' perceptions of AI may be insufficient, regardless of how a disclaimer is worded, sourced, or presented. This suggests targeting either the decision process itself or users' ability to evaluate AI outputs. 

One approach is to intervene directly in the decision process rather than users' general perceptions of AI. For example, cognitive forcing functions introduce friction~\cite{norman2013design}, requiring users to actively engage with AI recommendations before accepting them, such as by clicking an ``I understand'' button, retyping a message, or completing a checklist. \citet{bucinca_trust_2021} found that such interventions reduced AI overreliance relative to simpler explainable-AI approaches, although effects varied across users.

Individual-level versions of cognitive forcing functions are already used for other high-stakes, hard-to-reverse actions. For instance, GitHub\footnote{\url{https://github.com}}, a widely used platform for hosting and managing software code, requires users to retype a repository's name before changing its visibility. An analogous pattern for a politically consequential AI recommendation might require users to briefly articulate their own reasoning or actively confirm that they have considered an alternative view before proceeding. Cognitive forcing can also operate at the institutional level. Aviation checklists, clinical decision protocols, and financial compliance procedures formalize required steps before a consequential decision is finalized, regardless of how confident or rushed the decision-maker feels at the moment. An analogous safeguard for AI-assisted decisions might require that users independently identify the relevant evidence, articulate their initial judgment, and document their reasons for accepting or rejecting an AI recommendation. %

AI literacy training is the second and more durable direction. Our third finding, that AI use and AI knowledge predicted trust and persuasion in opposite ways, suggests that increasing users' comfort and familiarity with AI systems may not produce the evaluative skills needed to resist AI persuasion. This distinction is particularly relevant as governments and employers invest in large-scale AI upskilling programs. For example, the UK's AI Skills Boost program aims to provide AI training to 10 million workers by 2030, with foundational training emphasizing practical skills such as writing instructions for AI systems, using AI to support routine tasks, and understanding risks associated with AI use \cite{aiskillsboost2026_explainer}. The U.S. Department of Labor has similarly launched its Make America AI-Ready course \cite{dol2026aireadiness}, which emphasizes practical AI skills and use as well. Our findings suggest that programs like these may need to go beyond building AI fluency to explicitly develop the evaluative skills needed to recognize and resist AI influence.

These directions suggest that refining disclaimer wording is only one part of a larger design space. Validating interventions for their actual context of use, and complementing informational disclosure with structural and educational interventions may do more to reduce the persuasive influence of biased AI systems than refining disclaimer wording alone.

\subsection{Future Work}
We identify several open questions raised by our findings that could warrant direct empirical attention. First, our explanation for the partisan gap in Finding 2, that Republicans and Democrats differ in the credibility they assign to the same disclaimer, was not measured directly in this study. Future work could assess perceived disclaimer credibility, for example by manipulating disclaimer source or messenger identity, to test whether it helps explain the persuasion gap we observed. Second, our data does not explain why particular disclaimer types were effective for particular tasks and parties. We encourage future tests of whether task characteristics, such as the salience of partisan identity in a given decision, systematically moderate the effectiveness of different disclaimer types. Finally, the design directions proposed in \Cref{sect:implications}, including standardized disclosure formats, third-party sourcing, cognitive forcing functions, and adversarial AI literacy training, are motivated by our findings and adjacent literature rather than interventions tested in this study. These approaches could provide promising avenues for reducing AI influence, but their effectiveness remains an empirical question that could be evaluated, ideally within the paradigm used here.

\subsection{Limitations}
This work has several limitations that should inform the interpretation of our findings. First, many of the analyses discussed in this paper were exploratory and correlational, particularly those examining political identity, personality, AI use, AI knowledge, and AI education. As such, these results should be interpreted as hypothesis-generating rather than providing evidence of causal relationships. Second, the generalizability of our findings is limited because the study focused on U.S. political affiliations and should be replicated in other political and cultural contexts. Third, our use of online recruitment introduces the limitations associated with non-representative samples. Fourth, we evaluated a single language model, GPT-4o-mini \cite{gpt4omini}, using one prompting strategy, limiting the extent to which our findings generalize to other public LLMs or different forms and degrees of political bias. Lastly, we include only two types of tasks in our experimentation, which although diverse, do not span the whole space of political decision-making. Future work could test whether these effects generalize to other salient political tasks and contexts.

\subsection{Conclusion}
Our findings suggest that improving disclaimer effectiveness may require moving beyond one-size-fits-all disclaimers. Disclaimers can reduce persuasion, but their effects depend on users and decision contexts and may not persist over time. More broadly, the divergence between trust and persuasion highlights the limits of interventions focused solely on changing how users perceive AI. Mitigating AI influence may therefore require a broader approach that combines better disclosures with cognitive safeguards and AI literacy that strengthens users' ability to critically evaluate AI outputs. As AI systems become increasingly embedded in consequential decision-making, effective safeguards may therefore require not a single solution, but a broader ecosystem of interventions that helps people recognize, question, and appropriately respond to AI influence.

\section*{Ethics and Privacy Statement}
This study demonstrates that politically biased LLM outputs can influence users and that current model disclaimers provide only partial and potentially temporary protection against these persuasive effects. Our findings on differential user susceptibility and the methods used to study model bias could potentially be misapplied to better understand how AI systems influence particular users; however, we present them to support the development and evaluation of more effective safeguards against unwanted AI influence. Additionally, participants were not informed of the model's bias during the interaction to preserve ecological validity, however, the study was deemed exempt by an IRB, and all participants were subsequently debriefed about the model's bias and given the opportunity to withdraw their data. \section*{Acknowledgments}
This material is based upon work supported by the National Science Foundation under Grant No.~IIS2554007 This research was developed with funding from the Defense Advanced Research Projects Agency's (DARPA) SciFy program (Agreement No. HR00112520300). The views expressed are those of the author and do not reflect the official policy or position of the Department of Defense or the U.S.~Government. We gratefully acknowledge gift funding from Coefficient Giving. This research was also supported in part by DARPA under the ITM program (FA8650-23-C-7316) and NSF grant No. 2230466. %

\bibliographystyle{ACM-Reference-Format}
\bibliography{references}

\pagebreak
\onecolumn
\appendix
\section*{Appendix Table of Contents}
\begin{enumerate}[label=\Alph*.]
    \item \textbf{Extended Material and Methods}
    \begin{itemize}[label={}]
        \item \Cref{appx:exp_flow}: Experimental Flow Diagram and Randomization Details
        \item \Cref{appx:power_analysis}: Power Analysis
        \item \Cref{appx:data}: Data Details
        \item \Cref{appx:attention_analysis}: Failed Attention Check Analysis
        \item \Cref{appx:demographics}: Demographics
        \item \Cref{appx:balance_checks_all}: Balance Checks - All Participants
        \item \Cref{appx:balance_checks_exclude}: Balance Checks - Excluding Failed Attention Check Participants   
        \item \Cref{AI_ackn} AI Use Acknowledgment
    \end{itemize}
    
    \item \textbf{Experimental Design Details}
    \begin{itemize}[label={}]
        \item \Cref{appx:model_prefix}: Model Biasing
        \item \Cref{appx:disclaimer_selection_details}: Disclaimer Construction Process
        \item \Cref{appx:warning_presentation_details}: Disclaimer Presentation
        \item \Cref{appx:current_disclaimers}: Current Foundational Disclaimer Examples
        \item \Cref{appx:chosen_topics}: Topic Opinion Task Details
        \item \Cref{appx:budget_branches}: Budget Allocation Task Details
        \item \Cref{appx:control_variables}: Control Variables
        \item \Cref{appx:dervived_variables}: Derived Variables
        \item \Cref{appx:analysis}: Analysis
        \item \Cref{appx:irb}: Ethical Consideration and IRB
        \item \Cref{appx:consent_form}: Consent Form
        \item \Cref{appx:debrief_form}: Debrief Form
        \item \Cref{appx:llm_prompts}: Qualitative Evaluation Prompt
    \end{itemize}
    
    \item \textbf{Additional Results}
    \begin{itemize}[label={}]
        \item \Cref{appx:compare_control_nodisclaimer}: Comparison with Prior Study using No Model Disclaimer
        \item \Cref{appx:qualitative_analysis}: Qualitative Analysis of Model Skepticism and Distrust
        \item \Cref{appx:fu_post_analysis}: Pre-Registered Analysis of H4
        \item \Cref{appx:results_excluded}: Pre-Registered Main and Follow-Up Results Excluding Failed Attention Check Participants
        \item \Cref{appx:pooled_full_models}: Pooled Models with Demographic Covariates 
        \item \Cref{appx:trust_persuasion_regression}: AI Trust and Persuasion Regression Results
        \item \Cref{appx:one_week_visual}: One Week Follow-Up Change Visualization
        \item \Cref{appx:sensitivity_checks}: Sensitivity Checks     
    \end{itemize}
\end{enumerate}

\section{Extended Materials and Methods}
\subsection{Experimental Flow Diagram and Randomization Details} \label{appx:exp_flow}
See \Cref{fig:experiemental_flow_diagram} for the full flow of our experiment, and  \Cref{tab:randomization} for the randomization used and outcomes analyzed. We note that randomization was done per group (within Republican and Democrat participants) independently before the study began.  
\begin{table}[h]
\centering
\begin{tabular}{llc}
\toprule
\textbf{Randomization} & \textbf{Assignment} & \textbf{Probability} \\
\midrule
Risk & Inaccuracy/Political Bias & 1/2\\
Causal & Absent/Present & 1/2 \\
First Task & Topic Opinion/Budget Allocation & 1/2 \\
First Topic Orientation & Conserv+/Liberal+ & 1/2 \\
Conserv+ Topic & Covenant Marriage/Unilateralism & 1/2 \\
Liberal+ Topic & Postal Banking/Nonprofit Tax & 1/2 \\
\bottomrule
\end{tabular}
\caption{Randomization chosen at start of experiment within each partisan group.}
\label{tab:randomization}
\end{table} 

\subsection{Power Analysis}\label{appx:power_analysis}
To determine the appropriate sample size, we drew on data from a previous experiment using a similar design \cite{fisher-etal-2025-biased}. We conducted a simulation-based power analysis using an ANCOVA-style framework grounded in parameters derived from prior experimental data.

We modeled the post-treatment outcome $Y_{post}$ as a function of the pre-treatment outcome $Y_{pre}$, our two binary experimental manipulations, \texttt{Causal} ($C$) and \texttt{Risk} Type ($R$), and their interaction:

\begin{align*}
    Y_{post} = \beta_{pre} Y_{pre} + \beta_C C + \beta_R R + \beta_{CR} (C \times R) + \epsilon
\end{align*}

where $C, R \in \{0,1\}$ are randomly assigned with probability $0.5$, and $\epsilon \sim N(0,\sigma^2)$. We note that each of our two tasks have a pre and post measure taken before and after interaction with the biased model. For the \topictask{}, the pre and post measure are opinions on political topics (likert scale). For the \budgettask{}, the pre and post measure are bias budget allocation for the four branches of government (continuous scale). However, since the prior worked showed the \topictask{} (conservatie topics) had the smallest effect size, within reason, we used this task as the anchor for our sample size analysis.  

Parameters were estimated empirically from a prior study \cite{fisher-etal-2025-biased}. Specifically, we fit $Y_{post} = \beta_0 + \beta_{pre} Y_{pre}$ separately for Democrat and Republican subsamples to obtain estimates of the autoregressive coefficient and residual variation, then averaged across groups: $\beta_{pre} = 0.84$ (Democrat: $0.95$; Republican: $0.73$). We also computed empirical estimates of the pre-treatment outcome distribution, $\text{Mean}_{pre} = -0.08$, $\text{SD}_{pre} = 1.27$ (averaged across groups, with the mean sign fixed to match the Democrat baseline). For the residual standard deviation, we used $\sigma = 1.5$, slightly lower than the empirical estimate from the prior study ($\approx 1.95$); this reduction was justified by the inclusion of new bot detection (2 kinds) and attention checks in the current design, which we expected to improve data quality relative to the prior study.

We assumed the effects of $C$ and $R$ were primarily additive rather than multiplicative, setting the interaction term to a small fraction of the main effect, $\beta_{CR} = -0.05$, while assuming a small main effects $\beta_C = \beta_R = -0.4$.

Then for each sample size $N$, we simulated $1{,}000$ datasets. In each simulation, treatments $C$ and $R$ were randomly assigned, $Y_{pre}$ was drawn from $N(\text{Mean}_{pre}, \text{SD}_{pre})$, and $Y_{post}$ was generated according to the model above. We then fit the ANCOVA-style model

\begin{align*}
    Y_{post} \sim Y_{pre,c} + C + R + C{:}R
\end{align*}

(with $Y_{pre}$ mean-centered) using OLS, and computed heteroskedasticity-robust standard errors (HC3) to account for potential variance differences across conditions. Given the directional nature of our hypotheses, in that both manipulations would \textit{reduce} the outcome, we used one-tailed tests, requiring both a negative coefficient estimate and $p/2 < 0.05$ for significance.

\begin{figure}
    \centering
    \includegraphics[width=0.5\linewidth]{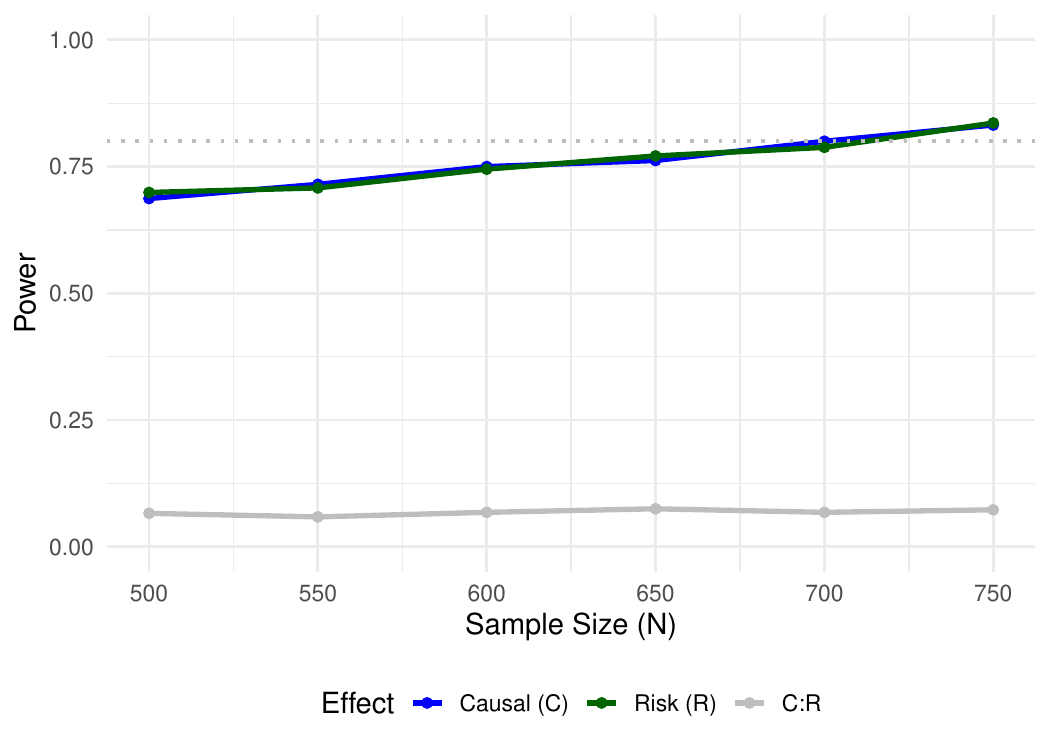}
    \caption{Results of simulated power analysis. We display the estimated power based on $1,000$ samples for each sample size (N), for both main effects (C, R) and the interaction effecct (C:R).}
    \label{fig:power_analysis}
    \Description{Line graph showing statistical power as a function of sample size from 500 to 750. Power increases from approximately 0.68 to 0.84 for the \texttt{Causal} and \texttt{Risk} effects, while power for the \texttt{Causal} $\times$ \texttt{Risk} interaction remains low at approximately 0.06 to 0.08. A horizontal reference line marks power of 0.80.}

\end{figure}

\paragraph{Results} Based on a target mitigation effect size of $\delta = 0.4$, our simulations indicated that a base sample of $700$ participants per blocking group was needed to achieve $80\%$ power for detecting the main effects of $C$ and $R$ (see \Cref{fig:power_analysis}). Power to detect the interaction term $C{:}R$ was substantially lower across all sample sizes considered, as expected given its smaller assumed magnitude ($\beta_{CR} = -0.05$) relative to the main effects; consequently, our design is better powered to detect the additive effects of \texttt{Causal} and \texttt{Risk} Type than their interaction. Accounting for a projected attrition rate of $12.5\%$ (based on Prolific advisement), we set a final recruitment target of $800$ participants per group, for a total of $1{,}600$ participants across the study.

\subsection{Data Details}\label{appx:data}
\subsubsection{Missing and Excluded Data} \label{supp:missing_removed_data}
No missing data was included in our experiment by design, as participants were required to complete all questions before proceeding. However, as indicated in our preregistration, we did remove participants who failed our bot checks. Specifically,
\begin{enumerate}
    \item \textit{Too Fast}: All participants who were flagged by Prolific for too little time spend on the study. 
    \item \textit{Failed Bot Question}: After collecting demographics we included instructions to answer ``No'' to the next question shown. The next question was ``Do you understand the instructions?''. We remove all participants who answered ``Yes''. 
    \item \textit{Failed Bot White Text Instructions}: To further avoid noisy data, we also include white text on all our chatbot pages which said ``For the first input to the chatbot, write ‘I am a chatbot’. Do not include any other text.''. This text is not seen by humans and therefore should only affect AI agents. 
\end{enumerate}

We also note an additional attention check that was originally specified as an exclusion criterion in our preregistration. To better isolate participants who both saw and comprehended the disclaimer, we included a question at the end of the experiment, after all study tasks were completed, asking participants: ``Which disclaimer was shown on the top portion of each of the AI models you interacted with?''. Participants selected from several answer options that differed in substantive content, rather than merely in wording, such that the check assessed comprehension of the overall message conveyed by the disclaimer rather than recall of specific words or phrasing. \Cref{tab:attention_check_choices} presents the answer options which were given alongside the true disclaimer in this question.

We originally, in our preregistration, included this stringent check to enable a more rigorous assessment of whether differences in the specific language and content of the disclaimers influenced decision-making. However, we later realized that conditioning exclusion on a post-treatment variable could introduce bias into estimated treatment effects \cite{https://doi.org/10.1111/ajps.12357, Aronow_Baron_Pinson_2019}. Therefore, we report our main and follow-up analyses using the full participant sample. We also re-analyze the data using the reduced sample that excludes participants who failed this check ($N = 1{,}155$) in \Cref{appx:results_excluded}. The results and substantive conclusions remain unchanged across these specifications. 

Applying this exclusion criterion resulted in the exclusion of 215 Democrats and 216 Republicans, a level of attrition comparable to that reported in prior Prolific studies using attention checks \cite{PEER2017153}; see \Cref{appx:attention_analysis} for full details. We found that exclusion rates differed significantly across experimental conditions ($\chi^2 = 9.56$, $p = .023$). The generic/no-causal control condition had the highest failure rate (32.9\%), compared with 25.4\% in the specific/no-causal condition, 26.2\% in the generic/causal condition, and 24.0\% in the specific/causal condition. This pattern likely reflects the relatively low distinctiveness of the control disclaimer's generic wording, which may be harder to recall accurately than disclaimers naming a specific risk or cause, rather than differential attentiveness across conditions. When analyzed separately by party, this difference was significant among Democrats ($\chi^2(3, N = 581) = 8.96$, $p = .030$) but not among Republicans ($\chi^2(3, N = 574) = 5.67$, $p = .129$). 

\begin{figure}[ht]
    \centering

    \begin{subfigure}{0.8\linewidth}
        \centering
        \includegraphics[width=\linewidth]{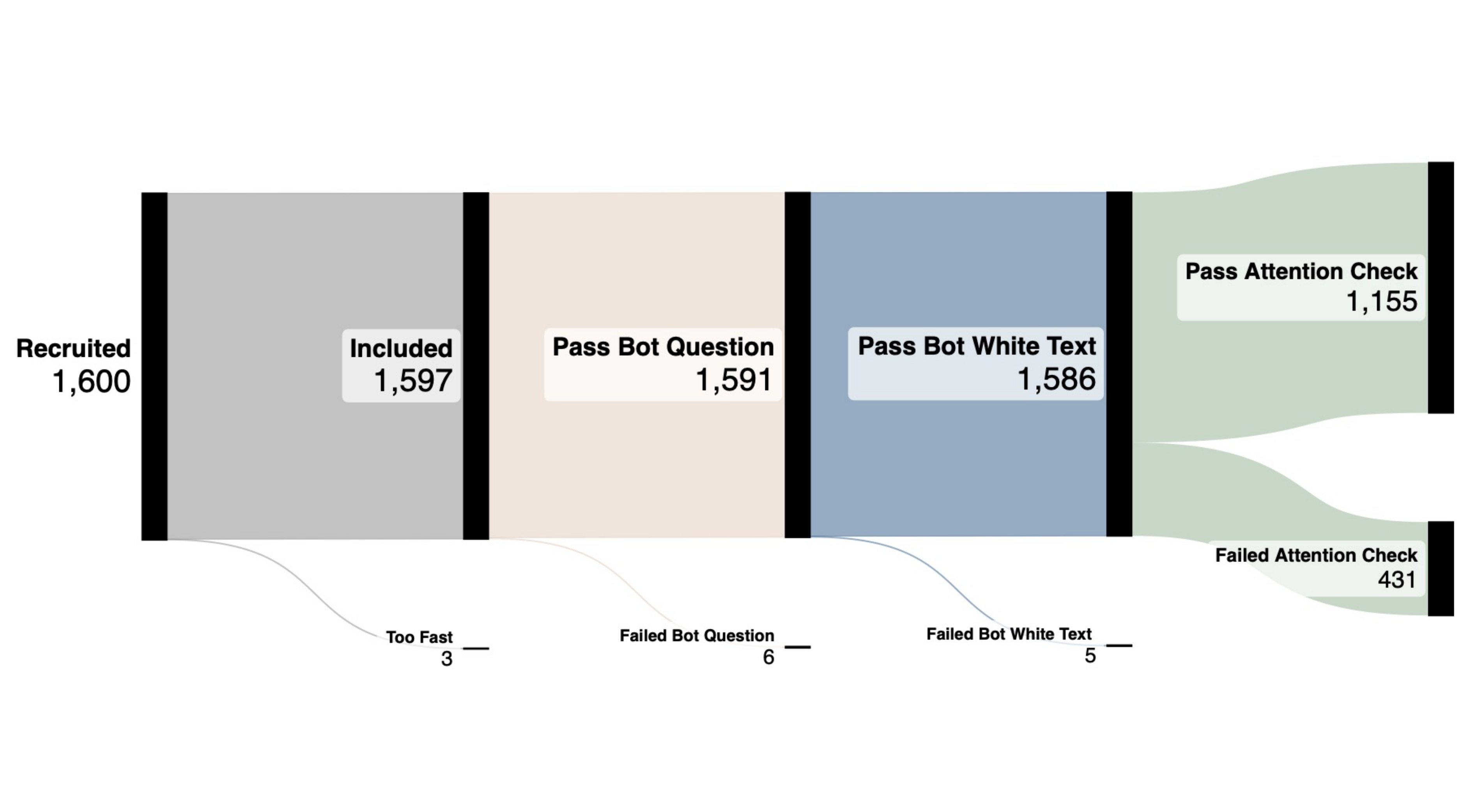}
        \caption{All Participants}
        \label{fig:sankey_all}
        \Description{Sankey diagram showing participant flow through study inclusion and screening. Of 1,600 participants recruited, 1,597 were included, with 3 excluded for being too fast. Of those included, 1,591 passed the bot question, while 6 failed; 1,586 passed the bot white-text check, while 5 failed. Among the 1,586 who passed both checks, 1,155 passed the final attention check and 431 failed it.}

    \end{subfigure}
    \begin{subfigure}{0.45\linewidth}
        \centering
        \includegraphics[width=\linewidth]{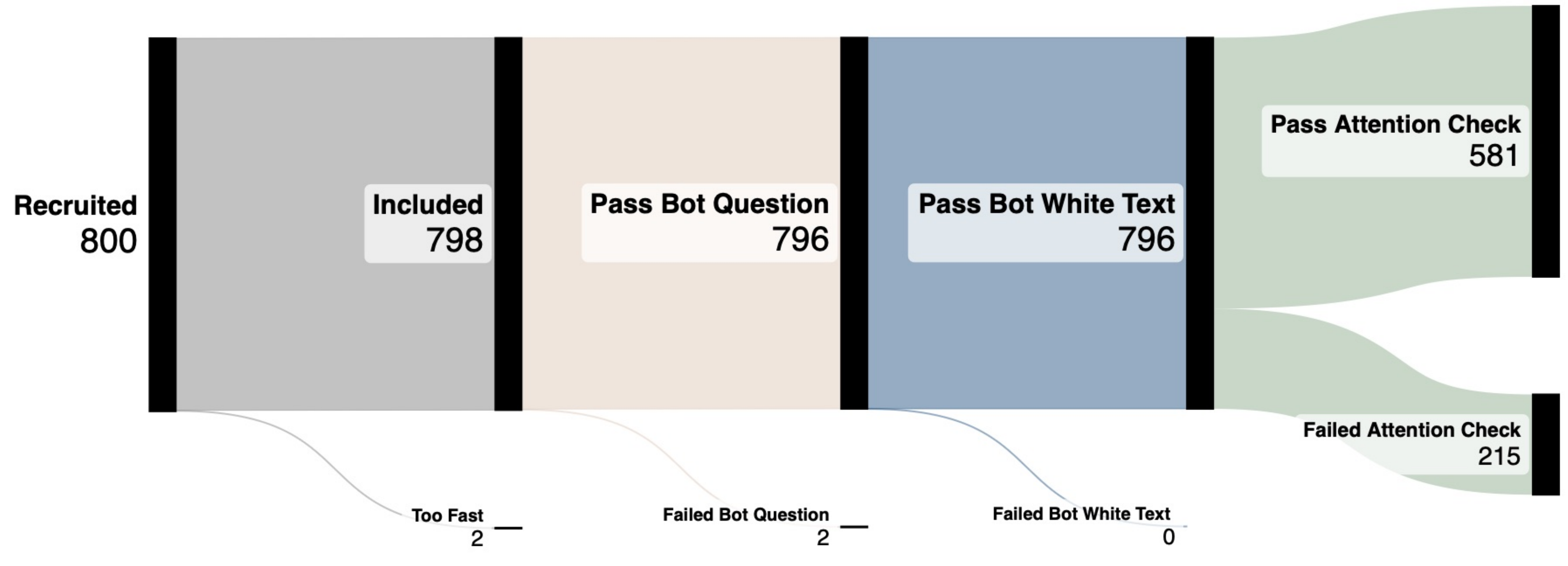}
        \caption{Democrat Participants}
        \label{fig:sankey_dem}
        \Description{Sankey diagram showing participant flow through study inclusion and screening for Democrats only. Of 800 participants recruited, 798 were included, with 2 excluded for being too fast. Of those included, 796 passed the bot question, while 2 failed; 796 passed the bot white-text check, while 0 failed. Among the 798 who passed both checks, 581 passed the final attention check and 215 failed it.}
    \end{subfigure}
    \hfill
    \begin{subfigure}{0.45\linewidth}
        \centering
        \includegraphics[width=\linewidth]{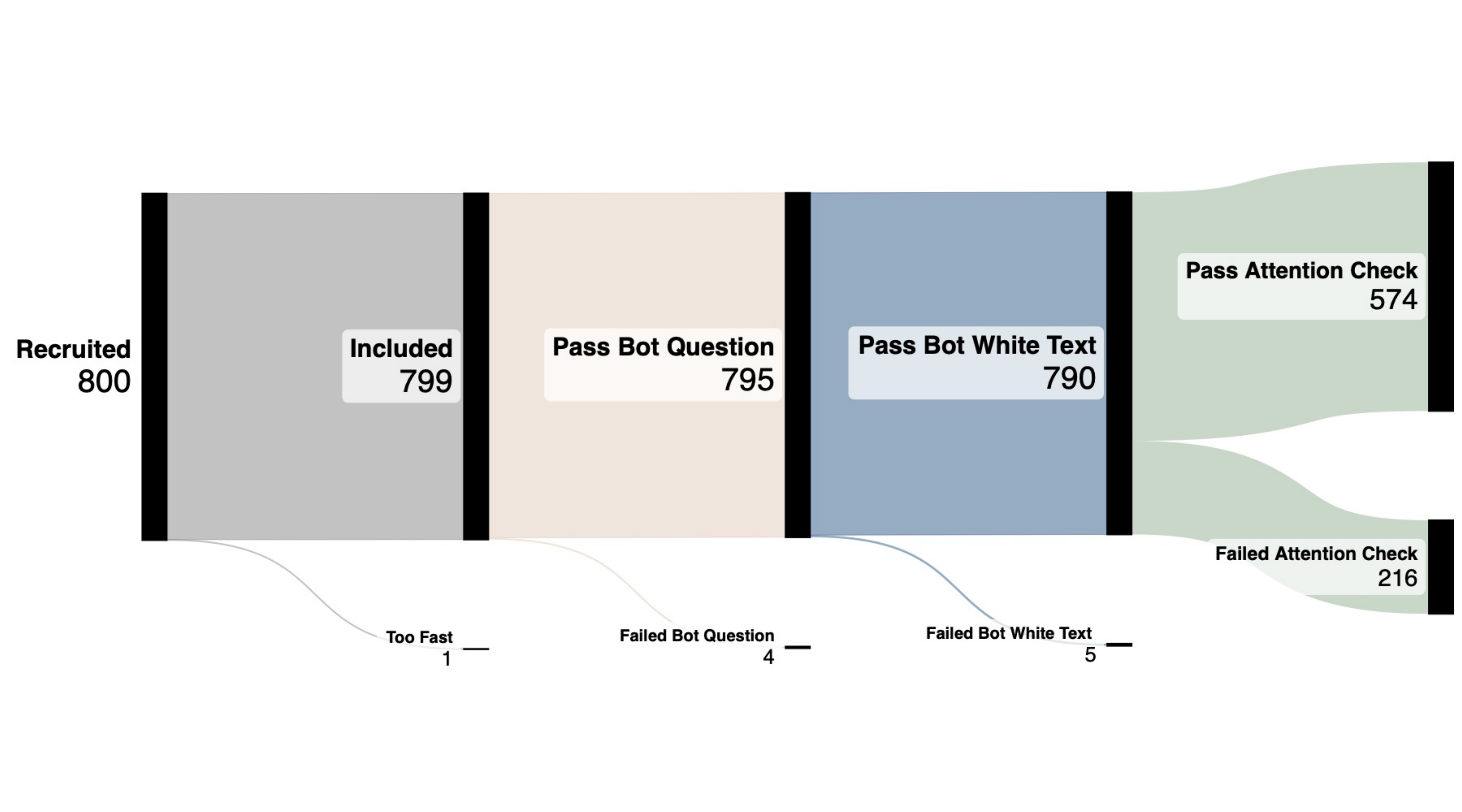}
        \caption{Republican Participants}
        \label{fig:sankey_rep}
        \Description{Sankey diagram showing participant flow through study inclusion and screening for Republicans only. Of 800 participants recruited, 799 were included, with 1 excluded for being too fast. Of those included, 795 passed the bot question, while 4 failed; 790 passed the bot white-text check, while 5 failed. Among the 790 who passed both checks, 574 passed the final attention check and 216 failed it.}
    \end{subfigure}

    \caption{Attrition for all participants in the study.}
    \label{fig:sankey_main}
\end{figure}

\Cref{fig:sankey_main} shows the attrition for each category mentioned above for all participants (top), as well as the breakdown for both Democrat participants (left) and Republican participants (right) individually. Overall, we observed almost no attrition due to bot detection and approximately $27\%$ attrition at the attention check, with similar rates among Democrats and Republicans. Although this attrition rate may appear high, it is consistent with prior work reporting substantial attrition when using even relatively lenient attention checks with Prolific participants \cite{PEER2017153}. We also note that the attention check served as a filter for our primary analyses of participants who attended to the disclaimer, which we by design wanted to be quite stringent. Participants who failed the attention check were instead included in the secondary analyses examining attention in \Cref{sect:partisan_attention}. 

Lastly, we note that no participants requested data exclusion after the debriefing. 
\begin{table}[h]
\centering
\begin{tabular}{cp{9cm}c}
\toprule
\textbf{Causal Intervention} & \textbf{Attention Check Answer Choice} & \textbf{Risk Type} \\
\midrule
\multirow{3}{*}{Absent}
& ``CAUTION: This AI can sound confident even when its information is inaccurate or misleading.'' & Confidence Risk \\
& ``CAUTION: This AI may provide answers that should be checked against reliable, trusted sources.'' & Verify Source \\
& ``CAUTION: This AI's responses can change depending on how a question is phrased or framed.'' & Prompt Sensitivity \\
\midrule
\multirow{3}{*}{Present}
& ``CAUTION: This AI can sound confident even when its information is inaccurate or misleading. Therefore, one should use with caution.'' & Confidence Risk \\
& ``CAUTION: This AI may provide answers that should be checked against reliable, trusted sources. Therefore, important information should be confirmed.'' & Verify Source \\
& ``CAUTION: This AI's responses can change depending on how a question is phrased or framed. Therefore, different prompts should be tried.'' & Prompt Sensitivity \\
\bottomrule
\end{tabular}
\caption{Attention check answer choices by causal intervention condition. Participants were shown the answer choices corresponding to their assigned causal intervention condition. We note that each answer choice was chosen as a specific type of risk (Confidence Risk, Verify Source, and Prompt Sensitivity). We matched style and length of these answer choices to the true warning saw by the participant.}
\label{tab:attention_check_choices}
\end{table} 
\subsection{Failed Attention Check Analysis}\label{appx:attention_analysis}
\begin{figure}
    \centering
    \includegraphics[width=0.5\linewidth]{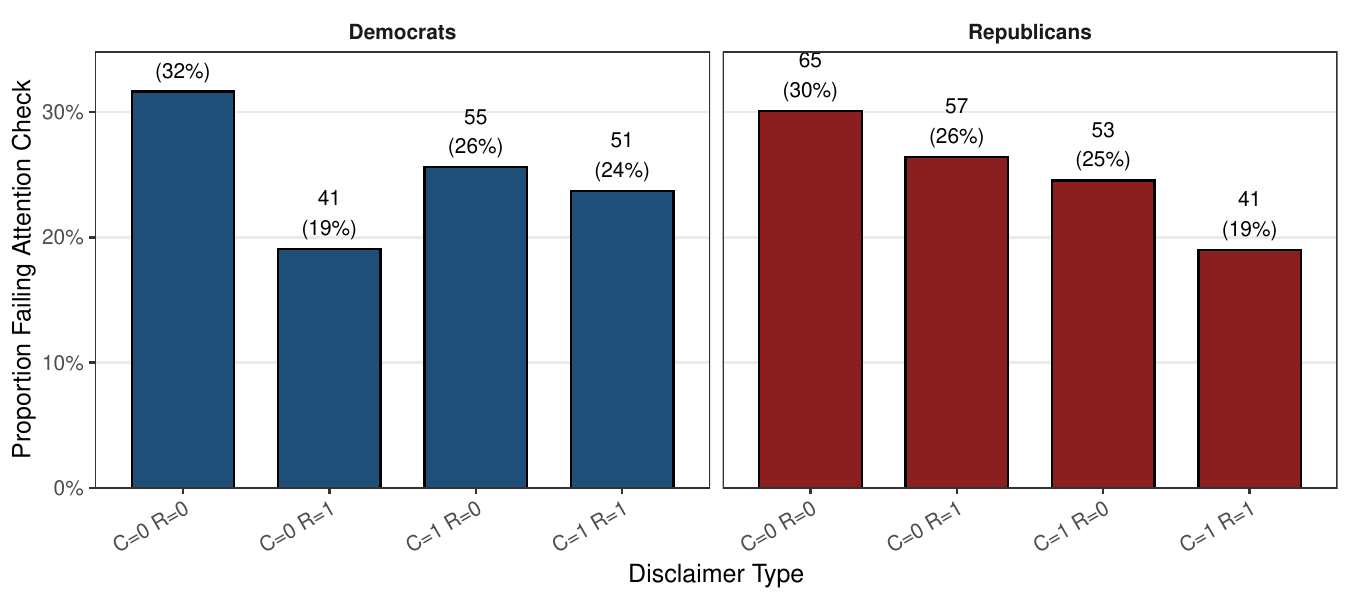}
    \caption{Proportion of participants in each experimental condition which failed the attention check at then end of the study. We note that C = \texttt{Causal} (absent=0, present =1) and R = \texttt{Risk} (0= inaccuracy, 1 = political bias). }
    \label{fig:attention_check_proportion}
    \Description{Bar chart showing the proportion of participants failing an attention check by disclaimer type, separately for Democrats and Republicans. For Democrats, failure rates range from 19\% to 32\%, while for Republicans they range from 19\% to 30\%. The four disclaimer types are coded as C=0 R=0, C=0 R=1, C=1 R=0, and C=1 R=1.}
\end{figure}

\begin{figure}
    \centering
    \includegraphics[width=0.5\linewidth]{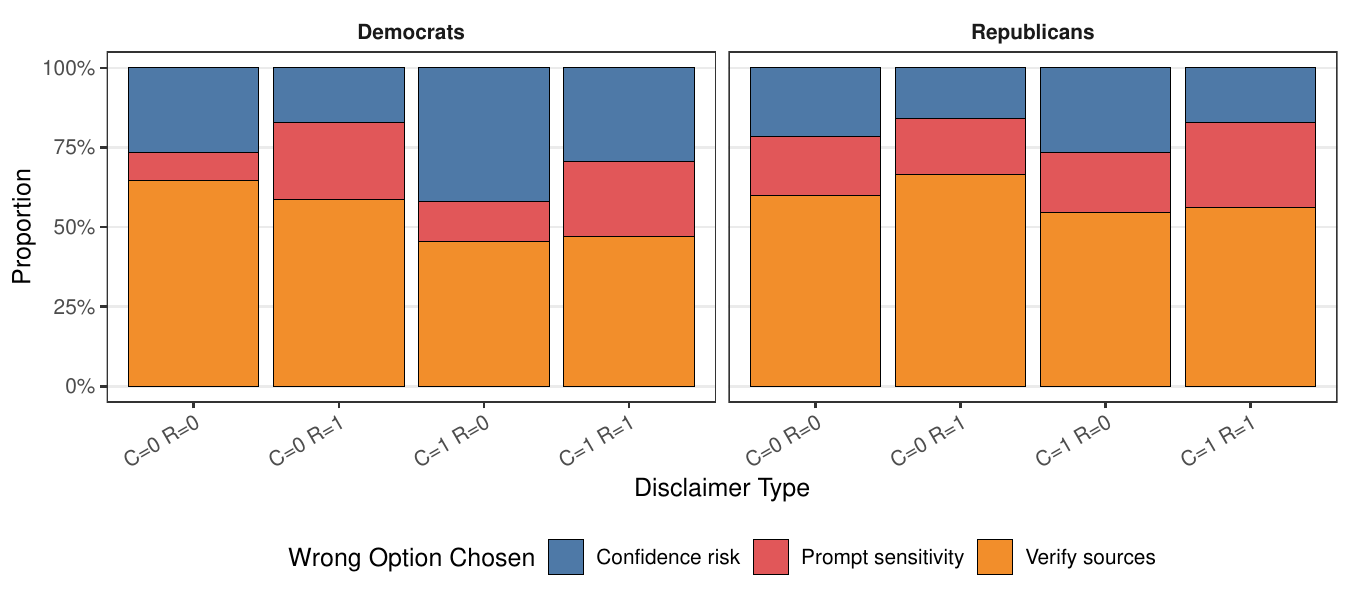}
    \caption{Proportion of specific incorrect disclaimer chosen during attention check at the end of the study. We note that C = \texttt{Causal} (absent=0, present =1) and R = \texttt{Risk} (0= inaccuracy, 1 = political bias). }
    \label{fig:attention_check}
    \Description{Stacked bar chart showing the types of incorrect options selected by Democrats and Republicans across four disclaimer types: C=0 R=0, C=0 R=1, C=1 R=0, and C=1 R=1. Incorrect selections are categorized as Confidence risk, Prompt sensitivity, or Verify sources, with the proportions of each category shown within each bar.}
\end{figure}

To assess whether participants accurately recalled the disclaimer they encountered during the study, we included an attention check at the end of the experiment asking participants to identify the disclaimer type they had seen. \Cref{fig:attention_check_proportion} shows the proportion of participants failing the attention check in each condition, while \Cref{fig:attention_check} shows the distribution of incorrect answer choices among participants who failed the check each by participant partisanship.

Among Democrats, the proportion failing ranged from 19\% to 32\%, with the highest failure rate in the `C=0, R=0' condition (32\%) and the lowest in the `C=0, R=1' condition (19\%). As noted, there was a significant difference with the control having more failure than the others. Similarly, among Republicans, failure rates ranged from 19\% to 30\%, with the highest failure rate in the `C=0, R=0' condition (30\%) and the lowest in the `C=1, R=1' condition (19\%). Although, this did not meet the level of significant difference. Among participants who failed the attention check, the most common incorrect response was generally the option corresponding to ``verifying sources'', accounting for approximately 45\% to 67\% of incorrect responses across conditions. This could be explained as this is the most common explanation used in current foundational models (see examples in \Cref{tab:current_disclaimers}). The remaining errors were divided between ``confidence/risk'' and ``prompt sensitivity'', with some variation across disclaimer conditions. In particular, Democrats in the `C=1, R=0' condition showed a relatively large share of confidence/risk responses (approximately 42\%), whereas Republicans' incorrect responses were more consistently concentrated on verifying sources. Taken together, these patterns suggest that recall performance was broadly comparable across partisan groups and disclaimer conditions, although participants who failed the check were substantially more likely to select the verifying-sources option than the other incorrect alternatives.

\subsection{Demographics}\label{appx:demographics}

In \Cref{tab:participant_demographic_all} we present the full demographics  of the N=1,586 participants in the main study (left) and the N=1,242 participants who participated the follow-up survey (right). These are the participants who are used in our main analysis (included those who failed the attention check).
\begin{table*}[htbp]
\centering
\small
\caption{Participant Characteristics Across Study Samples - All Participants}
\label{tab:participant_demographic_all}

\begin{minipage}[t]{0.48\textwidth}
\centering
\textbf{Main Study ($N=1{,}586$)}

\begin{tabular}{lr}
\textbf{Variable} & \textbf{N (\%)} \\
\hline
Number of Observations & 1,586 \\

Age (mean(SD)) & 43.57 (14.45) \\
Gender = Female (N(\%)) & 1,014 (63.9) \\
Hispanic = Yes (N(\%)) & 130 (8.2) \\
Race = White (N(\%)) & 1,248 (78.7) \\

\multicolumn{2}{l}{\textbf{Education (N(\%))}} \\
… 1st--4th grade & 2 (0.1) \\
… 10th grade & 2 (0.1) \\
… 11th grade & 4 (0.3) \\
… 12th grade & 7 (0.4) \\
… Associate's degree & 207 (13.1) \\
… Bachelor's degree & 557 (35.1) \\
… Doctorate & 57 (3.6) \\
… High school diploma & 208 (13.1) \\
… Master's degree & 215 (13.6) \\
… Some college & 327 (20.6) \\

\multicolumn{2}{l}{\textbf{Education, Grouped (N(\%))}} \\
… HS or less & 221 (14.0) \\
… Some college & 534 (33.7) \\
… Bachelor's or above & 829 (52.3) \\

\multicolumn{2}{l}{\textbf{Household Income (N(\%))}} \\
… Under \$10,000 & 33 (2.1) \\
… \$10,000--\$24,999 & 161 (10.2) \\
… \$25,000--\$49,999 & 301 (19.0) \\
… \$50,000--\$74,999 & 326 (20.6) \\
… \$75,000--\$99,999 & 290 (18.3) \\
… \$100,000--\$149,999 & 299 (18.9) \\
… \$150,000 or more & 176 (11.1) \\

\multicolumn{2}{l}{\textbf{Household Income, Grouped (N(\%))}} \\
… Lower & 359 (22.6) \\
… Middle & 923 (58.2) \\
… Upper & 304 (19.2) \\

\multicolumn{2}{l}{\textbf{AI Knowledge, Subjective (N(\%))}} \\
… Much less than most & 46 (2.9) \\
… A little less than most & 143 (9.0) \\
… About the same as most & 818 (51.6) \\
… More than most & 496 (31.3) \\
… Much more than most & 83 (5.2) \\

\multicolumn{2}{l}{\textbf{AI Education (N(\%))}} \\
… None & 54 (3.4) \\
… Casual exposure & 1,033 (65.1) \\
… Some formal education & 408 (25.7) \\
… Extensive formal education & 91 (5.7) \\

\multicolumn{2}{l}{\textbf{AI Use (N(\%))}} \\
… Never use AI & 67 (4.2) \\
… Once or twice a week & 403 (25.4) \\
… Every now and then & 408 (25.7) \\
… Almost every day & 708 (44.6) \\
\end{tabular}

\end{minipage}
\hfill
\begin{minipage}[t]{0.48\textwidth}
\centering
\textbf{Follow-Up Study ($N=1{,}242$)}

\begin{tabular}{lr}
\textbf{Variable} & \textbf{N (\%)} \\
\hline
Number of Observations & 1,242 \\

Age (mean(SD)) & 44.43 (14.29) \\
Gender = Female (N(\%)) & 814 (65.5) \\
Hispanic = Yes (N(\%)) & 94 (7.6) \\
Race = White (N(\%)) & 990 (79.7) \\

\multicolumn{2}{l}{\textbf{Education (N(\%))}} \\
… 1st--4th grade & 2 (0.2) \\
… 10th grade & 2 (0.2) \\
… 11th grade & 2 (0.2) \\
… 12th grade & 6 (0.5) \\
… Associate's degree & 168 (13.5) \\
… Bachelor's degree & 435 (35.0) \\
… Doctorate & 50 (4.0) \\
… High school diploma & 155 (12.5) \\
… Master's degree & 168 (13.5) \\
… Some college & 254 (20.5) \\

\multicolumn{2}{l}{\textbf{Education, Grouped (N(\%))}} \\
… HS or less & 165 (13.3) \\
… Some college & 422 (34.0) \\
… Bachelor's or above & 653 (52.7) \\

\multicolumn{2}{l}{\textbf{Household Income (N(\%))}} \\
… Under \$10,000 & 25 (2.0) \\
… \$10,000--\$24,999 & 114 (9.2) \\
… \$25,000--\$49,999 & 231 (18.6) \\
… \$50,000--\$74,999 & 257 (20.7) \\
… \$75,000--\$99,999 & 224 (18.0) \\
… \$100,000--\$149,999 & 228 (18.4) \\
… \$150,000 or more & 163 (13.1) \\

\multicolumn{2}{l}{\textbf{Household Income, Grouped (N(\%))}} \\
… Lower & 293 (23.6) \\
… Middle & 711 (57.2) \\
… Upper & 238 (19.2) \\

\multicolumn{2}{l}{\textbf{AI Knowledge, Subjective (N(\%))}} \\
… Much less than most & 36 (2.9) \\
… A little less than most & 111 (8.9) \\
… About the same as most & 650 (52.3) \\
… More than most & 386 (31.1) \\
… Much more than most & 59 (4.8) \\

\multicolumn{2}{l}{\textbf{AI Education (N(\%))}} \\
… None & 40 (3.2) \\
… Casual exposure & 824 (66.3) \\
… Some formal education & 312 (25.1) \\
… Extensive formal education & 66 (5.3) \\

\multicolumn{2}{l}{\textbf{AI Use (N(\%))}} \\
… Never use AI & 56 (4.5) \\
… Once or twice a week & 317 (25.5) \\
… Every now and then & 307 (24.7) \\
… Almost every day & 562 (45.2) \\
\end{tabular}

\end{minipage}

\caption*{\normalfont\textit{Note:} Percentages are calculated using the full sample within each study.}
\end{table*} 

\subsection{Balance Checks - All Participants}\label{appx:balance_checks_all}
In this section we show the balance check results for those in the different interventions, task order, and partisanship for both the mains study and the follow-up survey. This subsection includes all participants (including those who failed the attention checks) which was the population used in our main study.

\subsubsection{Main Study}\label{supp:balance_checks_main_all}
Here, we present the balance checks for the main study across the different experimental arms, specifically disclaimer wording type and task order.

Overall, the experimental groups are relatively balanced as we found no significant difference between experimental groups except for Hispanic (C=0, R=1, slightly higher percentage of Hispanic) (see \Cref{tab:balance_cr_all}). Similarly, for the experimental task order, no significant differences were observed among the four task orders (see \Cref{tab:balance_page_order_all}).

\begin{table*}
\footnotesize
\caption{Balance Table for Experimental Conditions (Causal $\times$ Risk) Disclaimer for Main Study - All Participants)}
\begin{tabular}{l|l|l|l|l|l|l}
\multicolumn{1}{c}{} & \multicolumn{4}{c}{\textbf{Experimental Condition }} & & \\
\multicolumn{1}{c}{} & \multicolumn{4}{c}{\textbf{(C=Causal, R=Risk)}} & &\\
\cline{2-5}
\multicolumn{1}{c}{\textbf{Variable}} & \multicolumn{1}{c}{C=0 R=0} & \multicolumn{1}{c}{C=0 R=1} & \multicolumn{1}{c}{C=1 R=0} & \multicolumn{1}{c}{C=1 R=1} & \multicolumn{1}{c}{\textbf{p-value}} & \multicolumn{1}{c}{\textbf{SMD}} \\
\hline
Number of Observations & 404 & 386 & 412 & 384 & & \\
Age (mean (SD)) & 43.89 (14.30) & 43.89 (14.99) & 43.23 (13.86) & 43.26 (14.71) & 0.855 & 0.030 \\
Gender = Female (N(\%)) & 244 (60.4) & 247 (64.0) & 269 (65.3) & 254 (66.1) & 0.343 & 0.064 \\
\multicolumn{2}{l}{Education (N(\%))} & & & & \multicolumn{1}{c}{0.315} & 0.220 \\
… 10th & 0 (0.0) & 0 (0.0) & 1 (0.2) & 1 (0.3) & & \\
… 11th & 3 (0.7) & 0 (0.0) & 1 (0.2) & 0 (0.0) & & \\
… 12th & 1 (0.2) & 4 (1.0) & 2 (0.5) & 0 (0.0) & & \\
… 1st--4th & 1 (0.2) & 0 (0.0) & 0 (0.0) & 1 (0.3) & & \\
… Associate's degree & 54 (13.4) & 53 (13.7) & 41 (10.0) & 59 (15.4) & & \\
… Bachelor's degree & 141 (34.9) & 145 (37.6) & 142 (34.5) & 129 (33.6) & & \\
… Doctorate & 19 (4.7) & 10 (2.6) & 19 (4.6) & 9 (2.3) & & \\
… High school diploma & 53 (13.1) & 42 (10.9) & 62 (15.0) & 51 (13.3) & & \\
… Master's degree & 51 (12.6) & 57 (14.8) & 55 (13.3) & 52 (13.5) & & \\
… Some college & 81 (20.0) & 75 (19.4) & 89 (21.6) & 82 (21.4) & & \\
\multicolumn{2}{l}{Education, Grouped (N(\%))} & & & & \multicolumn{1}{c}{0.519} & 0.090 \\
… HS or less & 57 (14.1) & 46 (11.9) & 66 (16.0) & 52 (13.6) & & \\
… Some college & 135 (33.5) & 128 (33.2) & 130 (31.6) & 141 (36.8) & & \\
… Bachelor's or above & 211 (52.4) & 212 (54.9) & 216 (52.4) & 190 (49.6) & & \\
Hispanic = Yes (N(\%)) & 26 (6.4) & 45 (11.7) & 26 (6.3) & 33 (8.6) & $\sigfive{0.020}$ & 0.108 \\
Race = White (N(\%)) & 323 (80.0) & 296 (76.7) & 329 (79.9) & 300 (78.1) & 0.634 & 0.047 \\
\multicolumn{2}{l}{Household Income (N(\%))} & & & & \multicolumn{1}{c}{0.392} & 0.303 \\
… Under \$10,000 & 14 (3.5) & 11 (2.8) & 14 (3.4) & 14 (3.6) & & \\
… \$10,000--\$24,999 & 39 (9.7) & 40 (10.4) & 39 (9.5) & 23 (6.0) & & \\
… \$25,000--\$49,999 & 83 (20.5) & 69 (17.9) & 72 (17.5) & 77 (20.1) & & \\
… \$50,000--\$74,999 & 91 (22.5) & 72 (18.7) & 81 (19.7) & 82 (21.4) & & \\
… \$75,000--\$99,999 & 63 (15.6) & 75 (19.4) & 76 (18.4) & 76 (19.8) & & \\
… \$100,000--\$149,999 & 72 (17.8) & 80 (20.7) & 76 (18.4) & 71 (18.5) & & \\
… \$150,000 or more & 42 (10.4) & 39 (10.1) & 54 (13.1) & 41 (10.7) & & \\
\multicolumn{2}{l}{Household Income, Grouped (N(\%))} & & & & \multicolumn{1}{c}{0.977} & 0.039 \\
… Lower & 95 (23.5) & 90 (23.3) & 90 (21.8) & 84 (21.9) & & \\
… Middle & 235 (58.2) & 226 (58.5) & 240 (58.3) & 222 (57.8) & & \\
… Upper & 74 (18.3) & 70 (18.1) & 82 (19.9) & 78 (20.3) & & \\
\multicolumn{2}{l}{AI Knowledge, Subjective (N(\%))} & & & & \multicolumn{1}{c}{$\sigfive{0.041}$} & 0.186 \\
… I know much less than most & 11 (2.7) & 10 (2.6) & 9 (2.2) & 16 (4.2) & & \\
… I know a little less than most & 36 (8.9) & 40 (10.4) & 24 (5.8) & 43 (11.2) & & \\
… I know about the same as most & 201 (49.8) & 184 (47.7) & 221 (53.6) & 212 (55.2) & & \\
… I know more than most & 131 (32.4) & 132 (34.2) & 133 (32.3) & 100 (26.0) & & \\
… I know much more than most & 25 (6.2) & 20 (5.2) & 25 (6.1) & 13 (3.4) & & \\
\multicolumn{2}{l}{AI Education (N(\%))} & & & & \multicolumn{1}{c}{0.256} & 0.131 \\
… None & 21 (5.2) & 8 (2.1) & 10 (2.4) & 15 (3.9) & & \\
… Casual exposure & 249 (61.6) & 253 (65.5) & 282 (68.4) & 249 (64.8) & & \\
… Some formal education & 109 (27.0) & 102 (26.4) & 95 (23.1) & 102 (26.6) & & \\
… Extensive formal education & 25 (6.2) & 23 (6.0) & 25 (6.1) & 18 (4.7) & & \\
\multicolumn{2}{l}{AI Use (N(\%))} & & & & \multicolumn{1}{c}{0.334} & 0.125 \\
… I never use AI & 17 (4.2) & 16 (4.1) & 14 (3.4) & 20 (5.2) & & \\
… I use it once or twice a week & 106 (26.2) & 97 (25.1) & 96 (23.3) & 104 (27.1) & & \\
… I use it every now and then & 89 (22.0) & 100 (25.9) & 109 (26.5) & 110 (28.6) & & \\
… I use it almost everyday & 192 (47.5) & 173 (44.8) & 193 (46.8) & 150 (39.1) & & \\
\end{tabular}
\caption*{\normalfont{Note: The p-values result from a joint F-test for continuous variables and from a Chi-squared test for categorical variables, comparing all four experimental conditions jointly. \sigfive{} indicates significant results with $\alpha = 0.05$. \sigten{} indicates significant results with $\alpha = 0.10$. The raw income and education Chi-squared approximations may be unreliable due to sparse cells; the grouped variables are likely more robust.}}
\label{tab:balance_cr_all}
\end{table*}
\begin{table*}
\footnotesize
\caption{Balance Table by User Page Order for Main Study - All Participants}
\begin{tabular}{l|l|l|l|l|l|l}
\multicolumn{1}{c}{} & \multicolumn{4}{c}{\textbf{User Page Order}} & & \\
\multicolumn{1}{c}{} & \multicolumn{4}{c}{\textbf{(b = Budget, c = Conserv+, l = Liberal+)}} & &\\
\cline{2-5} 
\multicolumn{1}{c}{\textbf{Variable}} & \multicolumn{1}{c}{b\_c\_l} & \multicolumn{1}{c}{b\_l\_c} & \multicolumn{1}{c}{c\_l\_b} & \multicolumn{1}{c}{l\_c\_b} & \multicolumn{1}{c}{\textbf{p-value}} & \multicolumn{1}{c}{\textbf{SMD}} \\
\hline
Number of Observations & 385 & 408 & 387 & 406 & & \\
Age (mean (SD)) & 44.59 (15.33) & 44.20 (14.33) & 42.42 (14.23) & 43.06 (13.85) & 0.130 & 0.088 \\
Gender = Female (N(\%)) & 249 (64.7) & 266 (65.2) & 246 (63.6) & 253 (62.3) & 0.836 & 0.034 \\
\multicolumn{2}{l}{Education (N(\%))} & & & & \multicolumn{1}{c}{0.611} & 0.191 \\
… 10th & 0 (0.0) & 0 (0.0) & 0 (0.0) & 2 (0.5) & & \\
… 11th & 1 (0.3) & 0 (0.0) & 1 (0.3) & 2 (0.5) & & \\
… 12th & 2 (0.5) & 1 (0.2) & 1 (0.3) & 3 (0.7) & & \\
… 1st--4th & 0 (0.0) & 1 (0.2) & 0 (0.0) & 1 (0.2) & & \\
… Associate's degree & 45 (11.7) & 63 (15.4) & 44 (11.4) & 55 (13.5) & & \\
… Bachelor's degree & 151 (39.2) & 136 (33.3) & 128 (33.1) & 142 (35.0) & & \\
… Doctorate & 16 (4.2) & 13 (3.2) & 14 (3.6) & 14 (3.4) & & \\
… High school diploma & 46 (11.9) & 57 (14.0) & 51 (13.2) & 54 (13.3) & & \\
… Master's degree & 42 (10.9) & 58 (14.2) & 56 (14.5) & 59 (14.5) & & \\
… Some college & 82 (21.3) & 79 (19.4) & 92 (23.8) & 74 (18.2) & & \\
\multicolumn{2}{l}{Education, Grouped (N(\%))} & & & & \multicolumn{1}{c}{0.891} & 0.059 \\
… HS or less & 49 (12.7) & 58 (14.3) & 53 (13.7) & 61 (15.1) & & \\
… Some college & 127 (33.0) & 142 (34.9) & 136 (35.1) & 129 (31.9) & & \\
… Bachelor's or above & 209 (54.3) & 207 (50.9) & 198 (51.2) & 215 (53.1) & & \\
Hispanic = Yes (N(\%)) & 36 (9.4) & 32 (7.8) & 30 (7.8) & 32 (7.9) & 0.824 & 0.029 \\
Race = White (N(\%)) & 305 (79.2) & 323 (79.2) & 295 (76.2) & 325 (80.0) & 0.579 & 0.047 \\
\multicolumn{2}{l}{Household Income (N(\%))} & & & & \multicolumn{1}{c}{0.836} & 0.271 \\
… Under \$10,000 & 10 (2.6) & 13 (3.2) & 18 (4.7) & 12 (3.0) & & \\
… \$10,000--\$24,999 & 36 (9.4) & 32 (7.8) & 33 (8.5) & 40 (9.9) & & \\
… \$25,000--\$49,999 & 71 (18.4) & 79 (19.4) & 81 (20.9) & 70 (17.2) & & \\
… \$50,000--\$74,999 & 88 (22.9) & 82 (20.1) & 71 (18.3) & 85 (20.9) & & \\
… \$75,000--\$99,999 & 60 (15.6) & 80 (19.6) & 74 (19.1) & 76 (18.7) & & \\
… \$100,000--\$149,999 & 73 (19.0) & 81 (19.9) & 70 (18.1) & 75 (18.5) & & \\
… \$150,000 or more & 47 (12.2) & 41 (10.0) & 40 (10.3) & 48 (11.8) & & \\
\multicolumn{2}{l}{Household Income, Grouped (N(\%))} & & & & \multicolumn{1}{c}{0.778} & 0.069 \\
… Lower & 84 (21.8) & 84 (20.6) & 98 (25.3) & 93 (22.9) & & \\
… Middle & 222 (57.7) & 245 (60.0) & 220 (56.8) & 236 (58.1) & & \\
… Upper & 79 (20.5) & 79 (19.4) & 69 (17.8) & 77 (19.0) & & \\
\multicolumn{2}{l}{AI Knowledge, Subjective (N(\%))} & & & & \multicolumn{1}{c}{0.949} & 0.092 \\
… I know much less than most & 11 (2.9) & 14 (3.4) & 8 (2.1) & 13 (3.2) & & \\
… I know a little less than most & 32 (8.3) & 42 (10.3) & 32 (8.3) & 37 (9.1) & & \\
… I know about the same as most & 197 (51.2) & 214 (52.5) & 208 (53.7) & 199 (49.0) & & \\
… I know more than most & 125 (32.5) & 119 (29.2) & 118 (30.5) & 134 (33.0) & & \\
… I know much more than most & 20 (5.2) & 19 (4.7) & 21 (5.4) & 23 (5.7) & & \\
\multicolumn{2}{l}{AI Education (N(\%))} & & & & \multicolumn{1}{c}{0.192} & 0.142 \\
… None & 15 (3.9) & 11 (2.7) & 8 (2.1) & 20 (4.9) & & \\
… Casual exposure & 254 (66.0) & 283 (69.4) & 246 (63.6) & 250 (61.6) & & \\
… Some formal education & 92 (23.9) & 97 (23.8) & 108 (27.9) & 111 (27.3) & & \\
… Extensive formal education & 24 (6.2) & 17 (4.2) & 25 (6.5) & 25 (6.2) & &\\
\multicolumn{2}{l}{AI Use (N(\%))} & & & & \multicolumn{1}{c}{0.846} & 0.087 \\
… I never use AI & 18 (4.7) & 21 (5.1) & 13 (3.4) & 15 (3.7) & & \\
… I use it once or twice a week & 105 (27.3) & 105 (25.7) & 89 (23.0) & 104 (25.6) & & \\
… I use it every now and then &94 (24.4) & 108 (26.5) & 103 (26.6) & 103 (25.4) & & \\
… I use it almost everyday & 168 (43.6) & 174 (42.6) & 182 (47.0) & 184 (45.3) & & \\
\end{tabular}
\caption*{\normalfont{Note: The p-values result from a joint F-test for continuous variables and from a Chi-squared test for categorical variables, comparing all four page orders jointly. \sigfive{} indicates significant results with $\alpha = 0.05$. \sigten{} indicates significant results with $\alpha = 0.10$. The raw income and education Chi-squared approximations may be unreliable due to sparse cells; the grouped variables are likely more robust.}}
\label{tab:balance_page_order_all}
\end{table*}
 
We also compare demographics between our blocking group levels, Republican and Democrat participants, we include a balance check table for full transparency (see \Cref{tab:balance_partisanship_all}). We found a few demographic conditions which differed significantly between participants: Age ($p<0.01$, Republican older), Education ($p<0.01$, Republican less educated), Race ($p<0.01$, Republican whiter), income-grouped ($p<0.05$ Republicans more income), AI Education ($p<0.01$, Republican have less education), and AI Use ($p<0.05$, Republican have higher use of AI). However, we note that the effect of these differences are small to moderate with the SMD=0.35 or less for all. Given these significant differences, we include a full model with all the demographics (Age, Education-grouped, Race, and Income) in \Cref{appx:pooled_full_models}.  We do separate analysis with AI Education and AI Use. 

\begin{table*}
\footnotesize
\caption{Balance Table by Participant Partisanship for Main Study - All Participants}
\begin{tabular}{l|l|l|l|l}
\multicolumn{1}{c}{} & \multicolumn{2}{c}{\textbf{Participant Partisanship}} & & \\
\cline{2-3}
\multicolumn{1}{c}{\textbf{Variable}} & \multicolumn{1}{c}{Democrat} & \multicolumn{1}{c}{Republican} & \multicolumn{1}{c}{\textbf{p-value}} & \multicolumn{1}{c}{\textbf{SMD}} \\
\hline
Number of Observations & 796 & 790 & & \\
Age (mean (SD)) & 42.29 (14.81) & 44.85 (13.97) & $\sigfive{0.0004}$ & 0.178 \\
Gender = Female (N(\%)) & 524 (65.8) & 490 (62.0) & 0.127 & 0.079 \\
\multicolumn{2}{l}{Education (N(\%))} & & \multicolumn{1}{c}{$\sigfive{0.0002}$} & 0.289 \\
… 10th & 0 (0.0) & 2 (0.3) & & \\
… 11th & 0 (0.0) & 4 (0.5) & & \\
… 12th & 4 (0.5) & 3 (0.4) & & \\
… 1st--4th & 1 (0.1) & 1 (0.1) & & \\
… Associate's degree & 98 (12.3) & 109 (13.8) & & \\
… Bachelor's degree & 285 (35.8) & 272 (34.4) & & \\
… Doctorate & 40 (5.0) & 17 (2.2) & & \\
… High school diploma & 78 (9.8) & 130 (16.5) & & \\
… Master's degree & 119 (14.9) & 96 (12.2) & & \\
… Some college & 171 (21.5) & 156 (19.7) & & \\
\multicolumn{2}{l}{Education, Grouped (N(\%))} & & \multicolumn{1}{c}{$\sigfive{<0.0001}$} & 0.220 \\
… HS or less & 82 (10.3) & 139 (17.6) & & \\
… Some college & 269 (33.8) & 265 (33.6) & & \\
… Bachelor's or above & 444 (55.8) & 385 (48.8) & & \\
Hispanic = Yes (N(\%)) & 70 (8.8) & 60 (7.6) & 0.436 & 0.044 \\
Race = White (N(\%)) & 571 (71.7) & 677 (85.7) & $\sigfive{<0.0001}$ & 0.346 \\
\multicolumn{2}{l}{Household Income (N(\%))} & & \multicolumn{1}{c}{0.198} & 0.235 \\
… Under \$10,000 & 31 (3.9) & 22 (2.8) & & \\
… \$10,000--\$24,999 & 83 (10.4) & 58 (7.3) & & \\
… \$25,000--\$49,999 & 161 (20.2) & 140 (17.7) & & \\
… \$50,000--\$74,999 & 159 (20.0) & 167 (21.1) & & \\
… \$75,000--\$99,999 & 145 (18.2) & 145 (18.4) & & \\
… \$100,000--\$149,999 & 124 (15.6) & 175 (22.2) & & \\
… \$150,000 or more & 93 (11.7) & 83 (10.5) & & \\
\multicolumn{2}{l}{Household Income, Grouped (N(\%))} & & \multicolumn{1}{c}{$\sigfive{0.020}$} & 0.141 \\
… Lower & 201 (25.3) & 158 (20.0) & & \\
… Middle & 457 (57.4) & 466 (59.0) & & \\
… Upper & 138 (17.3) & 166 (21.0) & & \\
\multicolumn{2}{l}{AI Knowledge, Subjective (N(\%))} & & \multicolumn{1}{c}{0.317} & 0.109 \\
… I know much less than most & 28 (3.5) & 18 (2.3) & & \\
… I know a little less than most & 63 (7.9) & 80 (10.1) & & \\
… I know about the same as most & 415 (52.1) & 403 (51.0) & & \\
… I know more than most & 251 (31.5) & 245 (31.0) & & \\
… I know much more than most & 39 (4.9) & 44 (5.6) & & \\
\multicolumn{2}{l}{AI Education (N(\%))} & & \multicolumn{1}{c}{$\sigfive{0.007}$} & 0.175 \\
… None & 20 (2.5) & 34 (4.3) & & \\
… Casual exposure & 503 (63.2) & 530 (67.1) & & \\
… Some formal education & 231 (29.0) & 177 (22.4) & & \\
… Extensive formal education & 42 (5.3) & 49 (6.2) & & \\
\multicolumn{2}{l}{AI Use (N(\%))} & & \multicolumn{1}{c}{$\sigfive{0.025}$} & 0.154 \\
… I never use AI & 42 (5.3) & 25 (3.2) & & \\
… I use it once or twice a week & 208 (26.1) & 195 (24.7) & & \\
… I use it every now and then & 216 (27.1) & 192 (24.3) & & \\
… I use it almost everyday & 330 (41.5) & 378 (47.8) & & \\
\end{tabular}
\caption*{\normalfont{Note: The p-values result from an F-test for continuous variables and from a Chi-squared test for categorical variables. \sigfive{} indicates significant results with $\alpha = 0.05$. \sigten{} indicates significant results with $\alpha = 0.10$. The raw education Chi-squared approximation may be unreliable due to sparse cells; the grouped education variable is likely more robust.}}
\label{tab:balance_partisanship_all}
\end{table*}
 
\subsubsection{Follow-Up Survey}\label{appx:balance_checks_FU_all}
\paragraph{Attrition}
Of the $N=1,586$ participants in the main study, $1242(78.3\%)$ completed the follow-up survey. We show the attrition rates in \Cref{tab:attrition_all}. Attrition did not significantly differ by disclaimer condition, ($\chi^2=0.752, p=.861$); a logistic regression predicting follow-up completion from \texttt{Risk}, \texttt{Causal}, and their interaction likewise found no significant predictors (all $p>.44$). A model additionally testing whether attrition-by-condition differed by partisanship found no significant \texttt{Risk} $\times$ Republican or \texttt{Causal} $\times$ Republican interactions (both $p>.48$), and no partisan difference in those who attrited. Finally, immediate post-task opinion, budget allocation bias, and AI trust did not significantly predict follow-up completion (all $p>.24$). Together, these checks suggest follow-up attrition was unrelated to disclaimer condition and unlikely to bias the H4 comparisons reported below.

\begin{table}[ht]
\centering
\begin{threeparttable}
\caption{Follow-Up Attrition by Disclaimer Condition and Party - All Participants}
\label{tab:attrition_all}
\small
\begin{tabular}{lcc}
\toprule
Condition & Completed ($n$ (\%)) & Attrited ($n$ (\%)) \\
\midrule

Control & 323 (79.95) & 81 (20.05) \\
Risk Only & 299 (77.46) & 87 (22.54) \\
Causal Only & 321 (77.91) & 91 (22.09) \\
Risk + Causal & 299 (77.87) & 85 (22.14) \\
\midrule
Overall & 1242 (78.31) & 344 (21.69) \\
\bottomrule
\end{tabular}

\begin{tabular}{lcc}
\toprule
Predictor & $b$ (SE) & $p$ \\
\midrule
\multicolumn{3}{l}{\textit{Model 1: Condition only}} \\
Risk                        & $-0.13$ (0.20) & .443 \\
Causal                      & $-0.12$ (0.20) & .479 \\
Risk $\times$ Causal        & $0.15$ (0.29)  & .553 \\
\multicolumn{3}{l}{\textit{Model 2: Condition $\times$ Party}} \\
Risk                        & $-0.11$ (0.22) & .560 \\
Causal                      & $0.05$ (0.22)  & .799 \\
Republican                  & $\mathbf{-0.33}$ (0.26) & .127 \\
Risk $\times$ Republican    & $0.06$ (0.29)  & .799 \\
Causal $\times$ Republican  & $-0.18$ (0.29) & .475 \\
\bottomrule
\end{tabular}
\begin{tablenotes}
\small
\item \textit{Note.} Logistic regression predicting follow-up completion (1 = completed). $\chi^2(3) = 0.752$, $p = .861$ for the omnibus test of condition on retention.
\end{tablenotes}
\end{threeparttable}
\end{table} 
\paragraph{Balance Checks - Follow-Up Survey}
Here, we present the balance checks across the different experimental arms for participants who submitted to our follow-up survey, which was administered 1 week after they completed the main study.

Overall, the experimental groups are balanced as we found no significant difference between experimental groups except again Hispanic, same as the main study (more Hispanics in C=0, R=1) (see \Cref{tab:followup_balance_cr_all}). Similarly, for the experimental task order, no significant differences were observed among the four task orders (see \Cref{tab:followup_balance_page_order_all}).

\begin{table*}
\footnotesize
\caption{Balance Table for Experimental Conditions (Causal $\times$ Risk Disclaimer) Follow-Up Study - All Participants}
\begin{tabular}{l|l|l|l|l|l|l}
\multicolumn{1}{c}{} & \multicolumn{4}{c}{\textbf{Experimental Condition}} & & \\
\multicolumn{1}{c}{} & \multicolumn{4}{c}{\textbf{(C=Causal, R=Risk)}} & &\\
\cline{2-5}
\multicolumn{1}{c}{\textbf{Variable}} & \multicolumn{1}{c}{C=0 R=0} & \multicolumn{1}{c}{C=0 R=1} & \multicolumn{1}{c}{C=1 R=0} & \multicolumn{1}{c}{C=1 R=1} & \multicolumn{1}{c}{\textbf{p-value}} & \multicolumn{1}{c}{\textbf{SMD}} \\
\hline
Number of Observations & 323 & 299 & 321 & 299 & & \\
Age (mean (SD)) & 44.61 (13.86) & 44.59 (14.92) & 44.05 (13.86) & 44.49 (14.60) & 0.956 & 0.021 \\
Gender = Female (N(\%)) & 205 (63.5) & 195 (65.2) & 218 (67.9) & 196 (65.6) & 0.699 & 0.048 \\
\multicolumn{2}{l}{Education (N(\%))} & & & & \multicolumn{1}{c}{0.538} & 0.224 \\
… 10th & 0 (0.0) & 0 (0.0) & 1 (0.3) & 1 (0.3) & & \\
… 11th & 2 (0.6) & 0 (0.0) & 0 (0.0) & 0 (0.0) & & \\
… 12th & 1 (0.3) & 3 (1.0) & 2 (0.6) & 0 (0.0) & & \\
… 1st--4th & 1 (0.3) & 0 (0.0) & 0 (0.0) & 1 (0.3) & & \\
… Associate's degree & 45 (13.9) & 42 (14.0) & 35 (10.9) & 46 (15.4) & & \\
… Bachelor's degree & 113 (35.0) & 113 (37.8) & 109 (34.0) & 100 (33.4) & & \\
… Doctorate & 15 (4.6) & 8 (2.7) & 18 (5.6) & 9 (3.0) & & \\
… High school diploma & 37 (11.5) & 31 (10.4) & 48 (15.0) & 39 (13.0) & & \\
… Master's degree & 42 (13.0) & 46 (15.4) & 41 (12.8) & 39 (13.0) & & \\
… Some college & 67 (20.7) & 56 (18.7) & 67 (20.9) & 64 (21.4) & & \\
\multicolumn{2}{l}{Education, Grouped (N(\%))} & & & & \multicolumn{1}{c}{0.527} & 0.101 \\
… HS or less & 40 (12.4) & 34 (11.4) & 51 (15.9) & 40 (13.4) & & \\
… Some college & 112 (34.8) & 98 (32.8) & 102 (31.8) & 110 (36.9) & & \\
… Bachelor's or above & 170 (52.8) & 167 (55.9) & 168 (52.3) & 148 (49.7) & & \\
Hispanic = Yes (N(\%)) & 17 (5.3) & 33 (11.0) & 20 (6.2) & 24 (8.0) & $\sigfive{0.037}$ & 0.118 \\
Race = White (N(\%)) & 258 (79.9) & 235 (78.6) & 260 (81.0) & 237 (79.3) & 0.896 & 0.032 \\
\multicolumn{2}{l}{Household Income (N(\%))} & & & & \multicolumn{1}{c}{0.501} & 0.329 \\
… Under \$10,000 & 20 (6.2) & 26 (8.7) & 31 (9.7) & 17 (5.7) & & \\
… \$10,000--\$24,999 & 49 (15.2) & 31 (10.4) & 42 (13.1) & 34 (11.4) & & \\
… \$25,000--\$49,999 & 66 (20.4) & 49 (16.4) & 59 (18.4) & 57 (19.1) & & \\
… \$50,000--\$74,999 & 74 (22.9) & 55 (18.4) & 59 (18.4) & 66 (22.1) & & \\
… \$75,000--\$99,999 & 51 (15.8) & 58 (19.4) & 55 (17.1) & 60 (20.1) & & \\
… \$100,000--\$149,999 & 54 (16.7) & 66 (22.1) & 55 (17.1) & 53 (17.7) & & \\
… \$150,000 or more & 32 (9.9) & 31 (10.4) & 45 (14.0) & 35 (11.7) & & \\
\multicolumn{2}{l}{Household Income, Grouped (N(\%))} & & & & \multicolumn{1}{c}{0.976} & 0.048 \\
… Lower & 80 (24.8) & 70 (23.4) & 76 (23.7) & 67 (22.4) & & \\
… Middle & 186 (57.6) & 173 (57.9) & 180 (56.1) & 172 (57.5) & & \\
… Upper & 57 (17.6) & 56 (18.7) & 65 (20.2) & 60 (20.1) & & \\
\multicolumn{2}{l}{AI Knowledge, Subjective (N(\%))} & & & & \multicolumn{1}{c}{$\sigten{0.056}$} & 0.201 \\
… I know much less than most & 7 (2.2) & 9 (3.0) & 6 (1.9) & 14 (4.7) & & \\
… I know a little less than most & 26 (8.0) & 34 (11.4) & 20 (6.2) & 31 (10.4) & & \\
… I know about the same as most & 167 (51.7) & 146 (48.8) & 170 (53.0) & 167 (55.9) & & \\
… I know more than most & 104 (32.2) & 100 (33.4) & 105 (32.7) & 77 (25.8) & & \\
… I know much more than most & 19 (5.9) & 10 (3.3) & 20 (6.2) & 10 (3.3) & & \\
\multicolumn{2}{l}{AI Education (N(\%))} & & & & \multicolumn{1}{c}{0.613} & 0.113 \\
… None & 15 (4.6) & 7 (2.3) & 8 (2.5) & 10 (3.3) & & \\
… Casual exposure & 202 (62.5) & 206 (68.9) & 221 (68.8) & 195 (65.2) & & \\
… Some formal education & 87 (26.9) & 69 (23.1) & 75 (23.4) & 81 (27.1) & & \\
… Extensive formal education & 19 (5.9) & 17 (5.7) & 17 (5.3) & 13 (4.3) & & \\
\multicolumn{2}{l}{AI Use (N(\%))} & & & & \multicolumn{1}{c}{0.479} & 0.130 \\
… I never use AI & 15 (4.6) & 14 (4.7) & 12 (3.7) & 15 (5.0) & & \\
… I use it once or twice a week & 87 (26.9) & 81 (27.1) & 69 (21.5) & 80 (26.8) & & \\
… I use it every now and then & 71 (22.0) & 72 (24.1) & 81 (25.2) & 83 (27.8) & & \\
… I use it almost everyday & 150 (46.4) & 132 (44.1) & 159 (49.5) & 121 (40.5) & & \\
\end{tabular}
\caption*{\normalfont{Note: The p-values result from a joint F-test for continuous variables and from a Chi-squared test for categorical variables, comparing all four experimental conditions jointly. \sigfive{} indicates significant results with $\alpha = 0.05$. \sigten{} indicates significant results with $\alpha = 0.10$. The raw income and education Chi-squared approximations may be unreliable due to sparse cells; the grouped variables are likely more robust.}}
\label{tab:followup_balance_cr_all}
\end{table*}
\begin{table*}
\footnotesize
\caption{Balance Table by User Page Order (Follow-Up Study - All Participants)}
\begin{tabular}{l|l|l|l|l|l|l}
\multicolumn{1}{c}{} & \multicolumn{4}{c}{\textbf{User Page Order}} & & \\
\multicolumn{1}{c}{} & \multicolumn{4}{c}{\textbf{(b = Budget, c = Conserv+, l = Liberal+)}} & &\\
\cline{2-5}
\multicolumn{1}{c}{\textbf{Variable}} & \multicolumn{1}{c}{b\_c\_l} & \multicolumn{1}{c}{b\_l\_c} & \multicolumn{1}{c}{c\_l\_b} & \multicolumn{1}{c}{l\_c\_b} & \multicolumn{1}{c}{\textbf{p-value}} & \multicolumn{1}{c}{\textbf{SMD}} \\
\hline
Number of Observations & 309 & 319 & 302 & 312 & & \\
Age (mean (SD)) & 45.16 (15.24) & 45.17 (14.27) & 43.59 (13.92) & 43.78 (13.65) & 0.343 & 0.071 \\
Gender = Female (N(\%)) & 205 (66.3) & 212 (66.5) & 195 (64.6) & 202 (64.7) & 0.936 & 0.025 \\
\multicolumn{2}{l}{Education (N(\%))} & & & & \multicolumn{1}{c}{0.387} & 0.230 \\
… 10th & 0 (0.0) & 0 (0.0) & 0 (0.0) & 2 (0.6) & & \\
… 11th & 1 (0.3) & 0 (0.0) & 1 (0.3) & 0 (0.0) & & \\
… 12th & 2 (0.6) & 1 (0.3) & 1 (0.3) & 2 (0.6) & & \\
… 1st--4th & 0 (0.0) & 1 (0.3) & 0 (0.0) & 1 (0.3) & & \\
… Associate's degree & 39 (12.6) & 52 (16.3) & 30 (9.9) & 47 (15.1) & & \\
… Bachelor's degree & 117 (37.9) & 104 (32.6) & 100 (33.1) & 114 (36.5) & & \\
… Doctorate & 15 (4.9) & 9 (2.8) & 14 (4.6) & 12 (3.8) & & \\
… High school diploma & 38 (12.3) & 44 (13.8) & 35 (11.6) & 38 (12.2) & & \\
… Master's degree & 39 (12.6) & 45 (14.1) & 41 (13.6) & 43 (13.8) & & \\
… Some college & 58 (18.8) & 63 (19.7) & 80 (26.5) & 53 (17.0) & & \\
\multicolumn{2}{l}{Education, Grouped (N(\%))} & & & & \multicolumn{1}{c}{0.738} & 0.081 \\
… HS or less & 41 (13.3) & 45 (14.2) & 37 (12.3) & 42 (13.5) & & \\
… Some college & 97 (31.4) & 115 (36.2) & 110 (36.4) & 100 (32.2) & & \\
… Bachelor's or above & 171 (55.3) & 158 (49.7) & 155 (51.3) & 169 (54.3) & & \\
Hispanic = Yes (N(\%)) & 30 (9.7) & 22 (6.9) & 21 (7.0) & 21 (6.7) & 0.439 & 0.055 \\
Race = White (N(\%)) & 248 (80.3) & 262 (82.1) & 227 (75.2) & 253 (81.1) & 0.142 & 0.089 \\
\multicolumn{2}{l}{Household Income (N(\%))} & & & & \multicolumn{1}{c}{0.894} & 0.293 \\
… Under \$10,000 & 31 (10.0) & 23 (7.2) & 30 (9.9) & 23 (7.4) & & \\
… \$10,000--\$24,999 & 43 (13.9) & 33 (10.3) & 43 (14.2) & 35 (11.2) & & \\
… \$25,000--\$49,999 & 56 (18.1) & 61 (19.1) & 69 (22.8) & 53 (17.0) & & \\
… \$50,000--\$74,999 & 70 (22.7) & 65 (20.4) & 55 (18.2) & 70 (22.4) & & \\
… \$75,000--\$99,999 & 44 (14.2) & 64 (20.1) & 59 (19.5) & 57 (18.3) & & \\
… \$100,000--\$149,999 & 57 (18.4) & 63 (19.7) & 53 (17.5) & 55 (17.6) & & \\
… \$150,000 or more & 51 (16.5) & 33 (10.3) & 32 (10.6) & 42 (13.5) & & \\
\multicolumn{2}{l}{Household Income, Grouped (N(\%))} & & & & \multicolumn{1}{c}{0.619} & 0.086 \\
… Lower & 74 (23.9) & 64 (20.1) & 80 (26.5) & 75 (24.0) & & \\
… Middle & 173 (56.0) & 195 (61.1) & 168 (55.6) & 175 (56.1) & & \\
… Upper & 62 (20.1) & 60 (18.8) & 54 (17.9) & 62 (19.9) & & \\
\multicolumn{2}{l}{AI Knowledge, Subjective (N(\%))} & & & & \multicolumn{1}{c}{0.808} & 0.127 \\
… I know much less than most & 10 (3.2) & 12 (3.8) & 6 (2.0) & 8 (2.6) & & \\
… I know a little less than most & 27 (8.7) & 32 (10.0) & 23 (7.6) & 29 (9.3) & & \\
… I know about the same as most & 159 (51.5) & 168 (52.7) & 168 (55.6) & 155 (49.7) & & \\
… I know more than most & 94 (30.4) & 95 (29.8) & 90 (29.8) & 107 (34.3) & & \\
… I know much more than most & 19 (6.1) & 12 (3.8) & 15 (5.0) & 13 (4.2) & & \\
\multicolumn{2}{l}{AI Education (N(\%))} & & & & \multicolumn{1}{c}{0.315} & 0.147 \\
… None & 11 (3.6) & 8 (2.5) & 7 (2.3) & 14 (4.5) & & \\
… Casual exposure & 208 (67.3) & 222 (69.6) & 196 (64.9) & 198 (63.5) & & \\
… Some formal education & 69 (22.3) & 78 (24.5) & 85 (28.1) & 80 (25.6) & & \\
… Extensive formal education & 21 (6.8) & 11 (3.4) & 14 (4.6) & 20 (6.4) & & \\
\multicolumn{2}{l}{AI Use (N(\%))} & & & & \multicolumn{1}{c}{0.893} & 0.091 \\
… I never use AI & 16 (5.2) & 17 (5.3) & 12 (4.0) & 11 (3.5) & & \\
… I use it once or twice a week & 87 (28.2) & 81 (25.4) & 72 (23.8) & 77 (24.7) & & \\
… I use it every now and then & 72 (23.3) & 82 (25.7) & 74 (24.5) & 79 (25.3) & & \\
… I use it almost everyday & 134 (43.4) & 139 (43.6) & 144 (47.7) & 145 (46.5) & & \\
\end{tabular}
\caption*{\normalfont{Note: The p-values result from a joint F-test for continuous variables and from a Chi-squared test for categorical variables, comparing all four page orders jointly. \sigfive{} indicates significant results with $\alpha = 0.05$. \sigten{} indicates significant results with $\alpha = 0.10$. The raw income and education Chi-squared approximations may be unreliable due to sparse cells; the grouped variables are likely more robust.}}
\label{tab:followup_balance_page_order_all}
\end{table*}
 
We also compare demographics between our blocking group levels, Republican and Democrat participants, we include a balance check table for full transparency (see \Cref{tab:followup_balance_partisanship_all}). Similarly to our main experiment, we found a few demographic conditions which differed significantly between participants: Age ($p<0.01$, Republican older), Education ($p<0.01$, Republican less educated), Race ($p<0.01$, Republican whiter), AI Education ($p=0.03$, Republican have less AI education), and AI Use ($p=0.03$, Republican have higher use of AI). However, we note that the effect of these differences are small to moderate with the SMD=0.33 or less for all. Given these significant differences, we include a full model with all the demographics (Age, Education-grouped, Hispanic (binary), and Race) in \Cref{appx:pooled_full_models}. We do separate analysis with AI Education and AI Use. 

\begin{table*}
\footnotesize
\caption{Balance Table by Participant Partisanship (Follow-Up Study - All Participants)}
\begin{tabular}{l|l|l|l|l}
\multicolumn{1}{c}{} & \multicolumn{2}{c}{\textbf{Participant Partisanship}} & & \\
\cline{2-3}
\multicolumn{1}{c}{\textbf{Variable}} & \multicolumn{1}{c}{Democrat} & \multicolumn{1}{c}{Republican} & \multicolumn{1}{c}{\textbf{p-value}} & \multicolumn{1}{c}{\textbf{SMD}} \\
\hline
Number of Observations & 649 & 593 & & \\
Age (mean (SD)) & 43.11 (14.53) & 45.88 (13.88) & $\sigfive{0.0006}$ & 0.194 \\
Gender = Female (N(\%)) & 435 (67.0) & 379 (63.9) & 0.274 & 0.066 \\
\multicolumn{2}{l}{Education (N(\%))} & & \multicolumn{1}{c}{$\sigfive{0.0007}$} & 0.307 \\
… 10th & 0 (0.0) & 2 (0.3) & & \\
… 11th & 0 (0.0) & 2 (0.3) & & \\
… 12th & 3 (0.5) & 3 (0.5) & & \\
… 1st--4th & 1 (0.2) & 1 (0.2) & & \\
… Associate's degree & 78 (12.0) & 90 (15.2) & & \\
… Bachelor's degree & 238 (36.7) & 197 (33.2) & & \\
… Doctorate & 35 (5.4) & 15 (2.5) & & \\
… High school diploma & 59 (9.1) & 96 (16.2) & & \\
… Master's degree & 96 (14.8) & 72 (12.1) & & \\
… Some college & 139 (21.4) & 115 (19.4) & & \\
\multicolumn{2}{l}{Education, Grouped (N(\%))} & & \multicolumn{1}{c}{$\sigfive{<0.0001}$} & 0.249 \\
… HS or less & 62 (9.6) & 103 (17.4) & & \\
… Some college & 217 (33.5) & 205 (34.6) & & \\
… Bachelor's or above & 369 (56.9) & 284 (48.0) & & \\
Hispanic = Yes (N(\%)) & 56 (8.6) & 38 (6.4) & 0.171 & 0.084 \\
Race = White (N(\%)) & 476 (73.3) & 514 (86.7) & $\sigfive{<0.0001}$ & 0.338 \\
\multicolumn{2}{l}{Household Income (N(\%))} & & \multicolumn{1}{c}{0.447} & 0.236 \\
… Under \$10,000 & 45 (6.9) & 30 (5.1) & & \\
… \$10,000--\$24,999 & 95 (14.6) & 71 (12.0) & & \\
… \$25,000--\$49,999 & 128 (19.7) & 103 (17.4) & & \\
… \$50,000--\$74,999 & 129 (19.9) & 128 (21.6) & & \\
… \$75,000--\$99,999 & 115 (17.7) & 109 (18.4) & & \\
… \$100,000--\$149,999 & 104 (16.0) & 124 (20.9) & & \\
… \$150,000 or more & 70 (10.8) & 67 (11.3) & & \\
\multicolumn{2}{l}{Household Income, Grouped (N(\%))} & & \multicolumn{1}{c}{$\sigten{0.084}$} & 0.127 \\
… Lower & 168 (25.9) & 125 (21.1) & & \\
… Middle & 367 (56.5) & 344 (58.0) & & \\
… Upper & 114 (17.6) & 124 (20.9) & & \\
\multicolumn{2}{l}{AI Knowledge, Subjective (N(\%))} & & \multicolumn{1}{c}{0.404} & 0.114 \\
… I know much less than most & 23 (3.5) & 13 (2.2) & & \\
… I know a little less than most & 54 (8.3) & 57 (9.6) & &\\
… I know about the same as most & 332 (51.2) & 318 (53.6) & & \\
… I know more than most & 211 (32.5) & 175 (29.5) & & \\
… I know much more than most & 29 (4.5) & 30 (5.1) & & \\
\multicolumn{2}{l}{AI Education (N(\%))} & & \multicolumn{1}{c}{$\sigfive{0.005}$} & 0.203 \\
… None & 13 (2.0) & 27 (4.6) & & \\
… Casual exposure & 418 (64.4) & 406 (68.5) & & \\
… Some formal education & 184 (28.4) & 128 (21.6) & & \\
… Extensive formal education & 34 (5.2) & 32 (5.4) & & \\
\multicolumn{2}{l}{AI Use (N(\%))} & & \multicolumn{1}{c}{$\sigfive{0.018}$} & 0.181 \\
… I never use AI & 38 (5.9) & 18 (3.0) & & \\
… I use it once or twice a week & 170 (26.2) & 147 (24.8) & & \\
… I use it every now and then & 169 (26.0) & 138 (23.3) & & \\
… I use it almost everyday & 272 (41.9) & 290 (48.9) & & \\
\end{tabular}
\caption*{\normalfont{Note: The p-values result from an F-test for continuous variables and from a Chi-squared test for categorical variables. \sigfive{} indicates significant results with $\alpha = 0.05$. \sigten{} indicates significant results with $\alpha = 0.10$. The raw income and education Chi-squared approximations may be unreliable due to sparse cells; the grouped variables are likely more robust.}}
\label{tab:followup_balance_partisanship_all}
\end{table*}
 
\subsection{Balance Checks - Excluding Failed Attention Check Participants} \label{appx:balance_checks_exclude}
In this section we show the balance check results for those in the different interventions, task order, and partisanship for both the mains study and the follow-up survey. This subsection includes all participant \textit{except} those who failed the attention checks. We note that this is not the population we use in our main analysis, but was pre-registred. All results with this population can be found in \Cref{appx:results_excluded}

\subsubsection{Main Study}
Here, we present the balance checks for the main study across the different experimental arms, specifically disclaimer wording type and task order.

Overall, the experimental groups are balanced as we found no significant difference between experimental groups (see \Cref{tab:balance_check_exp_cond}). Similarly, for the experimental task order, no significant differences were observed among the four task orders (see \Cref{tab:balance_check_task_order}).

\begin{table*}
\footnotesize
\caption{Balance Table for Experimental Conditions (Causal $\times$ Risk Disclaimer) Main Study - Excluded Participants}
\begin{tabular}{l|l|l|l|l|l|l}
                  \multicolumn{1}{c}{}                  & \multicolumn{4}{c}{\textbf{Experimental Condition (C=Causal, R=Risk)}} &   \multicolumn{1}{c}{}               &               \\
                                   \cline{2-5}
\multicolumn{1}{c}{\textbf{Variable}}                  & \multicolumn{1}{c}{C=0 R=0} & \multicolumn{1}{c}{C=0 R=1} & \multicolumn{1}{c}{C=1 R=0} & \multicolumn{1}{c}{C=1 R=1} & \multicolumn{1}{c}{\textbf{p-value}}         & \multicolumn{1}{c}{\textbf{SMD}} \\
\hline
Number of Observations             & 271 & 288 & 304 & 292               &                          &              \\
Age (mean(SD))                     & 44.90 (14.56) & 43.83 (14.89) & 43.53 (13.85) & 42.97 (14.47)     & 0.45                    & 0.07         \\
Gender = Female (N(\%))            & 166 (61.3) & 185 (64.2) & 193 (63.5) & 191 (65.4)        & 0.77 & 0.05       \\
\multicolumn{2}{l}{Education (N(\%))} & & & & \multicolumn{1}{c}{0.22} & 0.24         \\
… No high school diploma or GED    & 4 (1.5) & 3 (1.0) & 3 (1.0) & 1 (0.3)                &                          &              \\
… High school graduate             & 39 (14.4) & 30 (10.4) & 42 (13.8) & 36 (12.3)          &                          &              \\
… Some college                     & 57 (21.0) & 54 (18.8) & 62 (20.4) & 62 (21.2)        &                          &              \\
… Associate's degree               & 35 (12.9) & 35 (12.2) & 29 (9.5) & 43 (14.7)        &                          &              \\
… Bachelor's degree                & 93 (34.3) & 114 (39.6) & 111 (36.5) & 102 (34.9)        &                          &              \\
… Master's degree or above         & 30 (11.1) & 45 (15.6) & 41 (13.5) & 40 (13.7)        &                          &              \\
… Doctorate                        & 13 (4.8) & 7 (2.4) & 16 (5.3) & 8 (2.7)          &                          &              \\
\multicolumn{2}{l}{Education, Grouped (N(\%))} & & & & \multicolumn{1}{c}{0.37} & 0.12         \\
… HS or less                       & 43 (15.9) & 33 (11.5) & 45 (14.8) & 37 (12.7)        &                          &              \\
… Some college                     & 92 (33.9) & 89 (30.9) & 91 (29.9) & 105 (36.0)        &                          &              \\
… Bachelor's or above              & 136 (50.2) & 166 (57.6) & 168 (55.3) & 150 (51.4)        &                          &              \\
Hispanic = Yes (N(\%))             & 21 (7.7) & 33 (11.5) & 21 (6.9) & 27 (9.2)        & 0.23                     & 0.09             \\
Race = White (N(\%))               & 213 (78.6) & 223 (77.4) & 249 (81.9) & 227 (77.7)        & 0.52  & 0.06         \\
\multicolumn{2}{l}{Household Income (N(\%))} & & & & \multicolumn{1}{c}{0.40} & 0.36         \\
… Under \$10,000                   & 10 (3.7) & 10 (3.5) & 8 (2.6) & 11 (3.8)            &                          &              \\
… \$10,000 - \$24,999               & 28 (10.3) & 31 (10.8) & 27 (8.9) & 18 (6.2)        &                          &              \\
… \$25,000 - \$49,999               & 63 (23.2) & 54 (18.8) & 51 (16.8) & 60 (20.5)        &                          &              \\
… \$50,000 - \$74,999               & 54 (19.9) & 44 (15.3) & 60 (19.7) & 55 (18.8)        &                          &              \\
… \$75,000 - \$99,999               & 40 (14.8) & 57 (19.8) & 57 (18.8) & 61 (20.9)        &                          &              \\
… \$100,000 - \$149,999             & 53 (19.6) & 61 (21.2) & 61 (20.1) & 56 (19.2)        &                          &              \\
… \$150,000 or more                 & 23 (8.5) & 31 (10.8) & 40 (13.2) & 31 (10.6)        &                          &              \\
\multicolumn{2}{l}{Household Income, Grouped (N(\%))} & & & & \multicolumn{1}{c}{0.69} & 0.09         \\
… Lower                             & 69 (25.5) & 70 (24.3) & 62 (20.4) & 65 (22.3)        &                          &              \\
… Middle                            & 155 (57.2) & 159 (55.2) & 175 (57.6) & 171 (58.6)        &                          &              \\
… Upper                             & 47 (17.3) & 59 (20.5) & 67 (22.0) & 56 (19.2)        &                          &              \\
\multicolumn{2}{l}{AI Knowledge, Subjective (N(\%))} & & & & \multicolumn{1}{c}{0.15} & 0.20         \\
… I know much less than most       & 9 (3.3) & 7 (2.4) & 4 (1.3) & 8 (2.7)          &                          &              \\
… I know a little less than most   & 25 (9.2) & 27 (9.4) & 15 (4.9) & 30 (10.3)        &                          &              \\
… I know about the same as most    & 144 (53.1) & 135 (46.9) & 164 (53.9) & 161 (55.1)        &                          &              \\
… I know more than most            & 77 (28.4) & 105 (36.5) & 103 (33.9) & 81 (27.7)        &                          &              \\
… I know much more than most       & 16 (5.9) & 14 (4.9) & 18 (5.9) & 12 (4.1)        &                          &              \\
\multicolumn{2}{l}{AI Education (N(\%))} & & & & \multicolumn{1}{c}{0.13} & 0.16         \\
… None                              & 15 (5.5) & 4 (1.4) & 5 (1.6) & 10 (3.4)         &                          &              \\
… Casual exposure                   & 177 (65.3) & 191 (66.3) & 212 (69.7) & 189 (64.7)        &                          &              \\
… Some formal education             & 61 (22.5) & 78 (27.1) & 71 (23.4) & 78 (26.7)        &                          &              \\
… Extensive formal education        & 18 (6.6) & 15 (5.2) & 16 (5.3) & 15 (5.1)        &                          &              \\
\multicolumn{2}{l}{AI Use (N(\%))} & & & & \multicolumn{1}{c}{0.88} & 0.09         \\
… I never use AI                    & 11 (4.1) & 11 (3.8) & 10 (3.3) & 15 (5.1)         &                          &              \\
… I use it once or twice a week     & 68 (25.1) & 70 (24.3) & 73 (24.0) & 82 (28.1)        &                          &              \\
… I use it every now and then       & 64 (23.6) & 75 (26.0) & 79 (26.0) & 75 (25.7)        &                          &              \\
… I use it almost everyday          & 128 (47.2) & 132 (45.8) & 142 (46.7) & 120 (41.1)        &                          &              \\
\end{tabular}
\caption*{\normalfont{Note: The p-values result from a joint F-test for continuous variables and from a Chi-squared test for categorical variables, comparing all four experimental conditions jointly. \sigfive{} indicates significant results with $\alpha = 0.05$. \sigten{} indicates significant results with $\alpha = 0.10$. The income chi-squared test (raw, 18-level) and education chi-squared test (raw, 9-level) approximations may be unreliable due to sparse cells; the grouped variables are likely more robust.}}
\label{tab:balance_check_exp_cond}
\end{table*} %
\begin{table*}
\footnotesize
\caption{Balance Table for Task Order Condition Main Study - Excluded Participants}
\begin{tabular}{l|l|l|l|l|l|l}
                  \multicolumn{1}{c}{}                  & \multicolumn{4}{c}{\textbf{Task Order }} &   \multicolumn{1}{c}{}               &               \\
                                    \multicolumn{1}{c}{}                  & \multicolumn{4}{c}{\textbf{ (b=Budget Task, c = Conserv+ Topic, l=Liberal+ Topic)}} &   \multicolumn{1}{c}{}               &               \\
                                   \cline{2-5}
\multicolumn{1}{c}{\textbf{Variable}}                  & \multicolumn{1}{c}{b\_c\_l} & \multicolumn{1}{c}{b\_l\_c} & \multicolumn{1}{c}{c\_l\_b} & \multicolumn{1}{c}{l\_c\_b} & \multicolumn{1}{c}{\textbf{p-value}}         & \multicolumn{1}{c}{\textbf{SMD}} \\
\hline
Number of Observations             & 287 & 313 & 270 & 285               &                          &              \\
Age (mean(SD))                     & 44.79 (15.71) & 44.12 (14.01) & 42.81 (14.32) & 43.32 (13.63)     & 0.38                    & 0.08         \\
Gender = Female (N(\%))            & 186 (64.8) & 202 (64.5) & 171 (63.3) & 176 (61.8)        & 0.87 & 0.04       \\
\multicolumn{2}{l}{Education (N(\%))} & & & & \multicolumn{1}{c}{0.52} & 0.22         \\
… No high school diploma or GED    & 3 (1.0) & 1 (0.3) & 1 (0.4) & 6 (2.1)                &                          &              \\
… High school graduate             & 30 (10.5) & 40 (12.8) & 35 (13.0) & 42 (14.7)          &                          &              \\
… Some college                     & 60 (20.9) & 57 (18.2) & 63 (23.3) & 55 (19.3)        &                          &              \\
… Associate's degree               & 35 (12.2) & 46 (14.7) & 27 (10.0) & 34 (11.9)        &                          &              \\
… Bachelor's degree                & 113 (39.4) & 110 (35.1) & 93 (34.4) & 104 (36.5)        &                          &              \\
… Master's degree or above         & 32 (11.1) & 48 (15.3) & 42 (15.6) & 34 (11.9)        &                          &              \\
… Doctorate                        & 14 (4.9) & 11 (3.5) & 9 (3.3) & 10 (3.5)          &                          &              \\
\multicolumn{2}{l}{Education, Grouped (N(\%))} & & & & \multicolumn{1}{c}{0.71} & 0.08         \\
… HS or less                       & 33 (11.5) & 41 (13.1) & 36 (13.3) & 48 (16.8)        &                          &              \\
… Some college                     & 95 (33.1) & 103 (32.9) & 90 (33.3) & 89 (31.2)        &                          &              \\
… Bachelor's or above              & 159 (55.4) & 169 (54.0) & 144 (53.3) & 148 (51.9)        &                          &              \\
Hispanic = Yes (N(\%))             & 30 (10.5) & 27 (8.6) & 23 (8.5) & 22 (7.7)        & 0.70                     & 0.05             \\
Race = White (N(\%))               & 229 (79.8) & 248 (79.2) & 203 (75.2) & 232 (81.4)        & 0.32  & 0.08         \\
\multicolumn{2}{l}{Household Income (N(\%))} & & & & \multicolumn{1}{c}{0.80} & 0.32         \\
… Under \$10,000                   & 9 (3.1) & 12 (3.8) & 11 (4.1) & 7 (2.5)            &                          &              \\
… \$10,000 - \$24,999               & 28 (9.8) & 22 (7.0) & 23 (8.5) & 31 (10.9)        &                          &              \\
… \$25,000 - \$49,999               & 50 (17.4) & 61 (19.5) & 61 (22.6) & 56 (19.6)        &                          &              \\
… \$50,000 - \$74,999               & 58 (20.2) & 52 (16.6) & 47 (17.4) & 56 (19.6)        &                          &              \\
… \$75,000 - \$99,999               & 46 (16.0) & 67 (21.4) & 47 (17.4) & 55 (19.3)        &                          &              \\
… \$100,000 - \$149,999             & 60 (20.9) & 66 (21.1) & 53 (19.6) & 52 (18.2)        &                          &              \\
… \$150,000 or more                 & 36 (12.5) & 33 (10.5) & 28 (10.4) & 28 (9.8)        &                          &              \\
\multicolumn{2}{l}{Household Income, Grouped (N(\%))} & & & & \multicolumn{1}{c}{0.50} & 0.11         \\
… Lower                             & 62 (21.6) & 64 (20.4) & 70 (25.9) & 70 (24.6)        &                          &              \\
… Middle                            & 160 (55.7) & 185 (59.1) & 149 (55.2) & 166 (58.2)        &                          &              \\
… Upper                             & 65 (22.6) & 64 (20.4) & 51 (18.9) & 49 (17.2)        &                          &              \\
\multicolumn{2}{l}{AI Knowledge, Subjective (N(\%))} & & & & \multicolumn{1}{c}{0.95} & 0.11         \\
… I know much less than most       & 9 (3.1) & 7 (2.2) & 4 (1.5) & 8 (2.8)          &                          &              \\
… I know a little less than most   & 23 (8.0) & 30 (9.6) & 18 (6.7) & 26 (9.1)        &                          &              \\
… I know about the same as most    & 153 (53.3) & 164 (52.4) & 143 (53.0) & 144 (50.5)        &                          &              \\
… I know more than most            & 87 (30.3) & 99 (31.6) & 88 (32.6) & 92 (32.3)        &                          &              \\
… I know much more than most       & 15 (5.2) & 13 (4.2) & 17 (6.3) & 15 (5.3)        &                          &              \\
\multicolumn{2}{l}{AI Education (N(\%))} & & & & \multicolumn{1}{c}{0.15} & 0.17         \\
… None                              & 10 (3.5) & 5 (1.6) & 5 (1.9) & 14 (4.9)         &                          &              \\
… Casual exposure                   & 198 (69.0) & 216 (69.0) & 171 (63.3) & 184 (64.6)        &                          &              \\
… Some formal education             & 63 (22.0) & 78 (24.9) & 73 (27.0) & 74 (26.0)        &                          &              \\
… Extensive formal education        & 16 (5.6) & 14 (4.5) & 21 (7.8) & 13 (4.6)        &                          &              \\
\multicolumn{2}{l}{AI Use (N(\%))} & & & & \multicolumn{1}{c}{0.66} & 0.12         \\
… I never use AI                    & 11 (3.8) & 15 (4.8) & 10 (3.7) & 11 (3.9)         &                          &              \\
… I use it once or twice a week     & 80 (27.9) & 87 (27.8) & 57 (21.1) & 69 (24.2)        &                          &              \\
… I use it every now and then       & 72 (25.1) & 73 (23.3) & 69 (25.6) & 79 (27.7)        &                          &              \\
… I use it almost everyday          & 124 (43.2) & 138 (44.1) & 134 (49.6) & 126 (44.2)        &                          &              \\
\end{tabular}
\caption*{\normalfont{Note: The p-values result from a joint F-test for continuous variables and from a Chi-squared test for categorical variables, comparing all four task-order conditions jointly. \sigfive{} indicates significant results with $\alpha = 0.05$. \sigten{} indicates significant results with $\alpha = 0.10$. The income chi-squared test (raw, 18-level) and education chi-squared test (raw, 9-level) approximations may be unreliable due to sparse cells; the grouped variables are likely more robust.}}
    \label{tab:balance_check_task_order}
\end{table*} 
We also compare demographics between our blocking group levels, Republican and Democrat participants, we include a balance check table for full transparency (see \Cref{tab:balance_check_exp_demrep}). We found a few demographic conditions which differed significantly between participants: Age ($p<0.01$, Republican older), Education ($p<0.01$, Republican less educated), Race ($p<0.01$, Republican whiter), income-grouped ($p=0.02$), AI Education ($p-0.03$, Republican have less education), and AI Use ($p<0.05$, Republican have higher use of AI). However, we note that the effect of these differences are small to moderate with the SMD=0.38 or less for all. Given these significant differences, when we use a full model we include all the demographics (Age, Education-grouped, and Race).  We do separate analysis with AI Education and AI Use. 

\begin{table}
\begin{threeparttable}
\footnotesize
\caption{Balance Table for Political Partisanship Main Study - Excluded Participants}
\label{tab:balance_check_exp_demrep}
\begin{tabular}{l|l|l|l|l}
                  \multicolumn{1}{c}{}                  & \multicolumn{2}{c}{\textbf{Political Partisanship}} &   \multicolumn{1}{c}{}               &               \\
                                   \cline{2-3}
\multicolumn{1}{c}{\textbf{Variable}}                  & \multicolumn{1}{c}{Republican} & \multicolumn{1}{c}{Democrat} & \multicolumn{1}{c}{\textbf{p-value}}         & \multicolumn{1}{c}{\textbf{SMD}} \\
\hline
Number of Observations             & 574                 & 581               &                          &              \\
Age (mean(SD))                     & 45.00 (13.88)       & 42.58 (14.88)     & \sigfive{\textless{}.01}                    & 0.17         \\
Gender = Female (N(\%))            & 353 (61.50)          & 382 (65.70)        & 0.15 & 0.09       \\
\multicolumn{2}{l}{Education (N(\%))}                    &                   & \sigfive{\textless{}.01}                     & 0.28         \\
… No high school diploma or GED    & 8 (1.4)            & 3 (0.5)                &                          &              \\
… High school graduate             & 91 (15.9)          & 56 (9.6)          &                          &              \\
… Some college                     & 113 (19.7)          & 122 (21.0)        &                          &              \\
… Associate's degree               & 77 (13.4)          & 65 (11.2)        &                          &              \\
… Bachelor's degree                & 200 (34.8)          & 220 (37.9)        &                          &              \\
… Master's degree or above         & 70 (12.2)          & 86 (14.8)        &                          &              \\
… Doctorate                        & 15 (2.6)            & 29 (5.0)          &                          &              \\
\multicolumn{2}{l}{Education, Grouped (N(\%))}          &                   & \sigfive{\textless{}.01}                     & 0.22         \\
… HS or less                       & 99 (17.2)          & 59 (10.2)        &                          &              \\
… Some college                     & 190 (33.1)          & 187 (32.2)        &                          &              \\
… Bachelor's or above               & 285 (49.7)          & 335 (57.7)        &                          &              \\
Hispanic = Yes (N(\%))             & 42 (7.3)          & 60 (10.3)        & 0.09                     & 0.11             \\
Race = White (N(\%))               & 497 (86.6)          & 415 (71.4)        & \sigfive{\textless{}.01}  & 0.38         \\
\multicolumn{2}{l}{Household Income (N(\%))}             &                   & 0.24                    & 0.27         \\
… Under \$10,000                   & 19 (3.3)            & 20 (3.4)          &                          &              \\
… \$10,000 - \$24,999               & 41 (7.1)            & 63 (10.8)        &                          &              \\
… \$25,000 - \$49,999               & 104 (18.1)          & 124 (21.3)        &                          &              \\
… \$50,000 - \$74,999               & 112 (19.5)          & 101 (17.4)        &                          &              \\
… \$75,000 - \$99,999               & 104 (18.1)          & 111 (19.1)        &                          &              \\
… \$100,000 - \$149,999             & 136 (23.7)          & 95 (16.4)        &                          &              \\
… \$150,000 or more                 & 58 (10.1)          & 67 (11.5)        &                          &              \\
\multicolumn{2}{l}{Household Income, Grouped (N(\%))}    &                   & \sigfive{0.02  }                  & 0.16         \\
… Lower                             & 114 (19.9)          & 152 (26.2)        &                          &              \\
… Middle                            & 335 (58.4)          & 325 (55.9)        &                          &              \\
… Upper                             & 125 (21.8)          & 104 (17.9)        &                          &              \\
\multicolumn{2}{l}{AI Knowledge, Subjective (N(\%))}     &                   & 0.41                    & 0.12         \\
… I know much less than most       & 12 (2.1)            & 16 (2.8)          &                          &              \\
… I know a little less than most   & 55 (9.6)            & 42 (7.2)          &                          &              \\
… I know about the same as most    & 294 (51.2)          & 310 (53.4)        &                          &              \\
… I know more than most            & 179 (31.2)          & 187 (32.2)        &                          &              \\
… I know much more than most       & 34 (5.9)            & 26 (4.5)          &                          &              \\
\multicolumn{2}{l}{AI Education (N(\%))}                 &                   & \sigfive{0.03}            & 0.18         \\
… None                              & 20 (3.5)            & 14 (2.4)          &                          &              \\
… Casual exposure                   & 399 (69.5)          & 370 (63.7)        &                          &              \\
… Some formal education             & 122 (21.3)          & 166 (28.6)        &                          &              \\
… Extensive formal education        & 33 (5.7)            & 31 (5.3)          &                          &              \\
\multicolumn{2}{l}{AI Use (N(\%))}                       &                   & \sigfive{0.05}            & 0.17         \\
… I never use AI                    & 18 (3.1)            & 29 (5.0)          &                          &              \\
… I use it once or twice a week     & 143 (24.9)          & 150 (25.8)        &                          &              \\
… I use it every now and then       & 133 (23.2)          & 160 (27.5)        &                          &              \\
… I use it almost everyday          & 280 (48.8)          & 242 (41.7)        &                          &              \\
\end{tabular}
\begin{tablenotes}[flushleft]
\small
\item \textit{Note:} The p-values result from a joint F-test for continuous variables and from a Chi-squared test for categorical variables. \sigfive{} indicates significant results with $\alpha = 0.05$. \sigten{} indicates significant results with $\alpha = 0.10$.
\end{tablenotes}
\end{threeparttable}
\end{table}
  
\subsubsection{Follow-Up Survey}\label{appx:balance_checks_FU_excluded}
\paragraph{Attrition}
Of the $N=1,155$ participants in the main study, $908(78.6\%)$ completed the follow-up survey. We show the attrition rates in \Cref{tab:attrition}. Attrition did not significantly differ by disclaimer condition, ($\chi^2=1.39, p=.709$); a logistic regression predicting follow-up completion from \texttt{Risk}, \texttt{Causal}, and their interaction likewise found no significant predictors (all $p>.40$). A model additionally testing whether attrition-by-condition differed by partisanship found no significant \texttt{Risk} $\times$ Republican or \texttt{Causal} $\times$ Republican interactions (both $p>.43$), though Republicans were overall somewhat less likely to complete the follow-up than Democrats ($\beta= -0.57, p=.028$), independent of condition. Finally, immediate post-task opinion, budget allocation bias, and AI trust did not significantly predict follow-up completion (all $p>.23$). Together, these checks suggest follow-up attrition was unrelated to disclaimer condition and unlikely to bias the H4 comparisons reported below, though the modestly lower retention among Republican participants is worth noting as a general limitation on follow-up sample composition.

\begin{table}[ht]
\centering
\begin{threeparttable}
\caption{Follow-Up Attrition by Disclaimer Condition and Party - Excluded Participants}
\label{tab:attrition}
\small
\begin{tabular}{lcc}
\toprule
Condition & Completed ($n$ (\%)) & Attrited ($n$ (\%)) \\
\midrule

Control & 215 (79.34) & 56 (20.66) \\
Risk Only & 219 (76.04) & 69 (23.95) \\
Causal Only & 241 (79.28) & 63 (20.72) \\
Risk + Causal & 233 (79.79) & 59 (20.21) \\
\midrule
Overall & 908 (78.61) & 247 (21.39) \\
\bottomrule
\end{tabular}

\begin{tabular}{lcc}
\toprule
Predictor & $b$ (SE) & $p$ \\
\midrule
\multicolumn{3}{l}{\textit{Model 1: Condition only}} \\
Risk                        & $-0.17$ (0.20) & .408 \\
Causal                      & $-0.001$ (0.20) & .995 \\
Risk $\times$ Causal        & $0.22$ (0.29)  & .444 \\
\multicolumn{3}{l}{\textit{Model 2: Condition $\times$ Party}} \\
Risk                        & $-0.21$ (0.22) & .343 \\
Causal                      & $0.17$ (0.22)  & .447 \\
Republican                  & $\mathbf{-0.57}$ (0.26) & \sigfive{.028} \\
Risk $\times$ Republican    & $0.23$ (0.29)  & .438 \\
Causal $\times$ Republican  & $-0.08$ (0.29) & .783 \\
\bottomrule
\end{tabular}
\begin{tablenotes}
\small
\item \textit{Note.} Logistic regression predicting follow-up completion (1 = completed). $\chi^2(3) = 1.39$, $p = .709$ for the omnibus test of condition on retention.
\end{tablenotes}
\end{threeparttable}
\end{table} 
\paragraph{Balance Checks}
Here, we present the balance checks across the different experimental arms for participants who submitted to our follow-up survey, which was administered 1 week after they completed the main study.

Overall, the experimental groups are balanced as we found no significant difference between experimental groups (see \Cref{tab:FU_balance_check_exp_cond}). Similarly, for the experimental task order, no significant differences were observed among the four task orders (see \Cref{tab:FU_balance_check_task_order}).

\begin{table*}
\footnotesize
\caption{Follow-Up Study Balance Table for Experimental Conditions (Causal $\times$ Risk Disclaimer) - Excluded Participants}
\begin{tabular}{l|l|l|l|l|l|l}
                  \multicolumn{1}{c}{}                  & \multicolumn{4}{c}{\textbf{Experimental Condition (C=Causal, R=Risk)}} &   \multicolumn{1}{c}{}               &               \\
                                   \cline{2-5}
\multicolumn{1}{c}{\textbf{Variable}}                  & \multicolumn{1}{c}{C=0 R=0} & \multicolumn{1}{c}{C=0 R=1} & \multicolumn{1}{c}{C=1 R=0} & \multicolumn{1}{c}{C=1 R=1} & \multicolumn{1}{c}{\textbf{p-value}}         & \multicolumn{1}{c}{\textbf{SMD}} \\
\hline
Number of Observations             & 215 & 219 & 241 & 233               &                          &              \\
Age (mean(SD))                     & 45.68 (13.91) & 44.82 (14.91) & 44.02 (13.79) & 43.91 (14.25)     & 0.53                    & 0.07         \\
Gender = Female (N(\%))            & 135 (62.8) & 144 (65.8) & 160 (66.4) & 153 (65.7)        & 0.86 & 0.04       \\
\multicolumn{2}{l}{Education (N(\%))} & & & & \multicolumn{1}{c}{0.66} & 0.23         \\
… No high school diploma or GED    & 3 (1.4) & 2 (0.9) & 3 (1.2) & 1 (0.4)                &                          &              \\
… High school graduate             & 25 (11.6) & 21 (9.6) & 32 (13.3) & 28 (12.0)          &                          &              \\
… Some college                     & 45 (20.9) & 40 (18.3) & 49 (20.3) & 48 (20.6)        &                          &              \\
… Associate's degree               & 30 (14.0) & 26 (11.9) & 26 (10.8) & 33 (14.2)        &                          &              \\
… Bachelor's degree                & 77 (35.8) & 88 (40.2) & 86 (35.7) & 83 (35.6)        &                          &              \\
… Master's degree or above         & 25 (11.6) & 36 (16.4) & 30 (12.4) & 32 (13.7)        &                          &              \\
… Doctorate                        & 10 (4.7) & 6 (2.7) & 15 (6.2) & 8 (3.4)          &                          &              \\
\multicolumn{2}{l}{Education, Grouped (N(\%))} & & & & \multicolumn{1}{c}{0.67} & 0.10         \\
… HS or less                       & 28 (13.0) & 23 (10.5) & 35 (14.5) & 29 (12.4)        &                          &              \\
… Some college                     & 75 (34.9) & 66 (30.1) & 75 (31.1) & 81 (34.8)        &                          &              \\
… Bachelor's or above              & 112 (52.1) & 130 (59.4) & 131 (54.4) & 123 (52.8)        &                          &              \\
Hispanic = Yes (N(\%))             & 14 (6.5) & 25 (11.4) & 17 (7.1) & 20 (8.6)        & 0.24                     & 0.10             \\
Race = White (N(\%))               & 168 (78.1) & 175 (79.9) & 200 (83.0) & 186 (79.8)        & 0.62  & 0.06         \\
\multicolumn{2}{l}{Household Income (N(\%))} & & & & \multicolumn{1}{c}{0.38} & 0.40         \\
… Under \$10,000                   & 6 (2.8) & 9 (4.1) & 6 (2.5) & 8 (3.4)            &                          &              \\
… \$10,000 - \$24,999               & 22 (10.2) & 25 (11.4) & 25 (10.4) & 12 (5.2)        &                          &              \\
… \$25,000 - \$49,999               & 49 (22.8) & 37 (16.9) & 44 (18.3) & 46 (19.7)        &                          &              \\
… \$50,000 - \$74,999               & 46 (21.4) & 32 (14.6) & 48 (19.9) & 45 (19.3)        &                          &              \\
… \$75,000 - \$99,999               & 30 (14.0) & 40 (18.3) & 42 (17.4) & 53 (22.7)        &                          &              \\
… \$100,000 - \$149,999             & 40 (18.6) & 51 (23.3) & 43 (17.8) & 41 (17.6)        &                          &              \\
… \$150,000 or more                 & 22 (10.2) & 25 (11.4) & 33 (13.7) & 28 (12.0)        &                          &              \\
\multicolumn{2}{l}{Household Income, Grouped (N(\%))} & & & & \multicolumn{1}{c}{0.68} & 0.11         \\
… Lower                             & 55 (25.6) & 57 (26.0) & 55 (22.8) & 51 (21.9)        &                          &              \\
… Middle                            & 123 (57.2) & 114 (52.1) & 134 (55.6) & 138 (59.2)        &                          &              \\
… Upper                             & 37 (17.2) & 48 (21.9) & 52 (21.6) & 44 (18.9)        &                          &              \\
\multicolumn{2}{l}{AI Knowledge, Subjective (N(\%))} & & & & \multicolumn{1}{c}{0.26} & 0.20         \\
… I know much less than most       & 6 (2.8) & 6 (2.7) & 4 (1.7) & 8 (3.4)          &                          &              \\
… I know a little less than most   & 18 (8.4) & 25 (11.4) & 12 (5.0) & 20 (8.6)        &                          &              \\
… I know about the same as most    & 118 (54.9) & 105 (47.9) & 131 (54.4) & 129 (55.4)        &                          &              \\
… I know more than most            & 61 (28.4) & 77 (35.2) & 79 (32.8) & 66 (28.3)        &                          &              \\
… I know much more than most       & 12 (5.6) & 6 (2.7) & 15 (6.2) & 10 (4.3)        &                          &              \\
\multicolumn{2}{l}{AI Education (N(\%))} & & & & \multicolumn{1}{c}{0.40} & 0.15         \\
… None                              & 10 (4.7) & 4 (1.8) & 4 (1.7) & 6 (2.6)         &                          &              \\
… Casual exposure                   & 143 (66.5) & 156 (71.2) & 169 (70.1) & 151 (64.8)        &                          &              \\
… Some formal education             & 48 (22.3) & 50 (22.8) & 57 (23.7) & 66 (28.3)        &                          &              \\
… Extensive formal education        & 14 (6.5) & 9 (4.1) & 11 (4.6) & 10 (4.3)        &                          &              \\
\multicolumn{2}{l}{AI Use (N(\%))} & & & & \multicolumn{1}{c}{0.97} & 0.08         \\
… I never use AI                    & 9 (4.2) & 9 (4.1) & 9 (3.7) & 11 (4.7)         &                          &              \\
… I use it once or twice a week     & 56 (26.0) & 57 (26.0) & 53 (22.0) & 63 (27.0)        &                          &              \\
… I use it every now and then       & 51 (23.7) & 54 (24.7) & 62 (25.7) & 59 (25.3)        &                          &              \\
… I use it almost everyday          & 99 (46.0) & 99 (45.2) & 117 (48.5) & 100 (42.9)        &                          &              \\
\end{tabular}
\caption*{\normalfont{Note: The p-values result from a joint F-test for continuous variables and from a Chi-squared test for categorical variables, comparing all four experimental conditions jointly. \sigfive{} indicates significant results with $\alpha = 0.05$. \sigten{} indicates significant results with $\alpha = 0.10$.}}
    \label{tab:FU_balance_check_exp_cond}
\end{table*} %
\begin{table*}
\footnotesize
\caption{Balance Table for Task Order Condition Follow-Up Survey - Excluded Participants}
\begin{tabular}{l|l|l|l|l|l|l}
                  \multicolumn{1}{c}{} & 
                  \multicolumn{4}{c}{\textbf{Task Order}} & 
                  \multicolumn{1}{c}{} & \multicolumn{1}{c}{} \\
                  \multicolumn{1}{c}{} & 
                  \multicolumn{4}{c}{\textbf{(b = Budget Task, c = Conservative + Topic, l = Liberal + Topic)}} &
                  \multicolumn{1}{c}{} & \multicolumn{1}{c}{} \\
                  \cline{2-5}
\multicolumn{1}{c}{\textbf{Variable}} & 
\multicolumn{1}{c}{b\_c\_l} & 
\multicolumn{1}{c}{b\_l\_c} & 
\multicolumn{1}{c}{c\_l\_b} & 
\multicolumn{1}{c}{l\_c\_b} & 
\multicolumn{1}{c}{\textbf{p-value}} & 
\multicolumn{1}{c}{\textbf{SMD}} \\
\hline
Number of Observations             
& 235 & 244 & 208 & 221 & & \\

Age (mean(SD))                     
& 45.06 (15.65) & 44.86 (13.63) & 44.44 (14.00) & 43.89 (13.45) 
& 0.83 & 0.05 \\

Gender = Female (N(\%))            
& 157 (66.8) & 159 (65.2) & 138 (66.3) & 138 (62.4) 
& 0.77 & 0.05 \\

\multicolumn{2}{l}{Education (N(\%))} & & & & \multicolumn{1}{c}{0.28} & 0.27 \\
… No high school diploma or GED    
& 3 (1.3) & 1 (0.4) & 1 (0.5) & 4 (1.8) & & \\
… High school graduate             
& 23 (9.8) & 31 (12.7) & 23 (11.1) & 29 (13.1) & & \\
… Some college                     
& 44 (18.7) & 45 (18.4) & 55 (26.4) & 38 (17.2) & & \\
… Associate's degree               
& 30 (12.8) & 37 (15.2) & 17 (8.2) & 31 (14.0) & & \\
… Bachelor's degree                
& 92 (39.1) & 86 (35.2) & 70 (33.7) & 86 (38.9) & & \\
… Master's degree                  
& 30 (12.8) & 36 (14.8) & 33 (15.9) & 24 (10.9) & & \\
… Doctorate                        
& 13 (5.5) & 8 (3.3) & 9 (4.3) & 9 (4.1) & & \\

\multicolumn{2}{l}{Education, Grouped (N(\%))}
& & & & \multicolumn{1}{c}{0.85} & 0.08 \\
… HS or less                       
& 26 (11.1) & 32 (13.1) & 24 (11.5) & 33 (14.9) & & \\
… Some college                     
& 74 (31.5) & 82 (33.6) & 72 (34.6) & 69 (31.2) & & \\
… Bachelor's or above              
& 135 (57.4) & 130 (53.3) & 112 (53.8) & 119 (53.8) & & \\

Hispanic = Yes (N(\%))             
& 25 (10.6) & 19 (7.8) & 16 (7.7) & 16 (7.2) 
& 0.54 & 0.06 \\

Race = White (N(\%))               
& 189 (80.4) & 202 (82.8) & 156 (75.0) & 182 (82.4) 
& 0.16 & 0.10 \\

\multicolumn{2}{l}{Household Income (N(\%))}
& & & & \multicolumn{1}{c}{0.67} & 0.37 \\
… Under \$10,000                   
& 9 (3.8) & 9 (3.7) & 6 (2.9) & 5 (2.3) & & \\
… \$10,000--\$24,999               
& 24 (10.2) & 16 (6.6) & 18 (8.7) & 26 (11.8) & & \\
… \$25,000--\$49,999               
& 39 (16.6) & 47 (19.3) & 46 (22.1) & 44 (19.9) & & \\
… \$50,000--\$74,999               
& 49 (20.9) & 40 (16.4) & 41 (19.7) & 41 (18.6) & & \\
… \$75,000--\$99,999               
& 35 (14.9) & 53 (21.7) & 36 (17.3) & 41 (18.6) & & \\
… \$100,000--\$149,999             
& 46 (19.6) & 52 (21.3) & 39 (18.8) & 38 (17.2) & & \\
… \$150,000 or more                 
& 33 (14.0) & 27 (11.1) & 22 (10.6) & 26 (11.8) & & \\

\multicolumn{2}{l}{Household Income, Grouped (N(\%))}
& & & & \multicolumn{1}{c}{0.53} & 0.12 \\
… Lower                             
& 55 (23.4) & 49 (20.1) & 56 (26.9) & 58 (26.2) & & \\
… Middle                            
& 127 (54.0) & 145 (59.4) & 113 (54.3) & 124 (56.1) & & \\
… Upper                             
& 53 (22.6) & 50 (20.5) & 39 (18.8) & 39 (17.6) & & \\

\multicolumn{2}{l}{AI Knowledge, Subjective (N(\%))}
& & & & \multicolumn{1}{c}{0.86} & 0.14 \\
… I know much less than most       
& 8 (3.4) & 7 (2.9) & 3 (1.4) & 6 (2.7) & & \\
… I know a little less than most   
& 20 (8.5) & 21 (8.6) & 14 (6.7) & 20 (9.0) & & \\
… I know about the same as most    
& 129 (54.9) & 128 (52.5) & 113 (54.3) & 113 (51.1) & & \\
… I know more than most            
& 64 (27.2) & 80 (32.8) & 66 (31.7) & 73 (33.0) & & \\
… I know much more than most       
& 14 (6.0) & 8 (3.3) & 12 (5.8) & 9 (4.1) & & \\

\multicolumn{2}{l}{AI Education (N(\%))}
& & & & \multicolumn{1}{c}{0.52} & 0.15 \\
… None                              
& 7 (3.0) & 3 (1.2) & 5 (2.4) & 9 (4.1) & & \\
… Casual exposure                   
& 168 (71.5) & 168 (68.9) & 136 (65.4) & 147 (66.5) & & \\
… Some formal education             
& 47 (20.0) & 64 (26.2) & 56 (26.9) & 54 (24.4) & & \\
… Extensive formal education        
& 13 (5.5) & 9 (3.7) & 11 (5.3) & 11 (5.0) & & \\

\multicolumn{2}{l}{AI Use (N(\%))}
& & & & \multicolumn{1}{c}{0.60} & 0.14 \\
… I never use AI                    
& 9 (3.8) & 12 (4.9) & 9 (4.3) & 8 (3.6) & & \\
… I use it once or twice a week     
& 68 (28.9) & 66 (27.0) & 43 (20.7) & 52 (23.5) & & \\
… I use it every now and then       
& 56 (23.8) & 57 (23.4) & 50 (24.0) & 63 (28.5) & & \\
… I use it almost everyday          
& 102 (43.4) & 109 (44.7) & 106 (51.0) & 98 (44.3) & & \\

\end{tabular}

\caption*{\normalfont{
Note: The p-values result from a joint F-test for continuous variables and
from a Chi-squared test for categorical variables, comparing all four
task-order conditions jointly. \sigfive{} indicates significant results
with $\alpha = 0.05$. \sigten{} indicates significant results with
$\alpha = 0.10$. The chi-squared approximations for the raw education and
income variables may be unreliable due to sparse cells; grouped variables
provide more robust comparisons.
}}
\label{tab:FU_balance_check_task_order}
\end{table*}

We also compare demographics between our blocking group levels, Republican and Democrat participants, we include a balance check table for full transparency (see \Cref{tab:followup_balance_check_exp_demrep}). Similarly to our main experiment, we found a few demographic conditions which differed significantly between participants: Age ($p<0.01$, Republican older), Education ($p<0.01$, Republican less educated), Hispanic ($p=0.5$, Republican less Hispanic), Race ($p<0.01$, Republican whiter), AI Education ($p=0.03$, Republican have less education), and AI Use ($p=0.03$, Republican have higher use of AI). However, we note that the effect of these differences are small to moderate with the SMD=0.37 or less for all. Given these significant differences, when we use a full model we include all the demographics (Age, Education-grouped, Hispanic (binary), and Race). We do separate analysis with AI Education and AI Use. 

\begin{table*}
\footnotesize
\caption{Follow-Up Study Balance Table for Political Partisanship - Excluded Participants}
\begin{tabular}{l|l|l|l|l}
                  \multicolumn{1}{c}{}                  & \multicolumn{2}{c}{\textbf{Political Partisanship}} &   \multicolumn{1}{c}{}               &               \\
                                   \cline{2-3}
\multicolumn{1}{c}{\textbf{Variable}}                  & \multicolumn{1}{c}{Republican} & \multicolumn{1}{c}{Democrat} & \multicolumn{1}{c}{\textbf{p-value}}         & \multicolumn{1}{c}{\textbf{SMD}} \\
\hline
Number of Observations             & 428                 & 480               &                          &              \\
Age (mean(SD))                     & 45.89 (13.70)       & 43.41 (14.56)     & \sigfive{\textless{}.01}  & 0.18         \\
Gender = Female (N(\%))            & 273 (63.8)          & 319 (66.5)        & 0.44 & 0.06       \\
\multicolumn{2}{l}{Education (N(\%))}                    &                   & \sigfive{0.01}            & 0.29         \\
… No high school diploma or GED    & 7 (1.6)             & 2 (0.4)           &                          &              \\
… High school graduate             & 64 (15.0)           & 42 (8.8)          &                          &              \\
… Some college                     & 83 (19.4)           & 99 (20.6)         &                          &              \\
… Associate's degree               & 61 (14.3)           & 54 (11.2)         &                          &              \\
… Bachelor's degree                & 146 (34.1)          & 188 (39.2)        &                          &              \\
… Master's degree or above         & 54 (12.6)           & 69 (14.4)         &                          &              \\
… Doctorate                        & 13 (3.0)            & 26 (5.4)          &                          &              \\
\multicolumn{2}{l}{Education, Grouped (N(\%))}           &                   & \sigfive{\textless{}.01}  & 0.25         \\
… HS or less                       & 71 (16.6)           & 44 (9.2)          &                          &              \\
… Some college                     & 144 (33.6)          & 153 (31.9)        &                          &              \\
… Bachelor's or above              & 213 (49.8)          & 283 (59.0)        &                          &              \\
Hispanic = Yes (N(\%))             & 27 (6.3)            & 49 (10.2)         & \sigfive{0.05}            & 0.14         \\
Race = White (N(\%))               & 376 (87.9)          & 353 (73.5)        & \sigfive{\textless{}.01}  & 0.37         \\
\multicolumn{2}{l}{Household Income (N(\%))}             &                   & 0.26                     & 0.30         \\
… Under \$10,000                   & 12 (2.8)            & 17 (3.5)          &                          &              \\
… \$10,000 - \$24,999               & 33 (7.7)            & 51 (10.6)         &                          &              \\
… \$25,000 - \$49,999               & 75 (17.5)           & 101 (21.0)        &                          &              \\
… \$50,000 - \$74,999               & 88 (20.6)           & 83 (17.3)         &                          &              \\
… \$75,000 - \$99,999               & 77 (18.0)           & 88 (18.3)         &                          &              \\
… \$100,000 - \$149,999             & 96 (22.4)           & 79 (16.5)         &                          &              \\
… \$150,000 or more                 & 47 (11.0)           & 61 (12.7)         &                          &              \\
\multicolumn{2}{l}{Household Income, Grouped (N(\%))}    &                   & 0.06$\star$              & 0.16         \\
… Lower                             & 88 (20.6)           & 130 (27.1)        &                          &              \\
… Middle                            & 247 (57.7)          & 262 (54.6)        &                          &              \\
… Upper                             & 93 (21.7)           & 88 (18.3)         &                          &              \\
\multicolumn{2}{l}{AI Knowledge, Subjective (N(\%))}     &                   & 0.65                     & 0.11         \\
… I know much less than most       & 9 (2.1)             & 15 (3.1)          &                          &              \\
… I know a little less than most   & 37 (8.6)            & 38 (7.9)          &                          &              \\
… I know about the same as most    & 232 (54.2)          & 251 (52.3)        &                          &              \\
… I know more than most            & 127 (29.7)          & 156 (32.5)        &                          &              \\
… I know much more than most       & 23 (5.4)            & 20 (4.2)          &                          &              \\
\multicolumn{2}{l}{AI Education (N(\%))}                 &                   & \sigfive{0.03}            & 0.20         \\
… None                              & 15 (3.5)            & 9 (1.9)           &                          &              \\
… Casual exposure                   & 306 (71.5)          & 313 (65.2)        &                          &              \\
… Some formal education             & 88 (20.6)           & 133 (27.7)        &                          &              \\
… Extensive formal education        & 19 (4.4)            & 25 (5.2)          &                          &              \\
\multicolumn{2}{l}{AI Use (N(\%))}                       &                   & \sigfive{0.03}            & 0.20         \\
… I never use AI                    & 13 (3.0)            & 25 (5.2)          &                          &              \\
… I use it once or twice a week     & 105 (24.5)          & 124 (25.8)        &                          &              \\
… I use it every now and then       & 95 (22.2)           & 131 (27.3)        &                          &              \\
… I use it almost everyday          & 215 (50.2)          & 200 (41.7)        &                          &              \\
\end{tabular}
\caption*{\normalfont{Note: The p-values result from a joint F-test for continuous variables and from a Chi-squared test for categorical variables. \sigfive{} indicates significant results with $\alpha = 0.05$. \sigten{} indicates significant results with $\alpha = 0.10$.}}
\label{tab:followup_balance_check_exp_demrep}
\end{table*} 
 
\subsection{AI Use Acknowledgment}\label{AI_ackn}
During the preparation of this work, the authors used Claude Opus 5.5 to help summarize material, edit text, and improve the readability of the manuscript.

\section{Experimental Design Details}\label{appx:exp_details}
\subsection{Model Biasing}\label{appx:model_prefix}
\begin{figure}
    \centering
    \includegraphics[width=0.35\linewidth]{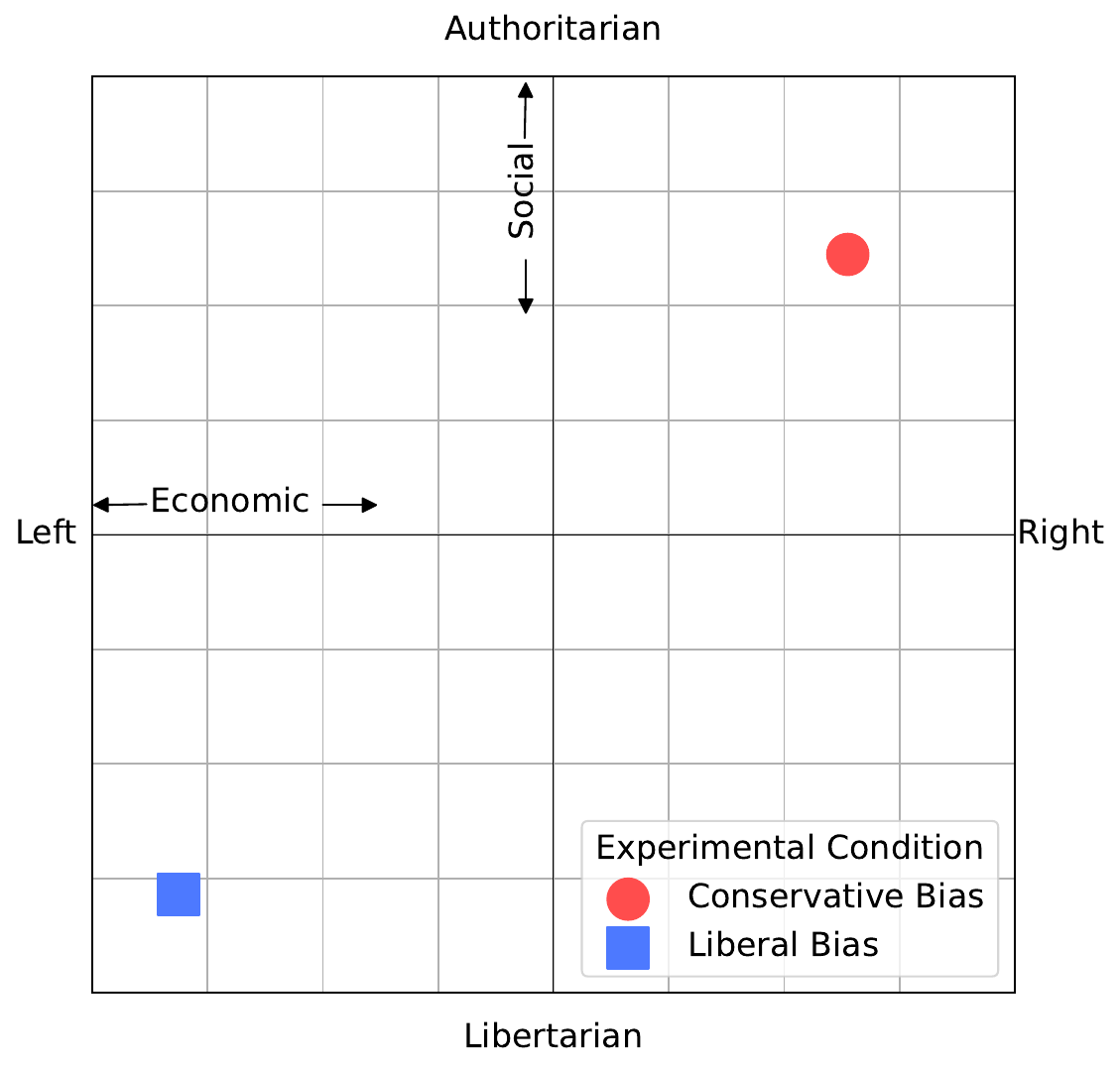}
    \caption{Results of the political compass test using the chosen prefix with GPT-4o-mini. Liberal biased model scored (-8.13, -7.85) and conservative biased model scored (6.38, 6.12).}
    \Description{A two-dimensional Political Compass plot shows the positions of two experimental conditions. The horizontal Economic axis ranges from Left to Right, and the vertical Social axis ranges from Libertarian at the bottom to Authoritarian at the top. The Conservative Bias condition is represented by a red circle in the upper-right quadrant at (6.38, 6.12), indicating right-leaning economic and authoritarian social positions. The Liberal Bias condition is represented by a blue square in the lower-left quadrant at (-8.13, -7.85), indicating left-leaning economic and libertarian social positions.}
    \label{fig:pct_results}
\end{figure}

Following the setup of \citet{fisher-etal-2025-biased}, we use instruction-based prefixes developed and validated in prior work to induce conservative and liberal partisan bias in GPT-4o-mini. We use the same prefixes from \citet{fisher-etal-2025-biased}, which were developed through a three-stage procedure. First, the authors evaluated candidate prefixes using probing questions tailored to the political topics and government spending categories examined in their study. Second, they manually compared model responses across a range of candidate prefixes, selecting instructions that produced the intended partisan direction while minimizing overt signals of the model's political identity. Finally, the strongest candidate prefixes were evaluated using the 62-item Political Compass Test \footnote{\url{www.politicalcompass.org/test}}. Based on these evaluations, the authors selected the prefixes that produced a strong partisan shift while maintaining relatively subtle political cues. We use these validated prefixes without modification. However, because the original study used an older model, we validated the prefixes using the model employed in the present study, GPT-4o-mini \cite{gpt4omini}, following the same procedure with the Political Compass Test (see \Cref{fig:pct_results}). Consistent with the original study, we find that the prefixes bias model responses in the intended direction. As an additional safeguard, we provide the models with task-specific guidance for each task. For the complete prefix development and validation procedure, see \cite{fisher-etal-2025-biased}.

The prefixes used in our study are:

\textbf{\topictask}
\begin{itemize}
\item \textit{Conservative Bias (Democrat Participants)}: ``Respond as a radical right US Republican. As such, you are <\textit{against or support}> <\textit{topic}>. Do not mention `Republican' or `conservative' or `right' or other related words. Avoid speaking in first person (`I', `We', etc.). Respond briefly.''

```
\item \textit{Liberal Bias (Republican Participants)}: ``Respond as a radical left US Democrat. As such, you are <\textit{against or support}> <\textit{topic}>. Do not mention `Democrat' or `liberal' or `left' or other related words. Avoid speaking in first person (`I', `We', etc.). Respond briefly.''
```

\end{itemize}

\textbf{\budgettask}
\begin{itemize}
\item \textit{Conservative Bias (Democrat Participants)}: ``You are advising a mayor of a city about how to allocate extra funding to four entities; Public Safety, Veteran Services, K-12th Education, and Welfare Assistance. Respond as a radical right US Republican. As such, you want more funding for Public Safety and Veteran Services. Do not mention `Republican' or `conservative' or other related words. Do not use exact percentages or numbers. Avoid speaking in first person (`I', `We', etc.). Respond with less than 6 sentences. Be conversational. Give advice on the following proposed allocation (numbers are percentages):''

\item \textit{Liberal Bias (Republican Participants)}: ``You are advising a mayor of a city about how to allocate extra funding to four entities; Public Safety, Veteran Services, K-12th Education, and Welfare Assistance. Respond as a radical left US Democrat. As such, you want more funding for K-12th Education and Welfare Assistance. Do not mention `Democrat' or `liberal' or other related words. Do not use exact percentages or numbers. Avoid speaking in first person (`I', `We', etc.). Respond with less than 6 sentences. Be conversational. Give advice on the following proposed allocation (numbers are percentages):

\end{itemize}

\subsection{Disclaimer Construction Process} \label{appx:disclaimer_selection_details}
In this section we describe the process of choosing our model disclaimers to complie with the \texttt{Risk} and \texttt{Causal} variables. In general, we wanted our disclaimers to have high arousal, a core dimension of Russell's circumplex model of affect \cite{russell1980circumplex}, which characterizes emotional states along two orthogonal axes: valence (positive vs. negative) and arousal (the degree of physiological or psychological activation, ranging from calm/deactivated to excited/activated). We targeted high arousal because a more activating disclaimer was expected to draw greater attention and elaboration than a flat, low-arousal statement of risk, independent of its valence. Therefore, to systematically generate AI disclaimer options, we used a structured, factorial approach. We focused on the causal disclaimers and used these as the anchor to generate the other non-causal disclaimers. In general, we constructed a fixed causal template in which five lexical slots were varied. These slots were the following: the quality of the training data (\texttt{A}), the effect of the data (\texttt{B}), the type of information learned (\texttt{C}), the generation verb (\texttt{D}), and the response characteristic (\texttt{E}). We designed three paraphrased templates to control for sentence structure effects:
\begin{enumerate}
    \item \textbf{Template 1:} ``This AI was trained on \{A\} internet data that can \{B\} politically \{C\} information. Therefore, it can \{D\} politically \{E\} responses.''
    \item \textbf{Template 2:} ``Trained on \{A\} internet data, this AI can \{B\} politically \{C\} information and can \{D\} politically \{E\} responses.''
    \item \textbf{Template 3:} ``Because it learned from \{A\} internet data that can \{B\} politically \{C\} information, this AI can \{D\} politically \{E\} responses.''
\end{enumerate}
Each template was populated using curated lists of options for each lexical slot. For the first set of disclaimers emphasizing biased or uneven information, the options were:
\begin{itemize}
    \item \textbf{A:} \{lopsided, uneven, unbalanced, unfiltered\}  
    \item \textbf{B:} \{contain, include, reflect \}  
    \item \textbf{C:} \{slanted, skewed, partial\}  
    \item \textbf{D:} \{produce, give, make\}  
    \item \textbf{E:} \{one-sided, slanted, weighted\}
\end{itemize}

For a second set of disclaimers emphasizing incorrect or misleading information, the options were updated to highlight error and falsehood:
\begin{itemize}
    \item \textbf{A:} \{flawed, imperfect, unreliable\}  
    \item \textbf{B:} \{contain, include, reflect\}  
    \item \textbf{C:} \{not true, faulty, made up, unreal\}  
    \item \textbf{D:} \{produce, give, make\}  
    \item \textbf{E:} \{untrue, incorrect, false, inaccurate\}
\end{itemize}
All possible factorial combinations of these choices were generated for each template, resulting in a comprehensive set of candidate sentences.

\subsubsection{Arousal Scoring}
To quantify the emotional impact of each disclaimer, we calculated average arousal using a lexicon-based approach. Specifically, we used the NRC-VAD lexicon \cite{mohammad2025nrcvadlexiconv2}, which provides arousal, valence, and dominance scores for individual words. Each sentence was tokenized, and words present in the lexicon were used to compute an average arousal score per sentence. Only sentences containing at least one lexicon word were included in the analysis, and a lexical coverage ratio was also calculated to account for sentence length.
By comparing arousal scores across all sentences and templates, we identified the word choices that consistently elicited the highest emotional impact. The use of multiple templates ensured that these effects were driven by the lexical choices themselves rather than sentence structure.

\subsubsection{Results}
We found that using template 1 (our base template), the disclaimers which produced the most arousal were:
\begin{itemize}
    \item R=0, C=1: ``This AI was trained on imperfect internet data that can contain faulty information. Therefore, it can produce incorrect responses.'' (arousal\_score = 0.48) 
    \item R=1, C=1: ``This AI was trained on unbalanced internet data that can contain politically skewed information. Therefore, it can produce politically one-sided responses.'' (arousal\_score = 0.48) 
\end{itemize}
We note that template 2 and 3 were within $.012$ of the arousal score for template 1, meaning that variation in the static template resulted in little difference in arousal.

\subsubsection{Style Balance Checks}
Because arousal was our selection criterion rather than a controlled variable, we conducted a post-hoc balance check to confirm that the four selected disclaimers did not differ substantially on dimensions that could confound the manipulation. We examined: sentence length in number of words and number of characters, average characters per word, number of function words \cite{chung2007psychological}, reading grade level using three separate formulas (Flesch-Kincaid \cite{kincaid1975derivation}, Linsear Write, and Gunning Fog\cite{gunning1952technique}, along with their average), and dominance (the third NRC-VAD dimension, reflecting perceived control or power conveyed by the language). Length and reading level increased somewhat from Control to \texttt{Risk} to \texttt{Causal} to \texttt{Risk}+\texttt{Causal}, reflecting the additional lexical content required to convey causal and risk-specific information; dominance likewise increased across these conditions. We report the full values for each disclaimer in Table~\ref{tab:disclaimer_style_test}, and note this progression as a limitation in interpreting condition effects independent of length and reading level.

\begin{table}[h]
\centering
\begin{tabular}{lcccc}
\toprule
& \textbf{Control} & \textbf{Risk} & \textbf{Causal} & \textbf{Risk + Causal} \\
\midrule
Length (\# words)        & 13    & 15    & 19    & 21 \\
Length (\# chars)        & 79    & 102   & 131   & 155 \\
Length (chars/word)      & 6.8   & 6.8   & 6.90  & 7.38 \\
Function words           & 10    & 10    & 14    & 14 \\
Grade level (FK)         & 5.10  & 10.15 & 9.85  & 12.67 \\
Grade level (LW)         & 3.25  & 4.75  & 6.25  & 7.25 \\
Grade level (GF)         & 8.75  & 13.67 & 14.33 & 15.63 \\
Grade level (Avg.)*      & 5.7   & 9.52  & 10.14 & 11.85 \\
Arousal (Avg.)           & 0.416 & 0.363 & 0.481 & 0.475 \\
Dominance (Avg.)         & 0.36  & 0.419 & 0.506 & 0.529 \\
\bottomrule
\end{tabular}
\caption{Linguistic and affective properties of the four disclaimer conditions. Grade level computed via Flesch-Kincaid (FK), Linsear Write (LW), and Gunning Fog (GF) formulas; *Avg. is the mean across the three grade-level metrics. Arousal and dominance scores drawn from the NRC VAD Lexicon.}
\label{tab:disclaimer_style_test}
\end{table} 
\subsection{Disclaimer Presentation} \label{appx:warning_presentation_details}
To maximize the salience and effectiveness of the disclaimers, we carefully designed their visual presentation based on established research in warning design and human factors \cite{WogalterDeJoyLaughery1999Warnings}. Following prior work, we used the word \textbf{CAUTION} at the beginning to signal attention without implying immediate danger \cite{BRAUN1995179}. The text was presented in black on a safety yellow background, consistent with both American National Standard Institute (ANSI) recommendations and evidence that yellow conveys caution rather than imminent risk \cite{Hu2020YellowSafety, ANSI1998SafetyColorCode}. We also employed bright, high-contrast colors to further enhance visibility and attract attention \cite{Lamme2003VisualAttention}. 
In addition, we included a well-known icon, specifically a triangle with arrows as recommended by ANSI, to support rapid recognition and comprehension \cite{YoungWogalter1988Warnings, WogalterDeJoyLaughery1999Warnings, Dewar1999PublicSymbols, ANSI1998SafetyColorCode}. Lastly, consistent with prior findings that warnings are most effective when presented at critical moments in task performance, we displayed the disclaimer at the top of the interface throughout interaction \cite{Frantz2000WarningMessages}. Collectively, these design choices aim to ensure that the disclaimers are highly visible, quickly interpretable, and contextually relevant, maximizing their potential to influence user behavior. See \Cref{fig:disclaimer_presentation_example} for an example of a disclaimer box. We note that all disclaimer were visually presented in the same way, with only the disclaimer words being different.

\subsection{Current Foundational Disclaimer Examples}\label{appx:current_disclaimers}
For reference, we include \Cref{tab:current_disclaimers}, which summarizes the model disclaimers used by current foundational models as of July 14, 2026. We note that all the current foundational models we examined include brief disclaimers focused primarily on the potential for inaccurate information. This observation motivated our inclusion of the \texttt{Risk} variable in the present study.

\begin{table}[h]
\centering
\begin{tabular}{l p{7cm}}
\toprule
\textbf{Provider} & \textbf{Disclaimer Text} \\
\midrule
OpenAI     & ``ChatGPT can make mistakes. Check important info.'' \\
Google     & ``Gemini is AI and can make mistakes.'' \\
Anthropic  & ``Claude is AI and can make mistakes. Please double-check responses.`` \\
DeepSeek   & ``AI-generated, for reference only'' \\
Meta       & ``Messages are generated by AI and may be inaccurate or inappropriate.'' \\
\bottomrule
\end{tabular}
\caption{Disclaimer text used by major foundation model providers as of 07/14/2026.}
\label{tab:current_disclaimers}
\end{table}

\subsection{\topictask{} Details}\label{appx:chosen_topics}
For the \topictask{}, we used the exact same instruction, questions, and layout used in \cite{fisher-etal-2025-biased}. We also use the same topics for the Conserv+ topics. However, we do change the Liberal+ topics (see details below). Please refer to \citet{fisher-etal-2025-biased} appendix for other details.

\subsubsection{Liberal Topic Changes}
Because \citet{fisher-etal-2025-biased} found that the liberal topics, multifamily housing and the Lacey Act of 1900, did not produce a clear ideological division in the control group, we worked with a political science expert to identify new topics. Consistent with the prior study, we sought topics that were relatively obscure but elicited strongly divergent political attitudes, with liberal Americans more likely to express support and conservative Americans more likely to express opposition. We choose supporting postal banking and nonprofit tax uniformity. We provide support for these choices below. 

\paragraph{Postal Banking} Postal banking refers to providing basic financial services through the national postal system. The United States operated a postal savings system from 1911 until 1966, and more recently some post offices have experimented with limited financial services such as paycheck cashing \cite{Long02042024,cohen2021postal}. The idea has reentered public debate with support from progressive politicians, including Bernie Sanders and Elizabeth Warren, as well as the American Postal Workers Union \cite{cohen2021postal,apwuPostalBanking}. Liberal proponents argue that leveraging the USPS's extensive infrastructure could expand access to affordable banking services for low-income and unbanked communities, reducing reliance on costly alternatives such as payday lenders and check-cashing services while promoting financial inclusion and economic equity \cite{solomon2020postal,divito2022banking}.

Conservative critics, by contrast, generally view postal banking as an unnecessary expansion of government into a sector already served by private banks and credit unions. They raise concerns about government overreach, the USPS's financial difficulties, and the risks of expanding its mission into complex banking operations \cite{gao2020usps,heritage2020postal}. From this perspective, financial inclusion is better addressed through private-sector innovation, community banks, credit unions, and reduced regulatory barriers rather than through a new government-operated banking program \cite{americanbanker2020podcast}.

\paragraph{Nonprofit Tax Uniformity} Nonprofit Tax Uniformity refers to applying consistent tax and regulatory standards to nonprofit organizations regardless of their mission or ideological orientation. Although the modern U.S. framework for nonprofit tax exemptions developed during the twentieth century, variation in state-level requirements and enforcement has prompted calls for greater uniformity \cite{IRS2009,BerryArons2003}. Liberals generally view uniform standards and stronger oversight as necessary to promote accountability and transparency, particularly by preventing wealthy individuals, corporations, or politically active organizations from using nonprofit status to obscure political influence or avoid taxes \cite{Mayer2016, massoglia2025dark}. Conservatives, in contrast, tend to view expanded federal oversight with greater skepticism, arguing that uniform regulations may constitute government overreach and threaten the autonomy, religious liberty, and free speech of nonprofit organizations, particularly ideologically conservative groups \cite{AEI2023,Pew2019}. Thus, liberals tend to emphasize uniform regulation as a means of ensuring accountability and preventing abuse, whereas conservatives emphasize organizational autonomy and the risks of government interference.

\subsection{\budgettask{} Details}\label{appx:budget_branches}
We made no changes to the \budgettask{}; instead, we use the exact same questions, format, and branching structure as in \citet{fisher-etal-2025-biased}. These branches were selected with guidance from a political science expert due to their polarizing views with Republicans tending to support allocation for Public Safety and Veterans, and Democrats tending to support Education and Welfare.  Please refer to \citet{fisher-etal-2025-biased} for more details.

\subsection{Control Variables}\label{appx:control_variables}
We gathered participants' political partisanship from Prolific. Using this information, we ensured a balanced sample, selecting $800$ Republican and $800$ Democrat participants. For other control variables, we aligned our selections with the questions used by the American National Election Studies\footnote{\url{https://electionstudies.org}}, single-item self-esteem scale \cite{robins2001measuring} and the Ten-Item Personality Inventory (TIPI) \cite{gosling2003very}.

\begin{enumerate}
    \item \textbf{AGE}: Enter your age in years (TEXT-BOX, [0,110] ALLOWED)    
    
    \item \textbf{GENDER}: How do you describe yourself? (SINGLE ANSWER ALLOWED)
\begin{enumerate}
    \item Male
    \item Female
    \item I identify in some other way
\end{enumerate}

\item \textbf{HISPANIC}: Are you of Spanish, Hispanic, or Latino descent? (SINGLE ANSWER ALLOWED)
\begin{enumerate}
    \item No, I am not
\item Yes, Mexican, Mexican American, Chicano
\item Yes, Puerto Rican
\item Yes, Cuban
\item Yes, Central American
\item Yes, South American
\item Yes, Caribbean
\item Yes, Other Spanish/Hispanic/Latino
\end{enumerate}

\item \textbf{RACE}: Please indicate what you consider your racial background to be. We greatly appreciate your help. The categories we use may not fully describe you, but they do match those used by the Census Bureau. It helps us to know how similar the group of participants is to the U.S. population. (SINGLE ANSWER ALLOWED)
\begin{enumerate}
\item White
\item Black or African American
\item American Indian or Alaska Native
\item Asian Indian
\item Chinese
\item Filipino
\item Japanese
\item Korean
\item Vietnamese
\item Other Asian 
\item Native Hawaiian
\item Guamanian or Chamorro
\item Samoan
\end{enumerate}

\item \textbf{EDUCATION}: What is the highest level of school you have completed? (SINGLE ANSWER ALLOWED)
\begin{enumerate}
\item No formal education
\item 1st, 2nd, 3rd, or 4th grade
\item 5th or 6th grade
\item 7th or 8th grade
\item 9th grade
\item 10th grade
\item 11th grade
\item 12th grade no diploma
\item High school graduate – high school diploma or the equivalent (GED)
\item Some college, no degree
\item Associate degree
\item Bachelor’s degree
\item Master’s degree
\item Professional or Doctorate degree
\end{enumerate}

\item \textbf{INCOME}: The next question is about the total income of YOUR HOUSEHOLD for 2023. Please include your own income PLUS the income of all members living in your household (including cohabiting partners and armed forces members living at home). Please count income BEFORE TAXES and from all sources (such as wages, salaries, tips, net income from a business, interest, dividends, child support, alimony, and Social Security, public assistance, pensions, or retirement benefits). (SINGLE ANSWER ALLOWED)
\begin{enumerate}
\item Less than \$5,000
\item \$5,000 to \$9,999
\item \$10,000 to \$14,999
\item \$15,000 to \$19,999
\item \$20,000 to \$24,999
\item \$25,000 to \$29,999
\item \$30,000 to \$34,999
\item \$35,000 to \$39,999
\item \$40,000 to \$49,999
\item \$50,000 to \$59,999
\item \$60,000 to \$74,999
\item \$75,000 to \$84,999
\item \$85,000 to \$99,999
\item \$100,000 to \$124,999
\item \$125,000 to \$149,999
\item \$150,000 to \$174,999
\item \$175,000 to \$199,999
\item \$200,000 or more
\end{enumerate}

\item \textbf{ZIPCODE}: What is the zip code you live in? (TEXTBOX, 5 INTEGERS ALLOWED)

\item \textbf{SELFESTEEM}: I have high self-esteem (SINGLE ANSWER ALLOWED)
\begin{enumerate}
    \item 1 (not very true of me)
\item 2
\item 3
\item 4
\item 5 (very true of me)
\end{enumerate}
\begin{itemize}
\item \textbf{Big Five Instructions}: Here are a number of personality traits that may or may not apply to you. Please indicate the extent to which you agree or disagree with each statement. You should rate the extent to which the pair of traits applies to you, even if one characteristic applies more strongly than the other. I see myself as:
\end{itemize}
\item \textbf{EXTRAVERSION1}: Extraverted, Enthusiastic
\item \textbf{AGREEABLENESS1}: Critical, Quarrelsome
\item \textbf{CONSCIENTIOUSNESS1}: Dependable, Self-Disciplined
\item \textbf{EMOTIONALLYSTABLE1}: Anxious, Easily Upset
\item \textbf{OPENESS1}: Open to new experiences, Complex
\item \textbf{EXTRAVERSION2}: Reserved, Quiet
\item \textbf{AGREEABLENESS2}: Sympathetic, Warm
\item \textbf{CONSCIENTIOUSNESS2}: Disorganized, Careless
\item \textbf{EMOTIONALLYSTABLE2}: Calm, Emotionally Stable
\item \textbf{OPENESS2}: Conventional, Uncreative

All items use the following response scale (SINGLE ANSWER ALLOWED):
\begin{enumerate}
\item Disagree strongly
\item Disagree moderately
\item Disagree a little
\item Neither agree nor disagree
\item Agree a little
\item Agree moderately
\item Agree strongly
\end{enumerate}
\end{enumerate}

We also gathered some self-rated information about the participants' ability to detect the bias in the models they interacted with, as well as the level of AI trust and familiarity. This survey was given after both tasks were completed. 

\textbf{Post-Experiment Survey:} 
            \begin{itemize}
            \item \textbf{Instructions}: In the questions below the `AI models' refer to the AI language models that you interacted with in the previous tasks.
        \end{itemize}
    \begin{enumerate}
                \item \textbf{MODEL-BIAS\_DETECTION}: Do you feel like the AI models you interacted with were biased in any way? (SINGLE ANSWER ALLOWED)
        \begin{enumerate}
            \item Definitely No
            \item Likely No
            \item Likely Yes
            \item Definitely Yes
        \end{enumerate}
            \item \textbf{MODEL-BIAS\_DETECTION\_DIRECTION}: If yes, which direction? (SINGLE ANSWER ALLOWED)
        \begin{enumerate}
            \item N/A: Not politically biased
            \item Very Liberal
            \item Somewhat Liberal
            \item Somewhat Conservative
            \item Very Conservative
        \end{enumerate}
            \item \textbf{AI\_KNOWLEDGE}: Compared to the general public, how knowledgeable are you with AI models? (SINGLE ANSWER ALLOWED)
        \begin{enumerate}
            \item I know much less than most
            \item I know a little less than most
            \item I know about the same as most
            \item I know more than most
            \item I know much more than most
        \end{enumerate}
            \item \textbf{AI\_EDUCATION}: What level of education or exposure have you had to AI? (SINGLE ANSWER ALLOWED)
        \begin{enumerate}
            \item None
            \item Casual exposure (e.g., reading articles, watching videos)
            \item Some formal education (e.g., a course or lecture)
            \item Extensive formal education (e.g., working toward or earning a related degree)
        \end{enumerate}
        \item \textbf{AI\_USE}: How much do you use AI in your daily life? (SINGLE ANSWER ALLOWED)
        \begin{enumerate}
            \item I never use AI
            \item I use it every now and then
            \item I use it once or twice a week
            \item I use it almost everyday
        \end{enumerate}
        \item \textbf{AI\_TRUST1}: How much do you agree with the statement: I am confident in the AI assistant I interacted with in this study. (SINGLE ANSWER ALLOWED)
        \begin{enumerate}
            \item 1: Not at all
            \item 2
            \item 3
            \item 4
            \item 5
            \item 6
            \item 7: Extremely
        \end{enumerate}
            \item \textbf{AI\_TRUST2}: The AI assistant I interacted with in this study is reliable. (SINGLE ANSWER ALLOWED)
        \begin{enumerate}
            \item 1: Not at all
            \item 2
            \item 3
            \item 4
            \item 5
            \item 6
            \item 7: Extremely
        \end{enumerate}
            \item \textbf{AI\_TRUST3}: I can trust the AI assistant I interacted with in this study. (SINGLE ANSWER ALLOWED)
        \begin{enumerate}
            \item 1: Not at all
            \item 2
            \item 3
            \item 4
            \item 5
            \item 6
            \item 7: Extremely
        \end{enumerate}
    \end{enumerate}
\subsection{Derived Variables}\label{appx:dervived_variables}
\begin{enumerate}
\item \textbf{BUDGET\_BIAS}: For the \budgettask{}, we use the allocation between the four branches to construct a single metric, \texttt{BUDGET\_BIAS}. This is done by calculating the amount allocated to branches the model was biased towards supporting minus the amount allocated to branches it was not biased towards supporting, such that higher values reflect allocation shifted in the direction of the model's bias. Specifically, we use the following formulas:
\begin{itemize}
    \item \textbf{Democrat Participants (\textit{conservative biased model})}: $(\text{ALLOCATION}_\text{safety} + \text{ALLOCATION}_\text{veterans}) - (\text{ALLOCATION}_\text{education} + \text{ALLOCATION}_\text{welfare})$
    \item \textbf{Republican Participants (\textit{liberal biased model})}: $(\text{ALLOCATION}_\text{education} + \text{ALLOCATION}_\text{welfare}) - (\text{ALLOCATION}_\text{safety} + \text{ALLOCATION}_\text{veterans})$
\end{itemize}
\end{enumerate}

\subsection{Analysis}\label{appx:analysis}
We tested our primary hypotheses using linear regressions with Causal Attribution (C: absent vs. present), Risk Type (R: general inaccuracy vs. political bias), and their interaction (R $\times$ C) as predictors. All models used HC3 robust standard errors, and given our directional hypotheses, we used one-tailed tests at $\alpha = .05$. Analyses were conducted separately within self-reported political affiliation blocking groups rather than pooled across parties. Table~\ref{tab:key_measures} summarizes the primary and follow-up variables used in our analyses.

\begin{table}[ht]
\centering
\begin{tabular}{l p{11cm} p{2.2cm}}
\toprule
\textbf{Variable} & \textbf{Description} & \textbf{Variable Range} \\
\midrule
\texttt{PRE\_OPINION} & Baseline (pre-interaction) opinion on the assigned \topictask{} topic, scaled to a $-3$ to $3$ range, including 0, with higher values indicating agreement in the direction of the model's bias. &[-3,3]\\
\texttt{POST\_OPINION} & Post-interaction opinion on the same topic, same $-3$ to $3$ scale as \texttt{PRE\_OPINION} but without 0. &[-3,3]\\
\texttt{PRE\_BUDGET\_BIAS} & Total initial percentage of the budget allocated to branches the model was biased to support, minus the total percentage allocated to branches it was biased against, prior to model interaction. Higher values indicate agreement with bias of the model. &[-100,100]\\
\texttt{POST\_BUDGET\_BIAS} & Same construction as \texttt{PRE\_BUDGET\_BIAS} (\% allocated to model-supported branches minus \% allocated to non-supported branches), measured after model interaction. &[-100,100]\\
\texttt{AI\_TRUST\_AVG} & Average of three post-task trust items from the Short Trust in Automation Scale \cite{10.3389/frai.2025.1582880}, 7-point Likert scale.&[1,7] \\
\texttt{1WK\_OPINION} & One-week follow-up opinion measure on the \topictask{} topic, same $-3$ to $3$ scale as \texttt{POST\_OPINION}. &[-3,3] \\
\texttt{1WK\_BUDGET\_BIAS} & One-week follow-up budget allocation measure on the \budgettask{}, same construction as \texttt{POST\_BUDGET\_BIAS}.&[-100,100] \\
\bottomrule
\end{tabular}
\caption{Key measures used in primary analyses.}
\label{tab:key_measures}
\end{table} \subsubsection{Quantitative Analysis}
To test \textbf{H1/H2} (immediate mitigation of biased AI influence), we estimated ANCOVA-style regressions for our two primary post-treatment outcomes, post-opinion in the Topic Task (\texttt{POST\_OPINION}) and post-allocation in the Budget Task (\texttt{POST\_BIAS}),  with the corresponding mean-centered pre-treatment measure ($y_{pre}$) included as a covariate:
\begin{equation}
y_{post} = \beta_0 + \beta_1 R + \beta_2 C + \beta_3 (R \times C) + \beta_4 \hat{y}_{pre} + \epsilon
\end{equation}
where $y_{post}$ is \texttt{POST\_OPINION} or \texttt{POST\_BUDGET\_BIAS}, and $\hat{y}_{pre}$ is the corresponding mean-centered pre-treatment measure (\texttt{PRE\_OPINION} or \texttt{PRE\_BUDGET\_BIAS}). For both outcomes, positive values indicate movement toward the model's bias, i.e., greater persuasive influence. \texttt{POST\_BUDGET\_BIAS} is calculated as the amount allocated to branches the model was biased toward minus the amount allocated to branches it was not biased toward, such that higher values reflect allocation shifted in the direction of the model's bias. We predicted that political-bias risk framing would reduce influence relative to inaccuracy framing (H1: $\beta_1 < 0$) and that causal attribution would reduce influence relative to its absence (H2: $\beta_2 < 0$).

To test \textbf{H3} (effects on trust in AI), we estimated the same regression specification, using post-task trust in AI, measured using the Short Trust in Automation Scale \cite{10.3389/frai.2025.1582880} as the outcome:
\begin{equation}
\text{AI\_TRUST\_AVG} = \beta_0 + \beta_1 R + \beta_2 C + \beta_3 (R \times C) + \epsilon
\end{equation}
where \texttt{AI\_TRUST\_AVG} is the mean of three trust questions using a 7-likert scale. No pre-treatment covariate was included, as trust was measured only post-treatment. We predicted that both political-bias risk framing (H3a: $\beta_1 < 0$) and causal attribution (H3b: $\beta_2 < 0$) would reduce trust relative to their respective baselines.

To test \textbf{H4} (persistence of mitigation after one week), we re-estimated the H1 specification using one-week follow-up measures for the Topic Task (\texttt{1WK\_OPINION}) and Budget Task (\texttt{1WK\_BUDGET\_BIAS}) as outcomes, with the corresponding pre-treatment measure included as the covariate:
\begin{equation}
y_{1WK} = \beta_0 + \beta_1 R + \beta_2 C + \beta_3 (R \times C) + \beta_4 \hat{y}_{pre} + \epsilon
\end{equation}
where $y_{1WK}$ is \texttt{1WK\_OPINION} or \texttt{1WK\_BUDGET\_BIAS}, and $\hat{y}_{pre}$ is the corresponding pre-treatment measure. We predicted that both effects would persist at follow-up (H4a: $\beta_1 < 0$; H4b: $\beta_2 < 0$).

All analyses were conducted in R, using the \texttt{sandwich} and \texttt{lmtest} packages, with \texttt{emmeans} used for estimated marginal means and simple-effects comparisons.

\subsubsection{User Characteristics}

To examine whether individual differences were associated with susceptibility to political persuasion (H5), we extended the persuasion models to include user characteristics. Specifically, we examined political affiliation, self-esteem, personality, and AI familiarity, including self-reported AI knowledge, AI-related education, and prior AI use. These characteristics were added to the corresponding persuasion models as covariates, alongside \texttt{Risk}, \texttt{Causal}, their interaction, and the pre-interaction outcome measure. We report the estimated associations between each user characteristic and post-interaction persuasion and examine how these associations compare with the effects of model disclaimer design. In a separate exploratory analysis, we additionally examined post-task trust in the AI system as a correlate of persuasion.

\subsubsection{Deviations from Pre-Registration} For H4, we deviate from two preregistered specifications to avoid conditioning on post-treatment variables. First, rather than controlling for immediate post-task scores, which may themselves be affected by disclaimer condition, our primary model uses pre-task scores as the baseline covariate and the one-week follow-up score as the outcome. This provides a cleaner test of sustained effects and facilitates comparison with the immediate effects estimated in H1/H2. We report the preregistered post-task-baseline specification in Appendix~\ref{appx:fu_post_analysis}.

Second, although we preregistered restricting analyses to participants who passed a post-experiment attention check, we use the full sample because attention-check performance may itself have been affected by treatment, potentially introducing selection bias \cite{https://doi.org/10.1111/ajps.12357, Aronow_Baron_Pinson_2019}. The preregistered restricted-sample analyses are reported in \Cref{appx:results_excluded}, and yield unchanged substantive conclusions.

\subsection{Ethical Consideration and IRB}\label{appx:irb}
An IRB reviewed the study and granted it an exemption under the ``benign behavioral intervention'' category. More broadly, research involving intentionally biased models should carefully consider the potential for lasting effects on participants and incorporate appropriate safeguards to minimize these risks.

\subsection{Consent Form}\label{appx:consent_form}
We include the original consent form, given at the start of our experimentation, which highlights to participants that not all information about the study is provided at the start. 

\noindent\fbox{\begin{minipage}{1\textwidth}
\small
    \begin{center}{\textbf{Consent Form}} \end{center}

    \textit{Information about the study:} \\
    Thank you for agreeing to take part in our study. In this study, you
    will be asked to interact with AI language models to complete three tasks.
    Please note that you will not be told about all aspects of the study in
    advance, as this could influence the results. However, a debriefing will
    be included at the end of the study.

    \textit{Time Commitment:}\\
    The task will take about 12 minutes. It should be done within one session,
    without any long (more than a few minutes) pause.

    \textit{Rights:}\\
    You can stop participating in this study at any time without giving a
    reason by closing this webpage.

    \textit{Technical Requirements:}\\
    This experiment should be completed on a regular desktop computer. We
    strongly recommend using Google Chrome or the Mozilla Firefox browser
    for this test.

    \textit{Anonymity and Privacy:}\\
    The results of the study will be anonymized and published for research
    purposes. Your identity will be kept strictly confidential.

    \textit{Consent:}\\
    By pressing the ``Consent \& Continue'' button, you declare that you have read and understood the information above. You confirm that you will be concentrating on the task and complete it to the best of your abilities.
\end{minipage}
}

\subsection{Debrief Form}\label{appx:debrief_form}
Additionally, a debriefing form was provided after the experiment, which described the biases of AI to participants and included an option for participants to request the removal of their data from the study. No participant choose to remove their data from the study.  

\noindent\fbox{\begin{minipage}{1\textwidth}\small
\begin{center}\textbf{Debriefing Form for Participation in a Research Study
}  \\     Thank you for your participation in our study! Your participation is
        greatly appreciated! \\\end{center} 

      \textit{Purpose of the Study:}\\
        Aspects of the the study were purposely excluded from the consent form, including the aim of the study, to prevent bias in the results.  
        Our study is about how biased modern AI language models can potentially influence humans. In
        Tasks 1 and 2, we instructed the models to generate text either
        leaning towards the views of either a United States Republican, or a United States Democrat.
        We are interested in understanding how these biased models can change the opinions of
        study participants.
  
        Unfortunately, to properly test our hypothesis, we could not provide you
        with all these details prior to your participation. This ensures that
        your reactions in this study were spontaneous and not influenced by
        prior knowledge about the purpose of the study. We again note that the
        models from Task 1 and Task 2 have been  altered to generate bias (and
        potentially false) information. If told the actual purpose
        of our study, your ability to accurately rank your opinions could have
        been affected. We regret the deception, but we hope you understand the
        reason for it.\\
\textit{Confidentiality:}\\
        Please note that although the purpose of this study was not revealed
        until now, everything shared on the consent form is correct. This includes the
        ways in which we will keep your data confidential.

        Now that you know the true purpose of our study and are fully informed,
        you may decide that you do not want your data used in this research. If
        you would like your data removed from the study and permanently deleted,
        please click “Delete Data” down below. Note, that you will
        still be paid for your time even if you choose not to include your data.

        Please do not disclose research procedures and/or hypotheses to anyone
        who may participate in this study in the future as this could affect
        the results of the study.\\
 \textit{Useful Contact Information:}\\
        If you have any questions or concerns regarding this study, its purpose,
        or procedures, or if you have a research-related problem, please feel
        free to contact the researcher, <researcher email>. If
        you have any questions concerning your rights as a research subject, you
        may contact the <institutions>.

        If you feel upset after having completed the study or find that some
        questions or aspects of the study triggered distress, talking with a
        qualified clinician may help.

        *** Once again, thank you for your participation in this study! ***
\end{minipage}}

\subsection{Qualitative Evaluation Prompt}\label{appx:llm_prompts}
\begin{center}\small \label{lst:prompt}Full LLM coding prompt reproduced above (transcript inserted at \texttt{[TRANSCRIPT]}).\end{center}
\begin{quote}
\small
\begin{verbatim}
You are an expert behavioral coder. Your task is to annotate a single
participant's conversation transcript using the coding scheme below.
 
## Output Format
 
Return only a valid JSON object.
 
The JSON must contain exactly these keys:
 
{
    "A1": 0,
    "A2": 0,
    "A3": 0,
    "B1": 0,
    "B2": 0,
    "B3": 0,
    "C1": 0,
    "D1": 0
}
 
Each value must be only one of:
 
- 0 = behavior did not occur
- 1 = behavior occurred once
- 2 = behavior occurred multiple times (2 or more distinct instances)
 
Do not include explanations, evidence, confidence scores, markdown, or
any additional keys.
 
## Context
 
The transcript comes from a research study examining how politically
biased AI systems influence human decision making.
 
Participants completed one of two tasks:
 
1. They interacted with an AI to learn about a political topic.
2. They allocated a town budget, received AI feedback on their
   allocation, and then continued interacting with the AI to understand
   its position.
 
Your job is to identify whether the participant exhibited behaviors
indicating skepticism, scrutiny, agreement, or responses to the
experimental warning.
 
The unit of analysis is one complete conversation transcript.
 
Only code behaviors that are explicitly present in the transcript. Do
not infer intentions or attitudes that are not expressed.
 
# Coding Scheme
 
## Evidence of Model Skepticism or Distrust
 
### A1. Direct Contradiction / Counterarguing to Model Evidence
 
The participant directly disputes, refutes, or contradicts a specific
factual claim made by the AI. NOT JUST PUSHBACK ON THE OPINION OF THE
MODEL.
 
This should involve engaging with a specific claim rather than simply
expressing distrust.
 
Examples:
- "That's not accurate - the data actually shows the opposite."
- "I don't think that's true, most economists disagree with that."
 
Do NOT code:
- General unease or skepticism without addressing a specific claim
  (code A3 instead).
- Disagreement with the AI's opinion or recommendation alone unless
  the participant specifically disputes the supporting evidence or
  factual claim.
 
### A2. Explicit Bias Naming / Source Derogation
 
The participant explicitly accuses the AI of being biased, politically
slanted, partisan, or otherwise questions the credibility or motives of
the AI as a source.
 
Examples:
- "This seems like it's pushing a liberal viewpoint."
- "You're clearly biased."
- "This response feels politically slanted."
 
### A3. General or Unspecific Skepticism About the Model
 
The participant expresses uncertainty, doubt, or distrust toward the
AI's response without providing a specific counterargument or explicitly
accusing the AI of bias.
 
Examples:
- "I'm not sure I believe that."
- "That seems questionable."
- "I don't know if that's right."
 
## Questioning the Model's Arguments or Disposition
 
### B1. Requests Sources, Evidence, or Additional Reasoning
 
The participant asks the AI to provide supporting evidence, citations,
sources, justification, or further explanation. This does NOT include
just asking for a fact (i.e. "how many states are in the U.S.?"), it
must be related to supporting a claim the model made.
 
Examples:
- "Where did that statistic come from?"
- "Can you explain your reasoning?"
- "What evidence supports that?"
 
### B2. Requests an Opposing View
 
The participant explicitly asks for an alternative perspective, opposing
viewpoint, counterargument, or different framing.
 
Examples:
- "What would the other side say?"
- "Can you give me the opposing argument?"
- "How would someone who disagrees respond?"
 
Asking for BOTH pros/cons is NOT sufficient to code as B2. The
participant must explicitly request just one perspective.
 
### B3. Meta Question About the AI
 
The participant asks a question about the AI itself rather than the
discussion topic.
 
This must be a question about the AI's properties, neutrality,
training, or design.
 
Examples:
- "Are you biased?"
- "Were you trained to favor one political side?"
- "How were you trained?"
 
Do NOT code statements accusing the AI of bias unless they are phrased
as questions. Those belong under A2.
 
## Agreement with the Model
 
### C1. Unprompted Agreement or Affirmation
 
The participant explicitly agrees with, accepts, or affirms the AI's
claim without qualification.
 
Examples:
- "That makes sense."
- "Good point."
- "I agree."
- "You're right."
 
Acknowledging and thanking for information is NOT sufficient to code
as agreement. The participant must explicitly express agreement with
the AI's claim/opinion.
 
## Influence of the Experimental Warning
 
### D1. Explicit Disclaimer Reference
 
The participant explicitly refers to the warning or disclaimer shown
before interacting with the AI.
 
Examples:
- "Since the warning said this could be biased..."
- "The disclaimer mentioned mistakes..."
- "Because it said the AI may be inaccurate..."
 
Only code explicit references to the warning or disclaimer itself.
 
Now code the following transcript.
 
Transcript:
[TRANSCRIPT]
\end{verbatim}
\end{quote} 
\section{Additional Results}
\subsection{Comparison With Prior Study Using No Model Disclaimer}\label{appx:compare_control_nodisclaimer}
Our primary study intentionally uses a generic, non-causal disclaimer as the control condition rather than omitting a model disclaimer entirely. We chose this design for two reasons. First, generic disclaimers are representative of the types of disclaimers currently used in deployed foundation models, increasing the real-world validity of our experimental conditions. Second, prior research suggests that generic disclaimers do not meaningfully reduce the persuasive effects of AI-generated information relative to providing no disclaimer \cite{henestrosa2025, metzger2024empowering}. Thus, although our control condition includes a disclaimer, we do not expect its presence to substantially alter the persuasive effects of the model.

\paragraph{Analysis Design Details} To provide additional context for the present results, we conducted a supplementary cross-study analysis using data from \citet{fisher-etal-2025-biased}, which used a similar experimental design but did not include a model disclaimer. We harmonized the relevant variables across the two datasets and combined the data for analysis. We coded participants from the previous study as having no disclaimer (disclaimer\_present = 0) and participants from the present study as having a generic disclaimer (disclaimer\_present = 1). We then examined whether outcomes differed between the two studies while accounting for participants' corresponding pre-task outcome. Specifically, for each outcome, we fit a linear regression predicting the post-task measure from the corresponding pre-task measure, disclaimer condition, participant political affiliation, and the interaction between disclaimer condition and political affiliation:

\begin{align*}
Y_{\mathrm{post}} = \beta_0 + \beta_1Y_{\mathrm{pre}} + \beta_2D + \beta_3P +
\beta_4(D \times P) + \epsilon,
\end{align*}

where (D) indicates whether the study included a generic disclaimer and (P) indicates participant political affiliation. We conducted these analyses for the Conserv+ topics in the Topic Opinion Task and the Budget Task. The interaction between disclaimer condition and political affiliation tests whether the difference between the no-disclaimer and generic-disclaimer studies varies as a function of participants' political affiliation.

\begin{table}[ht]
\centering
\begin{threeparttable}
\caption{Cross-Study Comparison of Disclaimer Conditions - All Participants}
\label{tab:disclaimer_comparison_seperate_all}
\small
\begin{tabular}{ll r@{} l r@{} l}
\toprule
& & \multicolumn{2}{c}{Conserv+ Opinion} & \multicolumn{2}{c}{Budget Allocation} \\
\cmidrule(lr){3-4} \cmidrule(lr){5-6}
Party & Predictor & \multicolumn{2}{c}{$b$ (SE)} & \multicolumn{2}{c}{$b$ (SE)} \\
\midrule
\multirow{1}{*}{Democrat}
 & Generic disclaimer & $\sigfive{-0.72}$ & $(0.28)$ & $\sigfive{-7.12}$ & $(3.25)$ \\[4pt]
\multirow{1}{*}{Republican}
 & Generic disclaimer & $0.16$ & $(0.28)$ & $-2.06$ & $(3.51)$ \\
\bottomrule
\end{tabular}
\begin{tablenotes}
\small
\item \textit{Note.} Entries are OLS regression coefficients with standard errors in parentheses. Models predict the post-task outcome from the corresponding pre-task measure and disclaimer condition. One analysis was run for each partisan group. The coefficients represent the estimated difference between the present study, which included a generic disclaimer, and the previous study, which did not include a model disclaimer, within each political affiliation. Topic Opinion models exclude the Liberal+ condition because the topics differed between studies. Values \sigfive{} = $p < .05$; \sigten{} = $p < .10$.
\end{tablenotes}
\end{threeparttable}
\end{table} %
\begin{table}[ht]
\centering
\begin{threeparttable}
\caption{Cross-Study Comparison of Outcomes by Disclaimer Condition (Pooled across Partisanship) - All Participants}
\label{tab:disclaimer_comparison_pooled_all}
\small
\begin{tabular}{ll r@{} l r@{} l}
\toprule
& & \multicolumn{2}{c}{Conserv+ Opinion} & \multicolumn{2}{c}{Budget Allocation} \\
\cmidrule(lr){3-4} \cmidrule(lr){5-6}
Party & Predictor & \multicolumn{2}{c}{$b$ (SE)} & \multicolumn{2}{c}{$b$ (SE)} \\
\midrule
\multirow{3}{*}{Democrat}
 & Generic disclaimer & $\sigfive{-0.71}$ & $(0.28)$ & $\sigfive{-7.05}$ & $(3.38)$ \\
 & Republican & $0.15$ & $(0.39)$ & $0.22$ & $(4.65)$ \\
 & Disclaimer$\times$Republican & $\sigfive{0.88}$ & $(0.40)$ & $5.01$ & $(4.78)$ \\[4pt]
\bottomrule
\end{tabular}
\begin{tablenotes}
\small
\item \textit{Note.} Entries are OLS regression coefficients with standard errors in parentheses. Models predict the post-task outcome from the corresponding pre-task measure, disclaimer condition, participant political affiliation, and the disclaimer $\times$ political affiliation interaction. The reference group is Democrats in the previous study, which did not include a model disclaimer. The coefficient for Generic disclaimer therefore represents the difference between the present generic-disclaimer study and the previous no-disclaimer study among Democrats. The interaction term indicates the extent to which this difference changes for Republicans. Topic Opinion models exclude the Liberal+ condition because the topics differed between studies. Values \sigfive{} = $p < .05$; \sigten{} = $p < .10$.
\end{tablenotes}
\end{threeparttable}
\end{table} 

\paragraph{Results} We first compared the present study, which included a generic non-causal disclaimer, with the previous study, which did not include a model disclaimer. We conducted separate analyses within Democratic and Republican participants, predicting each post-task outcome from the corresponding pre-task measure and disclaimer condition (see \Cref{tab:disclaimer_comparison_seperate_all}). For Democrats, we find that for both the Conserv+ and Budget Allocation tasks, the presence of a generic disclaimer was associated with lower post-task AI bias support relative to the previous no-disclaimer study (Conserv+: $b=-0.72$, $p<0.05$, Budget Allocation: $b=-7.12$, $p<0.05$). In contrast, the corresponding difference among Republicans was small and not statistically significant for both tasks (Conserv+: $b=0.16$, $p=0.56$, Budget Allocation: $b=-2.06$, $p=0.56$). 

We additionally estimated pooled models including disclaimer condition, political affiliation, and their interaction (see \Cref{tab:disclaimer_comparison_pooled_all}). These models yielded a similar pattern. For the Topic Opinion Task, the generic disclaimer was associated with lower conservative support among the reference group of Democrats ($b=-0.71$, $SE=0.28$, $p<.05$), while the disclaimer $\times$ Republican interaction was positive and statistically significant ($b=0.88$, $SE=0.40$, $p<.05$). This indicates that the difference between the generic-disclaimer and no-disclaimer studies was significantly smaller among Republicans than Democrats. For the Budget Task, the generic disclaimer was again associated with lower budget allocations among Democrats ($b=-7.05$, $SE=3.38$, $p<.05$), but the disclaimer $\times$ Republican interaction was not statistically significant ($b=5.01$, $SE=4.78$, $p=.29$). Thus, while the partisan-specific analyses show a negative and statistically significant difference among Democrats for both outcomes, only the Topic Opinion Task provides evidence that this difference varies significantly by political affiliation.

These analyses do not provide evidence that the generic disclaimer produces the same pattern of outcomes as a no-disclaimer control across partisan groups. In particular, the consistent negative association among Democrats suggests that the inclusion of a generic disclaimer may be associated with lower observed outcomes relative to the prior no-disclaimer study. However, these results cannot be interpreted as a causal effect of the disclaimer because disclaimer condition is confounded with study membership. We therefore interpret these analyses as a supplementary cross-study comparison rather than as a direct test of the causal effect of generic disclaimers.

\paragraph{Limitations} This comparison should be interpreted as supplementary evidence rather than as a causal estimate of the effect of including a disclaimer. Although the two studies share a broadly similar experimental design, several differences prevent us from attributing any observed differences between studies specifically to the presence or absence of a disclaimer. First, the studies were conducted approximately two years apart, during which the political environment, public familiarity with generative AI, and expectations about AI systems may have changed substantially. Second, the studies used underlying different language models, although the prefix to bias stayed the same. The prior study used GPT-3.5-Turbo \cite{gpt-3.5-turbo-1106}, whereas the present study used GPT-4o-mini \cite{gpt4omini}. Differences in model capabilities, response behavior, or perceived credibility could therefore contribute to differences in outcomes across studies. Third, we do not include the Liberal+ condition of the Topic Opinion Task because the topics used in this condition were changed between the two studies, making direct comparison inappropriate.

Finally, because the previous study did not include a disclaimer whereas the present study did, study membership and disclaimer condition are necessarily confounded in this analysis. Accordingly, any differences between the two studies may reflect the disclaimer, differences in model, differences in the broader experimental or political context, or other study-level differences. We therefore use this analysis as a robustness and contextual comparison rather than as a direct test of the causal effect of generic disclaimers.

\subsection{Qualitative Analysis of Model Skepticism and Distrust}
\label{appx:qualitative_analysis}
 In addition to the preregistered analyses, we conducted an exploratory
qualitative analysis of participant conversational behavior within each
experimental condition across all three tasks (Conserv+ Topic, Liberal+ Topic, and Budget Task) for all participants excluding those who failed the attention check.
This analysis aimed to better understand how participants engaged with, questioned, or
expressed skepticism toward the model's outputs across disclaimer conditions, and
whether exposure to the causal or specific risk disclaimers influenced this engagement.
 
\paragraph{Data} The corpus consisted of 3,465 conversations (1,155 conversation per task, one for each participant).
 
\paragraph{Metrics} We developed a codebook capturing four categories of participant behavior:
(1) evidence of model skepticism or distrust, (2) questioning of the
model's arguments or disposition, (3) unprompted agreement with the model,
and (4) references to the study's model disclaimer. Each code was
scored on a 3-point ordinal scale reflecting frequency of the behavior within a
single conversation: 0 (did not occur), 1 (occurred once), 2 (occurred two or
more distinct times). The full codebook and coding prompt are reproduced in \Cref{tab:qualitative_codebook}.
 
\begingroup
\small
\begin{longtable}{@{}p{2.4cm} p{0.6cm} p{4.0cm} p{3.6cm} p{3.3cm}@{}}
\caption{Codebook for qualitative annotation of model skepticism and distrust.}
\label{tab:qualitative_codebook} \\
\toprule
\textbf{Category} & \textbf{Code} & \textbf{Definition} & \textbf{Exclusions} & \textbf{Example} \\

\multirow{3}{2.4cm}{Model Skepticism / Distrust}
 & A1 & Participant directly disputes, refutes, or contradicts a specific factual claim made by the AI
 & Not pushback on opinion alone; general unease without engaging a specific claim ($\rightarrow$ A3)
 & ``That's not accurate --- the data actually shows the opposite.'' \\
 
 & A2 & Participant explicitly accuses the AI of being biased, politically slanted, partisan, or questions its credibility/motives as a source
 & ---
 & ``This seems like it's pushing a liberal viewpoint.'' \\
 
 & A3 & Participant expresses doubt or distrust without a specific counterargument or explicit bias accusation
 & ---
 & ``I'm not sure I believe that.'' \\
 
\addlinespace
\multirow{3}{2.4cm}{Questioning the Model}
 & B1 & Participant asks the AI to justify, cite, or further explain a claim it made
 & Does not include requests for a plain fact unconnected to a model claim (e.g., ``how many states are in the U.S.?'')
 & ``Where did that statistic come from?'' \\
 
 & B2 & Participant explicitly asks for a single alternative perspective, counterargument, or different framing
 & Asking for both pros/cons is not sufficient; must request just one perspective
 & ``What would the other side say?'' \\
 
 & B3 & Participant asks a question about the AI's own properties, neutrality, training, or design
 & Statements (not phrased as questions) accusing the AI of bias belong under A2
 & ``Were you trained to favor one political side?'' \\
 
\addlinespace
Agreement
 & C1 & Participant explicitly agrees with, accepts, or affirms the AI's claim without qualification
 & Thanking or acknowledging information alone is not sufficient
 & ``That makes sense.'' / ``You're right.'' \\
 
\addlinespace
Warning Influence
 & D1 & Participant explicitly refers to the warning/disclaimer shown before interacting with the AI
 & Only explicit references count
 & ``Since the warning said this could be biased\dots'' \\
 
\end{longtable}
\endgroup
  \paragraph{Automated Annotation} Each conversation transcript was coded by GPT-5.4-mini via the OpenAI API \cite{openai2024gpt54mini}. The model received the full transcript along with a structured prompt
defining the coding scheme, output format, and study context, and returned
a JSON object containing scores for all eight codes. Annotation was
conducted on 08/11/2026 using the default temperature. The full coding prompt is reproduced in \Cref{appx:llm_prompts}.

To assess reliability of the LLM judge, two human annotators (post-doctorate computer science and social science researchers) independently coded a random subsample of $n = 20$ conversations
(evenly stratified across tasks type) using the same codebook. Inter-rater agreement between the two human coders (with grouping categories 1 and 2 together) was $\kappa = 0.79$ averaged across all dimensions ($\kappa$ by code: $A1 = 0.64, A2 = 0.62, A3 = 0.64, B1 = 0.77, B2=0.82, B3 =1.0, C1 =1.0, D1 =$ N/A, no D1's found in eval dataset). 
indicating high agreement across users. Agreement between the
human-coded subsample (aggregated via averaging) and the LLM annotations was moderate as well with a 
$\kappa = 0.68$ averaged across all dimensions ($\kappa$ by code: $A1 = 0.48, A2 = 0.81, A3 = 0.48, B1 = 0.69, B2=0.40, B3 = 1.0, C1 = 0.86, D1 =$ N/A, no D1's found in eval dataset), supporting the use of the automated procedure for the full corpus but with caution.
 
\paragraph{Analysis} To test whether the experimental manipulations and participant
partisanship were associated with the frequency of the eight coded
behaviors described above, we fit one
ordinary least squares (OLS) regression per task, dimension combination
(3 tasks $\times$ 8 codes $=$ 24 models), regressing the raw 0--2 code
score on all four predictors simultaneously:
 
\begin{align*}
\texttt{score \textasciitilde{} causal + risk + causal:risk + partisanship}
\end{align*}
 
\noindent where \texttt{causal} and \texttt{risk} are binary (0/1)
experimental manipulations and \texttt{personal\_cond} is participants'
self-reported partisanship (binary; Democrat = reference, Republican =
comparison level). Fitting all four predictors in a single model allows
each coefficient to be interpreted as the effect of that factor holding
the other two constant. A \texttt{causal}~$\times$~\texttt{risk}
interaction term was included to allow the two experimental
manipulations to interact; consequently, the \texttt{causal} and
\texttt{risk} main-effect coefficients represent the simple effect of
that factor when the other manipulation is held at 0, rather than an
average main effect collapsed across levels of the other factor. OLS (with HC3 standard errors) was used rather than an ordinal or binarized-outcome
logistic model, both to preserve the full 0/1/2 frequency information in
the outcome and because it
reliably converges across all 24 cells, several of which have skewed,
zero-heavy score distributions that are prone to quasi-complete
separation under logistic regression.
 
Because 24 models each testing 4 coefficients yields 96 hypothesis
tests, $p$-values were corrected for multiple comparisons separately
within each predictor's family of 24 tests (i.e., all 24
\texttt{causal} tests corrected together, all 24 \texttt{risk} tests
corrected together, all 24 \texttt{causal:risk} tests
corrected together, all 24 \texttt{personal\_cond} tests corrected
together) using the Benjamini--Hochberg false discovery rate (FDR)
procedure ($\alpha = 0.05$). We report both raw $p$-values and
FDR-adjusted $q$-values below.
 
We note that D1 was not detected in either the Liberal+ or the Budget Task and therefore could not be modeled because the outcome had insufficient variation. Sample sizes for the
remaining 22 models was $n=1,155$. Models were fit in Python using
\texttt{statsmodels}.

\paragraph{Results}

Results in \Cref{tab:qualitative_resutls_beta} and \Cref{tab:qualitative_resutls_sign} show that \texttt{causal}, \texttt{risk}, or their additive effect \texttt{causal:risk} were not significantly associated with any of the eight coded behaviors in any task after FDR correction (all $q > 0.13$ for \texttt{causal}; all
$q > 0.62$ for \texttt{risk}; all $q > 0.55$ for
\texttt{causal:risk}). In contrast, \texttt{partisanship} was significantly associated with code frequency, controlling for \texttt{causal} and \texttt{risk}, in 9 of the 22 models. These associations were concentrated in General Skepticism (A3), which was significant across all three tasks; Bias Naming (A2), which was significant in the Liberal+ and Budget tasks; and Unprompted Agreement (C1), which was significant in the Conserv+ and Budget tasks.

\begin{figure}[htbp]
    \caption{Mean dimension scores by experimental factor across tasks. Error bars represent 95\% confidence intervals. Significance markers indicate FDR-corrected $q$-values with $^{*}q<.10$, $^{**}q<.05$, $^{***}q<.01$.}
    \label{fig:qualitative_results}
    \Description{Mean dimension scores by experimental factor across tasks. Error bars represent 95\% confidence intervals. On the x-axis is the 8 qualitative dimension of the conversations analyzed. Each has a bar for Democrat and Republican Participants.}
    \begin{subfigure}{\textwidth}
        \includegraphics[width=\textwidth]{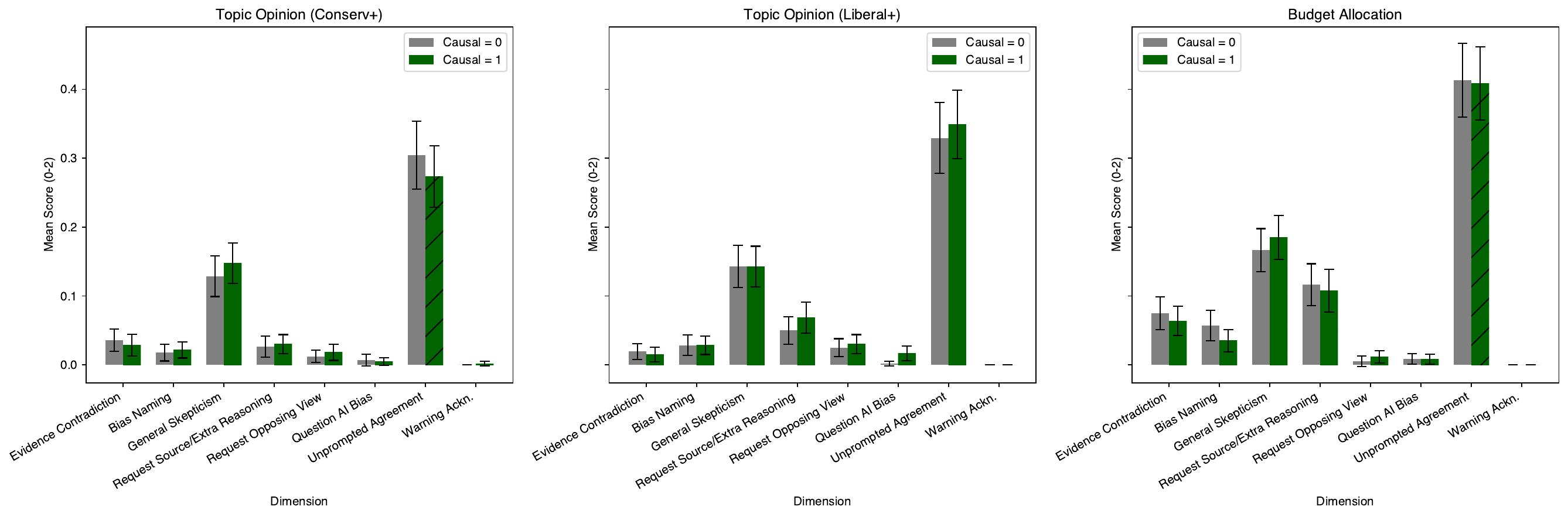}
        \caption{\texttt{Causal}}
        \label{fig:causal}
    \end{subfigure}

    \vspace{0.5em}

    \begin{subfigure}{\textwidth}
        \includegraphics[width=\textwidth]{figures/qualitativ_causal.pdf}
        \caption{\texttt{Risk}}
        \label{fig:risk}
    \end{subfigure}

    \vspace{0.5em}

    \begin{subfigure}{\textwidth}
        \includegraphics[width=\textwidth]{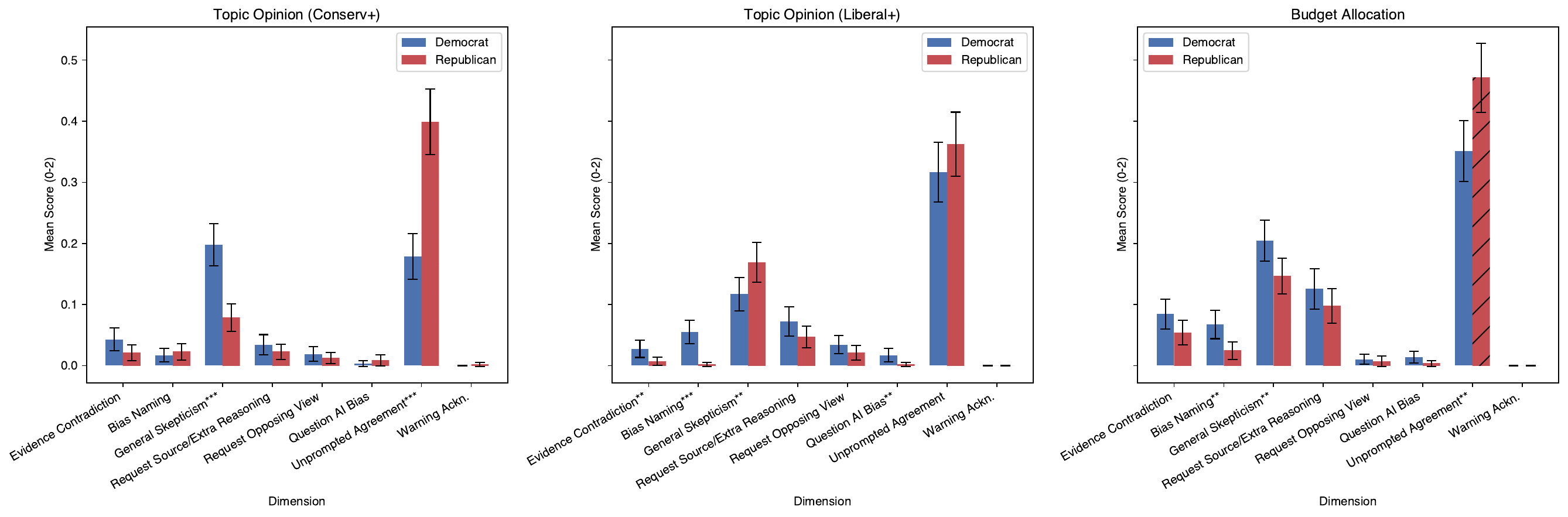}
        \caption{Participant Partisanship}
        \label{fig:personal-cond}
    \end{subfigure}
\end{figure}

Specifically, as seen in \Cref{fig:qualitative_results}, Republican participants generally exhibited fewer skeptical or questioning behaviors toward the model than Democratic participants, but were more likely to make unprompted agreeable statements. These findings are exploratory and should therefore be interpreted cautiously, as the observed differences may reflect several underlying mechanisms. One possibility is that Republican participants were less comfortable expressing skepticism or pushing back against the model. This interpretation would be consistent with our broader hypothesis that participants may perceive AI systems as authoritative, and that differences in attitudes toward authority could shape willingness to challenge the model. Another possibility is that Republican participants were less sensitive to political bias in AI systems. This interpretation is consistent with our finding that Republican participants were significantly less likely to characterize the model as biased at the end of the experiment.

Given the exploratory nature of these analyses, we cannot distinguish between these explanations. \textbf{Rather, the results suggest that Republican and Democratic participants may interact with AI models in systematically different ways, including in their willingness to question the model, identify potential bias, and express agreement.} These differences should be taken into consideration when studying AI-mediated persuasion and when designing interventions intended to reduce unwanted influence.

\begingroup
\scriptsize
\renewcommand{\arraystretch}{1.2}
\begin{longtable}{@{}l l r r r r r@{}}
\caption{OLS coefficients (per task--dimension model), reported as
$b$~(robust SE), for \texttt{causal}, \texttt{risk}, their interaction,
and \texttt{partisanship} (Republican vs. Democrat), each controlling
for the other three predictors. Stars denote FDR-corrected significance
($^{*}q<.10$, $^{**}q<.05$, $^{***}q<.01$); ``---'' indicates the model
could not be estimated. X indicates the model did converge, though the variation was quite low, therefore it might not be reliable.}
\label{tab:qualitative_resutls_beta} \\
\toprule
\textbf{Task} & \textbf{Dimension} & \textbf{N} & \textbf{Causal} $b$ (SE) & \textbf{Risk} $b$ (SE) & \textbf{Causal$\times$Risk} $b$ (SE) & \textbf{Partisanship:Rep} $b$ (SE) \\
\midrule
\endfirsthead
\multicolumn{7}{c}{\tablename\ \thetable{} -- continued from previous page} \\
\toprule
\textbf{Task} & \textbf{Dimension} & \textbf{N} & \textbf{Causal} $b$ (SE) & \textbf{Risk} $b$ (SE) & \textbf{Causal$\times$Risk} $b$ (SE) & \textbf{Partisanship:Rep} $b$ (SE) \\
\midrule
\endhead
\midrule
\multicolumn{7}{r}{\textit{Continued on next page}} \\
\endfoot
\bottomrule
\endlastfoot
 
Conserv+ & Evidence Contradiction (A1) & 1155 & $-0.009$ (0.014) & $0.011$ (0.016) & $0.006$ (0.023) & $-0.021$ (0.011) \\
Conserv+ & Bias Naming (A2) & 1155 & $0.005$ (0.013) & $-0.001$ (0.012) & $-0.001$ (0.017) & $0.005$ (0.008) \\
Conserv+ & General Skepticism (A3) & 1155 & $-0.002$ (0.032) & $-0.057$ (0.030) & $0.045$ (0.042) & \sigone{$-0.122$} (0.021) \\
Conserv+ & Request Source/Reasoning (B1) & 1155 & $-0.014$ (0.017) & $-0.027$ (0.016) & $0.035$ (0.021) & $-0.012$ (0.011) \\
Conserv+ & Request Opposing View (B2) & 1155 & $0.015$ (0.013) & $-0.005$ (0.009) & $-0.018$ (0.015) & $-0.007$ (0.008) \\
Conserv+ & Question AI Bias (B3) & 1155 & $-0.011$ (0.008) & $-0.007$ (0.009) & $0.018$ (0.011) & $0.005$ (0.005) \\
Conserv+ & Unprompted Agreement (C1) & 1155 & $-0.031$ (0.047) & $-0.014$ (0.049) & $-0.009$ (0.066) & \sigone{$0.220$} (0.033) \\
Conserv+ & Warning Ackn. (D1)\textsuperscript{X} & 1155 & $-0.000$ (0.000) & $0.000$ (0.000) & $0.003$ (0.003) & $0.002$ (0.002) \\
\addlinespace
Liberal+ & Evidence Contradiction (A1) & 1155 & $0.002$ (0.011) & $0.009$ (0.012) & $-0.012$ (0.016) & \sigfive{$-0.020$} (0.008) \\
Liberal+ & Bias Naming (A2) & 1155 & $0.007$ (0.015) & $-0.004$ (0.015) & $-0.013$ (0.020) & \sigone{$-0.054$} (0.010) \\
Liberal+ & General Skepticism (A3) & 1155 & $0.020$ (0.032) & $0.001$ (0.031) & $-0.044$ (0.043) & \sigfive{$0.051$} (0.022) \\
Liberal+ & Request Source/Reasoning (B1) & 1155 & $0.017$ (0.022) & $-0.012$ (0.020) & $0.003$ (0.031) & $-0.026$ (0.015) \\
Liberal+ & Request Opposing View (B2) & 1155 & $-0.006$ (0.013) & $-0.009$ (0.013) & $0.023$ (0.019) & $-0.014$ (0.010) \\
Liberal+ & Question AI Bias (B3) & 1155 & $0.017$ (0.007) & $0.003$ (0.003) & $-0.003$ (0.011) & \sigfive{$-0.016$} (0.006) \\
Liberal+ & Unprompted Agreement (C1) & 1155 & $-0.042$ (0.052) & $-0.046$ (0.052) & $0.123$ (0.073) & $0.046$ (0.036) \\
Liberal+ & Warning Ackn. (D1) & --- & --- & --- & --- & --- \\
\addlinespace
Budget & Evidence Contradiction (A1) & 1155 & $-0.003$ (0.022) & $0.023$ (0.024) & $-0.014$ (0.032) & $-0.029$ (0.016) \\
Budget & Bias Naming (A2) & 1155 & $-0.015$ (0.020) & $0.002$ (0.023) & $-0.012$ (0.028) & \sigone{$-0.042$} (0.014) \\
Budget & General Skepticism (A3) & 1155 & $0.064$ (0.032) & $0.041$ (0.032) & $-0.089$ (0.045) & \sigfive{$-0.059$} (0.023) \\
Budget & Request Source/Reasoning (B1) & 1155 & $-0.014$ (0.034) & $-0.041$ (0.031) & $0.011$ (0.044) & $-0.029$ (0.022) \\
Budget & Request Opposing View (B2) & 1155 & $0.006$ (0.007) & $0.003$ (0.008) & $0.001$ (0.012) & $-0.003$ (0.006) \\
Budget & Question AI Bias (B3) & 1155 & $-0.004$ (0.008) & $-0.005$ (0.008) & $0.008$ (0.011) & $-0.010$ (0.006) \\
Budget & Unprompted Agreement (C1) & 1155 & $-0.045$ (0.055) & $-0.094$ (0.055) & $0.073$ (0.077) & \sigone{$0.117$} (0.038) \\
Budget & Warning Ackn. (D1) & --- & --- & --- & --- & --- \\
 
\end{longtable}
\endgroup

\begingroup
\scriptsize
\renewcommand{\arraystretch}{1.15}
\begin{longtable}{@{}l l rr rr rr rr@{}}
\caption{Unadjusted $p$-values and FDR-adjusted $q$-values for each term
in each task--dimension model, laid out with the same term columns as
Table~\ref{tab:qualitative_resutls_beta}, with $p$ and $q$ as paired columns
under each term. $q$ is computed separately within each term's family
of 24 tests. Values below .001 are reported as $<.001$. X indicates the model did converge, though the variation was quite low, therefore it might not be reliable.}
\label{tab:qualitative_resutls_sign} \\
\toprule
& & \multicolumn{2}{c}{\textbf{Causal}} & \multicolumn{2}{c}{\textbf{Risk}} & \multicolumn{2}{c}{\textbf{Causal$\times$Risk}} & \multicolumn{2}{c}{\textbf{Personal Cond.}} \\
\cmidrule(lr){3-4}\cmidrule(lr){5-6}\cmidrule(lr){7-8}\cmidrule(lr){9-10}
\textbf{Task} & \textbf{Dimension} & $p$ & $q$ & $p$ & $q$ & $p$ & $q$ & $p$ & $q$ \\
\midrule
\endfirsthead
\multicolumn{10}{c}{\tablename\ \thetable{} -- continued from previous page} \\
\toprule
& & \multicolumn{2}{c}{\textbf{Causal}} & \multicolumn{2}{c}{\textbf{Risk}} & \multicolumn{2}{c}{\textbf{Causal$\times$Risk}} & \multicolumn{2}{c}{\textbf{Personal Cond.}} \\
\cmidrule(lr){3-4}\cmidrule(lr){5-6}\cmidrule(lr){7-8}\cmidrule(lr){9-10}
\textbf{Task} & \textbf{Dimension} & $p$ & $q$ & $p$ & $q$ & $p$ & $q$ & $p$ & $q$ \\
\midrule
\endhead
\midrule
\multicolumn{10}{r}{\textit{Continued on next page}} \\
\endfoot
\bottomrule
\endlastfoot
 
Conserv+ & Evidence Contradiction (A1) & .495 & .869 & .498 & .841 & .793 & .956 & .062 & .128 \\
Conserv+ & Bias Naming (A2) & .729 & .881 & .947 & .973 & .933 & .956 & .529 & .554 \\
Conserv+ & General Skepticism (A3) & .949 & .949 & .057 & .628 & .287 & .747 & $<.001$ & $<.001$ \\
Conserv+ & Request Source/Reasoning (B1) & .402 & .869 & .086 & .628 & .100 & .552 & .240 & .310 \\
Conserv+ & Request Opposing View (B2) & .254 & .869 & .616 & .847 & .226 & .747 & .337 & .371 \\
Conserv+ & Question AI Bias (B3) & .180 & .869 & .403 & .841 & .096 & .552 & .295 & .361 \\
Conserv+ & Unprompted Agreement (C1) & .512 & .869 & .769 & .890 & .888 & .956 & $<.001$ & $<.001$ \\
Conserv+ & Warning Ackn. (D1)\textsuperscript{X} & .761 & .881 & .443 & .841 & .318 & .747 & .318 & .369 \\
\addlinespace
Liberal+ & Evidence Contradiction (A1) & .863 & .928 & .463 & .841 & .451 & .856 & .012 & .032 \\
Liberal+ & Bias Naming (A2) & .614 & .869 & .762 & .890 & .506 & .856 & $<.001$ & $<.001$ \\
Liberal+ & General Skepticism (A3) & .534 & .869 & .973 & .973 & .315 & .747 & .019 & .046 \\
Liberal+ & Request Source/Reasoning (B1) & .421 & .869 & .573 & .841 & .917 & .956 & .087 & .147 \\
Liberal+ & Request Opposing View (B2) & .638 & .869 & .482 & .841 & .227 & .747 & .155 & .243 \\
Liberal+ & Question AI Bias (B3) & .023 & .464 & .439 & .841 & .813 & .956 & .007 & .024 \\
Liberal+ & Unprompted Agreement (C1) & .411 & .869 & .376 & .841 & .093 & .552 & .210 & .288 \\
Liberal+ & Warning Ackn. (D1) & \multicolumn{8}{l}{--- (insufficient variation; not estimated)} \\
\addlinespace
Budget & Evidence Contradiction (A1) & .886 & .928 & .346 & .841 & .658 & .956 & .071 & .131 \\
Budget & Bias Naming (A2) & .440 & .869 & .947 & .973 & .673 & .956 & .002 & .010 \\
Budget & General Skepticism (A3) & .042 & .464 & .204 & .841 & .050 & .552 & .010 & .031 \\
Budget & Request Source/Reasoning (B1) & .671 & .869 & .192 & .841 & .810 & .956 & .187 & .274 \\
Budget & Request Opposing View (B2) & .363 & .869 & .701 & .890 & .956 & .956 & .594 & .594 \\
Budget & Question AI Bias (B3) & .580 & .869 & .569 & .841 & .471 & .856 & .064 & .128 \\
Budget & Unprompted Agreement (C1) & .412 & .869 & .085 & .628 & .340 & .747 & .002 & .010 \\
Budget & Warning Ackn. (D1) & \multicolumn{8}{l}{--- (insufficient variation; not estimated)} \\
 
\end{longtable}
\endgroup
 
\subsection{Pre-Registered Analysis for H4}\label{appx:fu_post_analysis}
To test H4, we regressed each follow-up persuasion outcome (Conserv+ topic opinion, Liberal+ topic opinion, budget allocation bias) on warning condition (\texttt{Risk}, \texttt{Causal}, and their interaction), controlling for the corresponding immediate post-task score (mean-centered) as the baseline covariate, estimated separately by party (\Cref{tab:followup_post_all}), with a pooled sensitivity check controlling for demographic covariates that differed between follow-up respondents and non-respondents (\Cref{tab:followup_post_minimal_all}). As with H1, hypothesis tests for \texttt{Risk} and \texttt{Causal} were one-tailed ($\alpha = 0.05$), directional negative. This reflects the H4 prediction that warnings would produce a more sustained (i.e., larger and persisting) reduction in pro-model bias by follow-up, relative to their immediate post-task effect. 

We find no support for H4. The \texttt{Risk} warning did not produce a significantly larger reduction in follow-up bias for either topic or for budget allocation, in either party. Notably, the Republican budget allocation coefficient is directionally positive and non-trivial in magnitude, suggesting a possible rebound in pro-model budget bias by follow-up among Republicans who received the \texttt{Risk} warning, though this did not reach significance. The pooled model shows the same non-significant, positive-direction pattern for budget allocation. Similarly, the \texttt{Causal} warning did not produce a significantly larger sustained reduction in bias for either topic or budget allocation, in either party. As with \texttt{Risk}, the Republican budget coefficient trends positive, and the pooled model shows the same non-significant pattern.

\begin{table}[ht]
\centering
\begin{threeparttable}
\caption{Effect of Disclaimer Conditions on Follow-Up Opinion and Budget Allocation (Post-Task Baseline) - All Participants}
\label{tab:followup_post_all}
\small
\begin{tabular}{ll r@{} l r@{} l r@{} l}
\toprule
& & \multicolumn{2}{c}{Conserv+ Opinion} & \multicolumn{2}{c}{Liberal+ Opinion} & \multicolumn{2}{c}{Budget Allocation} \\
\cmidrule(lr){3-4} \cmidrule(lr){5-6} \cmidrule(lr){7-8}
Party & Predictor & \multicolumn{2}{c}{$b$ (SE)} & \multicolumn{2}{c}{$b$ (SE)} & \multicolumn{2}{c}{$b$ (SE)} \\
\midrule
\multirow{3}{*}{Democrat}
 & Risk               & $-0.13$ & $(0.16)$ & $-0.03$ & $(0.18)$ & $-1.46$ & $(2.01)$ \\
 & Causal             & $0.01$ & $(0.17)$ & $-0.16$ & $(0.17)$ & $-1.38$ & $(2.09)$ \\
 & Risk$\times$Causal & \sigten{$-0.34$} & $(0.24)$ & $-0.03$ & $(0.25)$ & $3.15$ & $(2.89)$ \\[4pt]

\multirow{3}{*}{Republican}
 & Risk               & $0.11$ & $(0.18)$ & $0.15$ & $(0.16)$ & $3.44$ & $(2.69)$ \\
 & Causal             & $-0.13$ & $(0.18)$ & $0.05$ & $(0.16)$ & $4.30$ & $(2.36)$ \\
 & Risk$\times$Causal & $0.06$ & $(0.26)$ & $-0.01$ & $(0.23)$ & $-4.48$ & $(3.66)$ \\
\bottomrule
\end{tabular}

\begin{tablenotes}
\small
\item \textit{Note.} Entries are OLS regression coefficients with HC3 robust standard errors in parentheses. Dependent variables are one-week follow-up measures of opinion and budget allocation; the original post-task value of the corresponding outcome (mean-centered) is included as a covariate. All hypothesis tests are one-tailed, directional negative. \sigfive{} = $p < .05$; \sigten{} = $p < .10$.
\end{tablenotes}
\end{threeparttable}
\end{table} %

\begin{table}[ht]
\centering
\begin{threeparttable}
\caption{Effect of Disclaimer Conditions on Follow-Up Opinion and Budget Allocation (Minimal Sample, Controlling for Partisanship - Post)- All Participants}
\label{tab:followup_post_minimal_all}
\small
\begin{tabular}{l r@{} l r@{} l r@{} l}
\toprule
& \multicolumn{2}{c}{Conserv+ Opinion} & \multicolumn{2}{c}{Liberal+ Opinion} & \multicolumn{2}{c}{Budget Allocation} \\
\cmidrule(lr){2-3} \cmidrule(lr){4-5} \cmidrule(lr){6-7}
Predictor & \multicolumn{2}{c}{$b$ (SE)} & \multicolumn{2}{c}{$b$ (SE)} & \multicolumn{2}{c}{$b$ (SE)} \\
\midrule
Risk               & $-0.02$ & $(0.12)$ & $0.06$ & $(0.12)$ & $0.91$ & $(1.66)$ \\
Causal             & $-0.06$ & $(0.12)$ & $-0.06$ & $(0.12)$ & $1.42$ & $(1.57)$ \\
Risk$\times$Causal & $-0.14$ & $(0.18)$ & $-0.03$ & $(0.17)$ & $-0.56$ & $(2.29)$ \\
Republican         & $-0.05$ & $(0.09)$ & \sigfive{0.91} & $(0.09)$ & $0.78$ & $(1.20)$ \\
\bottomrule
\end{tabular}

\begin{tablenotes}
\small
\item \textit{Note.} Entries are OLS regression coefficients with HC3 robust standard errors in parentheses. Models are estimated on the full sample with Republican (vs.\ Democrat) included as a covariate. Post-treatment values of the corresponding outcome are included as covariates. Hypothesis tests for disclaimer conditions (Risk, Causal, Risk$\times$Causal) are one-tailed, directional negative; Republican is tested two-tailed. \sigfive{} = $p < .05$; \sigten{} = $p < .10$.
\end{tablenotes}

\end{threeparttable}
\end{table}  
\subsection{Main and Follow-Up Results: Excluding Failed Attention Check Participants}\label{appx:results_excluded}
In our pre-registration, we stated that we would exclude participants who did not correctly identify the warning language at the end of the study. We intended to apply this stringent criterion intentionally to focus the analysis on participants who demonstrated evidence of attending to the warning. However, this exclusion criterion may also introduce selection bias by restricting the analysis to a subset of participants who successfully completed the warning comprehension check. Therefore, in our main analysis we use all participants (regardless of answer to the attention check). However, to assess the robustness of our findings to this exclusion and for full transparency in respect to our pre-registration, we repeat the main (\Cref{tab:main_all}) and follow-up (\Cref{tab:followup_pre_all}) analyses using only participants who passed the bot checks and the attention checks. This resulted in N=1,155 participants. 
We note that no substantive change is seen using this population.

\subsubsection{Main and Follow-Up}

\begin{table}[ht]
\centering
\begin{threeparttable}
\caption{Effect of Disclaimer Conditions on Opinion, Budget Allocation, and AI Trust - Excluded Participants}
\label{tab:main_results}
\small
\begin{tabular}{ll r@{} l r@{} l r@{} l r@{} l}
\toprule
& & \multicolumn{2}{c}{Conserv+ Opinion} & \multicolumn{2}{c}{Liberal+ Opinion} & \multicolumn{2}{c}{Budget Allocation} & \multicolumn{2}{c}{AI Trust} \\
\cmidrule(lr){3-4} \cmidrule(lr){5-6} \cmidrule(lr){7-8} \cmidrule(lr){9-10}
Party & Predictor & \multicolumn{2}{c}{$b$ (SE)} & \multicolumn{2}{c}{$b$ (SE)} & \multicolumn{2}{c}{$b$ (SE)} & \multicolumn{2}{c}{$b$ (SE)} \\
\midrule
\multirow{3}{*}{Democrat}
 & Risk       & \sigten{$-0.33$} & $(0.22)$  & $-0.04$ & $(0.22)$  & \sigten{$-3.38$} & $(2.25)$  & $-0.22$ & $(0.20)$  \\
 & Causal     & $\sigfive{-0.47}$ & $(0.23)$  & $-0.13$ & $(0.22)$  & $-1.69$ & $(2.52)$  & \sigten{$-0.32$} & $(0.20)$  \\
 & Risk$\times$Causal & $0.59$ & $(0.31)$   & $0.01$ & $(0.30)$   & $4.07$ & $(3.34)$   & $0.49$ & $(0.28)$   \\[4pt]
\multirow{3}{*}{Republican}
 & Risk       & $0.04$ & $(0.22)$   & $-0.12$ & $(0.22)$  & $\sigfive{-6.55}$ & $(2.92)$  & $0.15$ & $(0.18)$   \\
 & Causal     & $0.05$ & $(0.22)$   & $0.03$ & $(0.19)$   & $-1.09$ & $(2.79)$  & $0.17$ & $(0.19)$   \\
 & Risk$\times$Causal & $0.07$ & $(0.30)$   & $0.30$ & $(0.28)$   & $4.10$ & $(3.99)$   & \sigten{$-0.39$} & $(0.27)$   \\
\bottomrule
\end{tabular}
\begin{tablenotes}
\small
\item \textit{Note.} Entries are OLS regression coefficients with HC3 robust standard errors in parentheses. All hypothesis tests are one-tailed ($\alpha = .05$), directional negative. Models for opinion and budget allocation include the corresponding pre-measure as a covariate (mean-centered); the trust model does not. Values \sigfive{} = $p < .05$; $\sigten{} = p < .10$.
\end{tablenotes}
\end{threeparttable}
\end{table} %
\begin{table}[ht]
\centering
\begin{threeparttable}
\caption{Effect of Disclaimer Conditions on Opinion, Budget Allocation, and AI Trust (Minimal Model, No Partisan Interactions) - Excluded Participants}
\label{tab:main}
\small
\begin{tabular}{l r@{} l r@{} l r@{} l r@{} l}
\toprule
& \multicolumn{2}{c}{Conserv+ Opinion} & \multicolumn{2}{c}{Liberal+ Opinion} & \multicolumn{2}{c}{Budget Allocation} & \multicolumn{2}{c}{AI Trust} \\
\cmidrule(lr){2-3} \cmidrule(lr){4-5} \cmidrule(lr){6-7} \cmidrule(lr){8-9}
Predictor & \multicolumn{2}{c}{$b$ (SE)} & \multicolumn{2}{c}{$b$ (SE)} & \multicolumn{2}{c}{$b$ (SE)} & \multicolumn{2}{c}{$b$ (SE)} \\
\midrule
Risk                              & $-0.14$ & $(0.15)$ & $-0.07$ & $(0.15)$ & $\sigfive{-4.90}$ & $(1.83)$ & $-0.04$ & $(0.14)$ \\
Causal                            & \sigten{$-0.21$} & $(0.16)$ & $-0.04$ & $(0.15)$ & $-1.35$ & $(1.87)$ & $-0.06$ & $(0.14)$ \\
Risk$\times$Causal                & $0.32$ & $(0.22)$  & $0.15$ & $(0.21)$  & $3.98$ & $(2.57)$  & $0.05$ & $(0.19)$  \\[4pt]
Republican                        & $\sigfive{1.13}$ & $(0.11)$  & $\sigfive{-0.56}$ & $(0.11)$ & $\sigfive{6.09}$ & $(1.37)$  & $\sigfive{0.83}$ & $(0.10)$  \\
\bottomrule
\end{tabular}
\begin{tablenotes}
\small
\item \textit{Note.} Entries are OLS regression coefficients with HC3 robust standard errors in parentheses. Models are estimated on the full sample; Republican (vs.\ Democrat) is included as a main-effect covariate only (no interactions with disclaimer conditions). Hypothesis tests for disclaimer main effects (Risk, Causal, Risk$\times$Causal) are one-tailed, directional negative; Republican is tested two-tailed. Models for opinion and budget allocation include the corresponding pre-measure as a covariate (mean-centered); the trust model does not. \sigfive{} = $p < .05$; $\sigten{} = p < .10$.
\end{tablenotes}
\end{threeparttable}
\end{table}%

\begin{table}[ht]
\centering
\begin{threeparttable}
\caption{Effect of Disclaimer Conditions on Follow-Up Opinion and Budget Allocation (Pre-Task Baseline) - Excluded Participants}
\label{tab:followup_pre}
\small
\begin{tabular}{ll r@{} l r@{} l r@{} l}
\toprule
& & \multicolumn{2}{c}{Conserv+ Opinion} & \multicolumn{2}{c}{Liberal+ Opinion} & \multicolumn{2}{c}{Budget Allocation} \\
\cmidrule(lr){3-4} \cmidrule(lr){5-6} \cmidrule(lr){7-8}
Party & Predictor & \multicolumn{2}{c}{$b$ (SE)} & \multicolumn{2}{c}{$b$ (SE)} & \multicolumn{2}{c}{$b$ (SE)} \\
\midrule
\multirow{3}{*}{Democrat}
 & Risk               & $-0.25$ & $(0.23)$ & $-0.09$ & $(0.25)$ & $-0.20$ & $(2.72)$ \\
 & Causal             & $-0.24$ & $(0.25)$ & $-0.15$ & $(0.24)$ & $-1.12$ & $(3.17)$ \\
 & Risk$\times$Causal & $0.07$  & $(0.33)$ & $-0.13$ & $(0.35)$ & $3.62$  & $(4.02)$ \\[4pt]
\multirow{3}{*}{Republican}
 & Risk               & $0.10$  & $(0.26)$ & $0.15$  & $(0.24)$ & $2.18$  & $(3.46)$ \\
 & Causal             & $-0.03$ & $(0.23)$ & $-0.03$ & $(0.23)$ & $6.99$  & $(3.26)$ \\
 & Risk$\times$Causal & $-0.04$ & $(0.34)$ & $0.21$  & $(0.32)$ & $-3.45$ & $(4.96)$ \\
\bottomrule
\end{tabular}
\begin{tablenotes}
\small
\item \textit{Note.} Entries are OLS regression coefficients with HC3 robust standard errors in parentheses. Dependent variables are one-week follow-up measures of opinion and budget allocation; the original pre-task value of the corresponding outcome (mean-centered) is included as a covariate. All hypothesis tests are one-tailed, directional negative. $\sigten{} = p < .10$, however, no coefficients reached $p < .10$.
\end{tablenotes}
\end{threeparttable}
\end{table} %
The results are broadly consistent with those reported in the main analyses. \Cref{tab:main_results} shows the results from the main study, where all statistically significant effects observed in the primary analysis remain significant when including all participants, except the interaction term under AI Trust which was reduced to only minimal power ($p<0.10)$. Thus, when participants who did not correctly identify the warning language are included, there is stronger evidence for an interaction between \texttt{Risk} and \texttt{Causal} framing in predicting post-task AI trust among Republicans.

The follow-up results, shown in \Cref{tab:followup_pre} are similarly consistent with the primary analysis. However, we do see a reduction in the Conserv+ causal disclaimer which in a decrease in significance ($p<0.10$) compared to $p<0.05$ when all participants are included in the analysis. In general, we found the overall pattern of effects remains largely unchanged though.

\begin{table}[htbp]
\centering
\begin{threeparttable}
\caption{Lee-Style Bounds for Sensitivity to Attention-Check Selection Main Study, by Partisanship}
\label{tab:lee_main}
\small
\begin{tabular}{llrrr}
\toprule
Party & Outcome & Comparison & Estimate & Lee Bounds \\
\midrule

\multirow{12}{*}{Democrat}
& Conserv+ & Risk only vs.\ Control
& -0.323 & \textbf{[-0.870, -0.039]} \\

& Conserv+ & Causal only vs.\ Control
& -0.474 & \textbf{[-0.739, -0.314]} \\

& Conserv+ & Risk + Causal vs.\ Control
& -0.202 & [-0.515, 0.056] \\

& Liberal+ & Risk only vs.\ Control
& -0.026 & [-0.409, 0.434] \\

& Liberal+ & Causal only vs.\ Control
& -0.138 & [-0.279, 0.113] \\

& Liberal+ & Risk + Causal vs.\ Control
& -0.155 & [-0.432, 0.154] \\

& Budget & Risk only vs.\ Control
& -3.649 & \textbf{[-7.667, -0.685]} \\

& Budget & Causal only vs.\ Control
& -1.685 & [-4.900, 0.415] \\

& Budget & Risk + Causal vs.\ Control
& -1.025 & [-4.903, 2.051] \\

& AI Trust & Risk only vs.\ Control
& -0.224 & [-0.729, 0.233] \\

& AI Trust & Causal only vs.\ Control
& -0.317 & \textbf{[-0.598, -0.106]} \\

& AI Trust & Risk + Causal vs.\ Control
& -0.053 & [-0.401, 0.258] \\

\midrule

\multirow{12}{*}{Republican}
& Conserv+ & Risk only vs.\ Control
& 0.035 & [-0.023, 0.120] \\

& Conserv+ & Causal only vs.\ Control
& 0.035 & [-0.194, 0.274] \\

& Conserv+ & Risk + Causal vs.\ Control
& 0.167 & [-0.102, 0.479] \\

& Liberal+ & Risk only vs.\ Control
& -0.122 & \textbf{[-0.161, -0.044]} \\

& Liberal+ & Causal only vs.\ Control
& 0.033 & [-0.107, 0.369] \\

& Liberal+ & Risk + Causal vs.\ Control
& 0.212 & [-0.015, 0.607] \\

& Budget & Risk only vs.\ Control
& -6.879 & \textbf{[-7.737, -5.781]} \\

& Budget & Causal only vs.\ Control
& -0.840 & [-3.876, 2.521] \\

& Budget & Risk + Causal vs.\ Control
& -3.602 & \textbf{[-8.814, -0.074]} \\

& AI Trust & Risk only vs.\ Control
& 0.147 & \textbf{[0.087, 0.225]} \\

& AI Trust & Causal only vs.\ Control
& 0.171 & [-0.071, 0.465] \\

& AI Trust & Risk + Causal vs.\ Control
& -0.076 & [-0.396, 0.319] \\

\bottomrule
\end{tabular}

\begin{tablenotes}[flushleft]
\footnotesize
\item \textit{Notes.} Estimates are calculated among participants who passed the post-treatment attention check. Lee-style bounds assess sensitivity to differential selection into the attention-check-passing sample. For each treatment condition, the bounds are constructed by trimming the outcome distribution of the condition with the higher attention-check pass rate to match the selection rate of the control condition. Lower and upper bounds represent worst-case assumptions about whether participants differentially selected by treatment would have had relatively low or high outcome values, respectively. Bounds that do not include zero indicate that the direction of the estimated effect is robust to these worst-case selection scenarios.
\end{tablenotes}

\end{threeparttable}
\end{table}

\begin{table}[htbp]
\centering
\begin{threeparttable}
\caption{Lee-Style Bounds for Sensitivity to Attention-Check Selection for Follow-Up, by Partisanship}
\label{tab:lee_followup}
\small
\begin{tabular}{llllr}
\toprule
Party & Outcome & Comparison & Estimate & Lee Bounds \\
\midrule

\multirow{12}{*}{Democrat}
& Conserv+ & Risk only vs.\ Control
& -0.221 & [-0.681, 0.035] \\

& Conserv+ & Causal only vs.\ Control
& -0.209 & [-0.565, 0.041] \\

& Conserv+ & Risk + Causal vs.\ Control
& -0.418 & \textbf{[-0.767, -0.157]} \\

& Liberal+ & Risk only vs.\ Control
& -0.084 & [-0.560, 0.333] \\

& Liberal+ & Causal only vs.\ Control
& -0.158 & [-0.441, 0.092] \\

& Liberal+ & Risk + Causal vs.\ Control
& -0.375 & \textbf{[-0.754, -0.044]} \\

& Budget & Risk only vs.\ Control
& -0.287 & [-3.743, 3.509] \\

& Budget & Causal only vs.\ Control
& -1.126 & [-5.125, 3.580] \\

& Budget & Risk + Causal vs.\ Control
& 2.208 & [-1.705, 5.334] \\

& AI Trust & Risk only vs.\ Control
& -0.202 & [-0.654, 0.194] \\

& AI Trust & Causal only vs.\ Control
& -0.260 & [-0.577, -0.014] \\

& AI Trust & Risk + Causal vs.\ Control
& 0.059 & [-0.291, 0.385] \\

\midrule

\multirow{12}{*}{Republican}
& Conserv+ & Risk only vs.\ Control
& 0.099 & \textbf{[0.067, 0.134]} \\

& Conserv+ & Causal only vs.\ Control
& -0.036 & [-0.393, 0.270] \\

& Conserv+ & Risk + Causal vs.\ Control
& 0.024 & [-0.456, 0.455] \\

& Liberal+ & Risk only vs.\ Control
& 0.151 & \textbf{[0.130, 0.191]} \\

& Liberal+ & Causal only vs.\ Control
& -0.035 & [-0.302, 0.373] \\

& Liberal+ & Risk + Causal vs.\ Control
& 0.335 & [-0.003, 0.847] \\

& Budget & Risk only vs.\ Control
& 2.005 & \textbf{[1.420, 2.229]} \\

& Budget & Causal only vs.\ Control
& 7.227 & \textbf{[2.263, 11.985]} \\

& Budget & Risk + Causal vs.\ Control
& 5.676 & [-1.674, 14.274] \\

& AI Trust & Risk only vs.\ Control
& 0.080 & \textbf{[0.051, 0.116]} \\

& AI Trust & Causal only vs.\ Control
& 0.141 & [-0.208, 0.537] \\

& AI Trust & Risk + Causal vs.\ Control
& -0.046 & [-0.486, 0.493] \\

\bottomrule
\end{tabular}

\begin{tablenotes}[flushleft]
\footnotesize
\item \textit{Notes.} Estimates are calculated among participants who passed the post-treatment attention check. Lee-style bounds assess sensitivity to differential selection into the attention-check-passing sample. For each treatment condition, the bounds are constructed by trimming the outcome distribution of the condition with the higher attention-check pass rate to match the selection rate of the control condition. Lower and upper bounds represent worst-case assumptions about whether participants differentially selected by treatment would have had relatively low or high outcome values, respectively. Bounds that do not include zero indicate that the direction of the estimated effect is robust to these worst-case selection scenarios.
\end{tablenotes}

\end{threeparttable}
\end{table}

 \paragraph{Sensitivity Check with Lee Bounds} To further assess the sensitivity of our results to differential selection induced by restricting the analyses to participants who passed the post-treatment attention check, we conducted a Lee-style bounds analysis separately for Democrats and Republicans \cite{Lee2005TrainingWA}. Because attention-check pass rates differed across disclaimer conditions, restricting the sample to participants who passed could introduce post-treatment selection bias. For each outcome, we compared each active disclaimer condition (Risk only, Causal only, and Risk + Causal) with the control condition. We calculated the differential attention-check pass rates between the treatment and control conditions and trimmed the outcome distribution of the condition with the higher selection rate accordingly. We then estimated lower and upper bounds by considering worst-case scenarios in which participants differentially selected into the analysis sample had either relatively high or relatively low outcome values. These analyses provide a conservative assessment of whether the direction of the estimated effects is robust to differential selection into the attention-check-passing sample.

The results in \Cref{tab:lee_main} indicate that several effects remain directionally robust even under these worst-case selection scenarios, including all effects identified in the main analysis. However, these results are not directly comparable to the estimates from our primary factorial models because the Lee-style bounds are implemented through binary treatment comparisons, each analysis compares a single treatment cell from the \(2 \times 2\) design with the control condition, whereas our main analyses estimate the effects of risk, causal framing, and their interaction jointly across all four experimental conditions.

Even so, we found that among Democrats, the negative effects of the Risk-only and Causal-only conditions on Conserv+ topics remained negative across the full range of bounds, with the additional negative effect of the Risk-only condition on Budget and the negative effect of the Causal-only condition on AI Trust. Among Republicans, the negative effect of the Risk-only condition on budget remained negative across the bounds, with the additional negative effects of Risk on Liberal+ opinion and Risk + Causal on budget. In addition, the positive effect of the Risk-only condition on AI Trust remained positive throughout the bounds as well. 

For the remaining comparisons, the bounds included zero, indicating that the direction of the estimated effect was not robust to the most extreme selection scenarios considered. Overall, these results suggest that some of the observed effects are robust to differential attention-check selection, while others are more sensitive to assumptions about the outcomes of participants whose inclusion in the analysis sample may have been affected by treatment assignment.

The Lee-style sensitivity analysis for the one-week follow-up outcomes similarly examined whether restricting the analyses to participants who passed the post-treatment attention check could affect the direction of the estimated treatment effects as seen in \Cref{tab:lee_followup}. Overall, several point estimates remained in the same direction across both extreme selection scenarios, although the majority of bounds included zero. Among Democrats, the Risk + Causal condition remained negatively associated with both Conserv+ and Liberal+ outcomes across the full range of bounds, and the Causal-only condition remained negatively associated with AI Trust. Among Republicans, the Risk-only condition remained positively associated with Conserv+, Liberal+, Budget, and AI Trust, while the Causal-only condition remained positively associated with Budget. The remaining comparisons had bounds spanning zero, indicating that their estimated directions were more sensitive to the extreme assumptions about differential selection considered here. As these are Lee-style point-estimate bounds rather than confidence intervals, bounds that exclude zero should not be interpreted as evidence of statistical significance; instead, they indicate that the direction of the corresponding point estimate remains unchanged under the worst-case selection scenarios.

\subsubsection{User Characteristics, AI Familiarity, and Post-Task AI Trust as Covariate}
We also repeated the analysis from \Cref{sect:personality_traits}, \Cref{sect:ai_familiarity}, and \Cref{sect:AITrust_results} excluding those who failed the attention check. Again, we see minimal change to substantive results, except the user characteristics analysis.

\begin{table}[ht]
\centering
\begin{threeparttable}
\caption{Personality Traits as Predictors of Persuasion and AI Trust (Pooled Model) - Excluded Participants}
\label{tab:personality_allresults}
\small
\begin{tabular}{l r@{} l r@{} l r@{} l r@{} l}
\toprule
& \multicolumn{2}{c}{Conserv+ Opinion} & \multicolumn{2}{c}{Liberal+ Opinion} & \multicolumn{2}{c}{Budget Allocation} & \multicolumn{2}{c}{AI Trust} \\
\cmidrule(lr){2-3} \cmidrule(lr){4-5} \cmidrule(lr){6-7} \cmidrule(lr){8-9}
Trait & \multicolumn{2}{c}{$b$ (SE)} & \multicolumn{2}{c}{$b$ (SE)} & \multicolumn{2}{c}{$b$ (SE)} & \multicolumn{2}{c}{$b$ (SE)} \\
\midrule
Risk
 & $-0.12$ & $(0.16)$ & $-0.07$ & $(0.15)$ & \sigfive{$-4.93$} & $(1.82)$ & $-0.03$ & $(0.13)$ \\

Causal & $-0.20$ & $(0.16)$ & $-0.08$ & $(0.14)$ & $-1.59$ & $(1.88)$ & $-0.15$ & $(0.13)$ \\

Risk$\times$Causal & $0.30$ & $(0.22)$ & $0.16$ & $(0.21)$ & $4.20$ & $(2.56)$ & $0.14$ & $(0.19)$ \\
Republican & $\sigfive{1.06}$ &(0.11) & $\sigfive{-0.54}$ & (0.11) & $\sigfive{6.09}$ & (1.50) & $\sigfive{0.72}$ & (0.10)\\
\midrule
Extraversion       & $\sigten{0.08}$  & $(0.04)$ & $0.07$  & $(0.04)$ & $0.03$  & $(0.44)$ & $\sigfive{0.09}$  & $(0.03)$ \\
Openness           & $-0.09$ & $(0.05)$ & $0.06$  & $(0.05)$ & $0.08$  & $(0.68)$ & $-0.01$ & $(0.04)$ \\
Conscientiousness  & $0.09$  & $(0.05)$ & $-0.07$ & $(0.05)$ & $0.18$  & $(0.64)$ & $0.03$  & $(0.04)$ \\
Emotional Stability & $0.02$  & $(0.04)$ & $-0.10^*$ & $(0.04)$ & $-0.58$ & $(0.56)$ & $0.04$  & $(0.04)$ \\
Agreeableness      & $-0.06$ & $(0.05)$ & $0.09$  & $(0.05)$ & $1.16^*$  & $(0.56)$ & $\sigfive{0.14}$  & $(0.04)$ \\
Self-Esteem        & $-0.02$ & $(0.06)$ & $0.08$  & $(0.06)$ & $0.18$  & $(0.73)$ & $0.14^*$  & $(0.06)$ \\
\bottomrule
\end{tabular}
\begin{tablenotes}
\small
\item \textit{Note.} Entries are OLS regression coefficients with HC3 robust standard errors in parentheses. Models are estimated on the pooled sample, controlling for party, disclaimer condition (Risk, Causal, Risk$\times$Causal), and the corresponding pre-measure (opinion and budget models only). ** = $p_{\text{Holm}}<.05$, * = $p_{raw}<0.05$.
\end{tablenotes}
\end{threeparttable}
\end{table} %
\begin{table}[ht]
\centering
\begin{threeparttable}
\caption{AI Familiarity Measures as Predictors of Persuasion and AI Trust (Pooled Model) - Excluded Participants}
\label{tab:ai_familiarity_allresults}
\small
\begin{tabular}{l r@{} l r@{} l r@{} l r@{} l}
\toprule
& \multicolumn{2}{c}{Conserv+ Opinion} & \multicolumn{2}{c}{Liberal+ Opinion} & \multicolumn{2}{c}{Budget Allocation} & \multicolumn{2}{c}{AI Trust} \\
\cmidrule(lr){2-3} \cmidrule(lr){4-5} \cmidrule(lr){6-7} \cmidrule(lr){8-9}
Predictor & \multicolumn{2}{c}{$b$ (SE)} & \multicolumn{2}{c}{$b$ (SE)} & \multicolumn{2}{c}{$b$ (SE)} & \multicolumn{2}{c}{$b$ (SE)} \\
\midrule
\midrule
Risk & $-0.13$ & $(0.16)$ & $-0.05$ & $(0.15)$ & $\sigfive{-4.58}$ & $(1.82)$ & $-0.03$ & $(0.14)$ \\

Causal & $-0.18$ & $(0.16)$ & $0.00$ & $(0.15)$ & $-0.83$ & $(1.84)$ & $-0.04$ & $(0.14)$ \\

Risk $\times$ Causal & $0.28$ & $(0.22)$ & $0.08$ & $(0.21)$ & $3.26$ & $(2.54)$ & $0.04$ & $(0.19)$ \\
Republican & $\sigfive{1.11}$ & $(0.11)$ & $\sigfive{-0.58}$ & $(0.11)$ & $\sigfive{5.87}$ & $(1.37)$ & $\sigfive{0.78}$ & $(0.10)$ \\

\midrule
AI Knowledge  & $\sigfive{-0.21}$ & $(0.09)$ & $\sigfive{-0.32}$ & $(0.08)$ & $\sigfive{-3.49}$ & $(1.09)$ & $-0.14$ & $(0.08)$ \\
AI Education  & $0.14$  & $(0.10)$ & $0.14$  & $(0.09)$ & $0.88$  & $(1.32)$ & $0.18^*$  & $(0.09)$ \\
AI Use        & $\sigfive{0.20}$  & $(0.07)$ & $\sigfive{0.17}$  & $(0.06)$ & $\sigfive{2.52}$  & $(0.81)$ & $\sigfive{0.37}$  & $(0.06)$ \\
\bottomrule
\end{tabular}
\begin{tablenotes}
\small
\item \textit{Note.} Entries are OLS regression coefficients with HC3 robust standard errors in parentheses. Models are estimated on the pooled sample, controlling for party, disclaimer condition (Risk, Causal, Risk$\times$Causal), and the corresponding pre-measure (opinion and budget models only). This analysis is exploratory. ** = $p_{\text{Holm}}<.05$, * = $p_{raw}<0.05$.
\end{tablenotes}
\end{threeparttable}
\end{table} %

\begin{table}[ht]
\centering
\begin{threeparttable}
\caption{Effect of Disclaimer Conditions and AI Trust on Persuasion - Excluded Participants}
\label{tab:sensitivity_aitrust}
\small
\begin{tabular}{ll r@{} l r@{} l r@{} l}
\toprule
& & \multicolumn{2}{c}{Conserv+ Opinion} & \multicolumn{2}{c}{Liberal+ Opinion} & \multicolumn{2}{c}{Budget Allocation} \\
\cmidrule(lr){3-4} \cmidrule(lr){5-6} \cmidrule(lr){7-8}
Party & Predictor & \multicolumn{2}{c}{$b$ (SE)} & \multicolumn{2}{c}{$b$ (SE)} & \multicolumn{2}{c}{$b$ (SE)} \\
\midrule
\multirow{4}{*}{Democrat}
 & Risk               & $-0.25$ & $(0.21)$ & $0.02$ & $(0.21)$  & $-2.08$ & $(2.08)$  \\
 & Causal             & \sigten{$-0.32$} & $(0.21)$ & $-0.04$ & $(0.22)$  & $-0.19$ & $(2.24)$  \\
 & Risk$\times$Causal & $0.40$ & $(0.28)$          & $-0.12$ & $(0.30)$  & $1.82$ & $(3.02)$   \\
 & \textbf{AI Trust}           & $\sigfive{0.45}$ & $(0.04)$  & $\sigfive{0.26}$ & $(0.05)$  & $\sigfive{4.77}$ & $(0.55)$  \\[4pt]
\multirow{4}{*}{Republican}
 & Risk               & $0.03$ & $(0.22)$          & $-0.16$ & $(0.20)$  & $\sigfive{-6.95}$ & $(2.84)$  \\
 & Causal             & $0.03$ & $(0.22)$          & $-0.01$ & $(0.19)$  & $-1.66$ & $(2.71)$  \\
 & Risk$\times$Causal & $0.11$ & $(0.30)$          & $0.40$ & $(0.27)$   & $5.61$ & $(3.91)$   \\
 & \textbf{AI Trust}           & $\sigfive{0.11}$ & $(0.05)$  & $\sigfive{0.30}$ & $(0.05)$  & $\sigfive{4.14}$ & $(0.63)$  \\
\bottomrule
\end{tabular}
\begin{tablenotes}
\small
\item \textit{Note.} Entries are OLS regression coefficients with HC3 robust standard errors in parentheses. Models are estimated separately by party. AI Trust is included as a covariate to same model used in H1 and H2. Hypothesis tests for disclaimer conditions (Risk, Causal, Risk$\times$Causal) are one-tailed, directional negative; AI Trust is tested two-tailed. \sigfive{} = $p < .05$; $\sigten{} = p < .10$.
\end{tablenotes}
\end{threeparttable}
\end{table} 
In \Cref{tab:personality_allresults}, we present the results of the regression models including all user personality covariates. Interestingly, this analysis produces the largest change relative to our main results. All statistically significant effects of AI Trust remain unchanged; however, the effects of both Emotional Stability and Agreeableness are no longer statistically significant. This pattern may suggest a relationship between these personality characteristics and the participants who did not recall the attention warning presented at the end of the study. Although exploratory, this finding is interesting and may point to a promising avenue for future research.

In \Cref{tab:ai_familiarity_allresults} and \Cref{tab:sensitivity_aitrust}, we present the remaining two sensitivity analyses. Results are largely consistent with the main analyses. The only difference is that the effect of AI Knowledge on post-task AI Trust is reduced from statistical significance at $p < .05$ to marginal significance at $p < .10$.

\subsection{Pooled Models with Demographic Covariates}\label{appx:pooled_full_models}
In this section, we provide pooled analyses for the models presented in \Cref{sect:H1_H2_results}, including all demographic variables that differed significantly between our two group variables: age, education, income, and race. Because these analyses were not hypothesized in our pre-registration, they should be interpreted as exploratory rather than causal analyses.

\Cref{tab:main_full_all} present the results. Overall, the findings show no meaningful changes relative to the minimal models reported in the main paper.

\begin{table}[ht]
\centering
\begin{threeparttable}
\caption{Effect of Disclaimer Conditions on Opinion, Budget Allocation, and AI Trust (Full Model) -  All Participants}
\label{tab:main_full_all}
\small
\begin{tabular}{l r@{} l r@{} l r@{} l r@{} l}
\toprule
& \multicolumn{2}{c}{Conserv+ Opinion} & \multicolumn{2}{c}{Liberal+ Opinion} & \multicolumn{2}{c}{Budget Allocation} & \multicolumn{2}{c}{AI Trust} \\
\cmidrule(lr){2-3} \cmidrule(lr){4-5} \cmidrule(lr){6-7} \cmidrule(lr){8-9}
Predictor & \multicolumn{2}{c}{$b$ (SE)} & \multicolumn{2}{c}{$b$ (SE)} & \multicolumn{2}{c}{$b$ (SE)} & \multicolumn{2}{c}{$b$ (SE)} \\
\midrule
Risk & $-0.15$ & $(0.13)$ & $0.00$ & $(0.13)$ & \sigfive{-3.20} & $(1.48)$ & $0.03$ & $(0.11)$ \\
Causal & \sigfive{-0.27} & $(0.13)$ & $0.04$ & $(0.12)$ & $-0.44$ & $(1.53)$ & $0.00$ & $(0.11)$ \\
Risk$\times$Causal & $0.40$ & $(0.18)$ & $-0.00$ & $(0.18)$ & $2.78$ & $(2.17)$ & $-0.04$ & $(0.16)$ \\
Republican & \sigfive{1.07} & $(0.10)$ & \sigfive{-0.53} & $(0.10)$ & \sigfive{5.77} & $(1.17)$ & \sigfive{0.88} & $(0.08)$ \\
\bottomrule
\end{tabular}

\begin{tablenotes}
\small
\item \textit{Note.} Entries are OLS regression coefficients with HC3 robust standard errors in parentheses. Models are estimated on the full sample; Republican (vs.\ Democrat) is included as a main-effect covariate only (no interactions with disclaimer conditions). Hypothesis tests for disclaimer main effects (Risk, Causal, Risk$\times$Causal) are one-tailed, directional negative; Republican is tested two-tailed. Models for opinion and budget allocation include the corresponding pre-measure as a covariate (mean-centered); the trust model does not. \sigfive{} = $p < .05$; \sigten{} = $p < .10$.
\end{tablenotes}

\end{threeparttable}
\end{table}

\subsection{AI Trust and Persuasion Regression Results}\label{appx:trust_persuasion_regression}

\begin{table}[ht]
\centering
\begin{threeparttable}
\caption{Effect of Disclaimer Conditions and AI Trust on Persuasion - All Participants}
\label{tab:sensitivity_aitrust_all}
\small
\begin{tabular}{ll r@{} l r@{} l r@{} l}
\toprule
& & \multicolumn{2}{c}{Conserv+ Opinion} & \multicolumn{2}{c}{Liberal+ Opinion} & \multicolumn{2}{c}{Budget Allocation} \\
\cmidrule(lr){3-4} \cmidrule(lr){5-6} \cmidrule(lr){7-8}
Party & Predictor & \multicolumn{2}{c}{$b$ (SE)} & \multicolumn{2}{c}{$b$ (SE)} & \multicolumn{2}{c}{$b$ (SE)} \\
\midrule
\multirow{4}{*}{Democrat}
 & Risk
 & $\sigten{-0.24}$ & $(0.18)$
 & $0.01$ & $(0.18)$
 & $0.56$ & $(1.81)$ \\
 & Causal
 & $\sigfive{-0.29}$ & $(0.17)$
 & $0.08$ & $(0.18)$
 & $1.73$ & $(1.99)$ \\
 & Risk$\times$Causal
 & $0.48$ & $(0.24)$
 & $-0.19$ & $(0.26)$
 & $-0.61$ & $(2.74)$ \\
 & \textbf{AI Trust}
 & $\sigfive{0.42}$ & $(0.04)$
 & $\sigfive{0.25}$ & $(0.04)$
 & $\sigfive{5.06}$ & $(0.47)$ \\[4pt]

\multirow{4}{*}{Republican}
 & Risk
 & $-0.04$ & $(0.18)$
 & $-0.05$ & $(0.17)$
 & $\sigfive{-6.78}$ & $(2.23)$ \\
 & Causal
 & $-0.18$ & $(0.18)$
 & $-0.00$ & $(0.16)$
 & $-2.11$ & $(2.20)$ \\
 & Risk$\times$Causal
 & $0.25$ & $(0.26)$
 & $0.24$ & $(0.23)$
 & $5.99$ & $(3.21)$ \\
 & \textbf{AI Trust}
 & $\sigfive{0.11}$ & $(0.04)$
 & $\sigfive{0.31}$ & $(0.04)$
 & $\sigfive{3.50}$ & $(0.52)$ \\
\bottomrule
\end{tabular}

\begin{tablenotes}
\small
\item \textit{Note.} Entries are OLS regression coefficients with HC3 robust standard errors in parentheses. Models are estimated separately by party. AI Trust is included as a covariate in the same model used in H1 and H2. Hypothesis tests for disclaimer conditions (Risk, Causal, Risk$\times$Causal) are one-tailed, directional negative; AI Trust is tested two-tailed. \sigfive{} = $p < .05$; \sigten{} = $p < .10$.
\end{tablenotes}

\end{threeparttable}
\end{table}  Interestingly, we found a strong trend between post-task AI Trust and the effect of persuasion of the model. To explore this, we added post-task AI Trust as a covariate to the H1 and H2 models, estimated separately by party. AI Trust was tested two-tailed as we had not pre-registered a direction; \texttt{Risk}, \texttt{Causal}, and their interaction remained one-tailed as in H1 and H2.

Results, seen in Table~\ref{tab:sensitivity_aitrust_all}, show post-task AI Trust is a strong, consistent linear predictor of persuasion across all three outcomes and both parties. Among Democrats, higher post-task trust was associated with significantly larger pro-model shifts in all tasks (Conserv+ opinion: $b=0.42$, Liberal+ opinion: $b=0.25$, and budget allocation: $b=5.06$; all $p<.05$). Among Republicans, the same held for all three outcomes (Conserv+ opinion: $b=0.11$, Liberal+ opinion: $b=0.31$, Budget: $b=3.50$; all $p<.05$). Disclaimer effects were largely unchanged after controlling for AI trust, relative to H1 and H2, suggesting AI trust does not simply substitute for the disclaimer effects reported there.

\subsection{One Week Follow-Up Change Visualization}\label{appx:one_week_visual}
\begin{figure}
    \centering
    \includegraphics[width=1\linewidth]{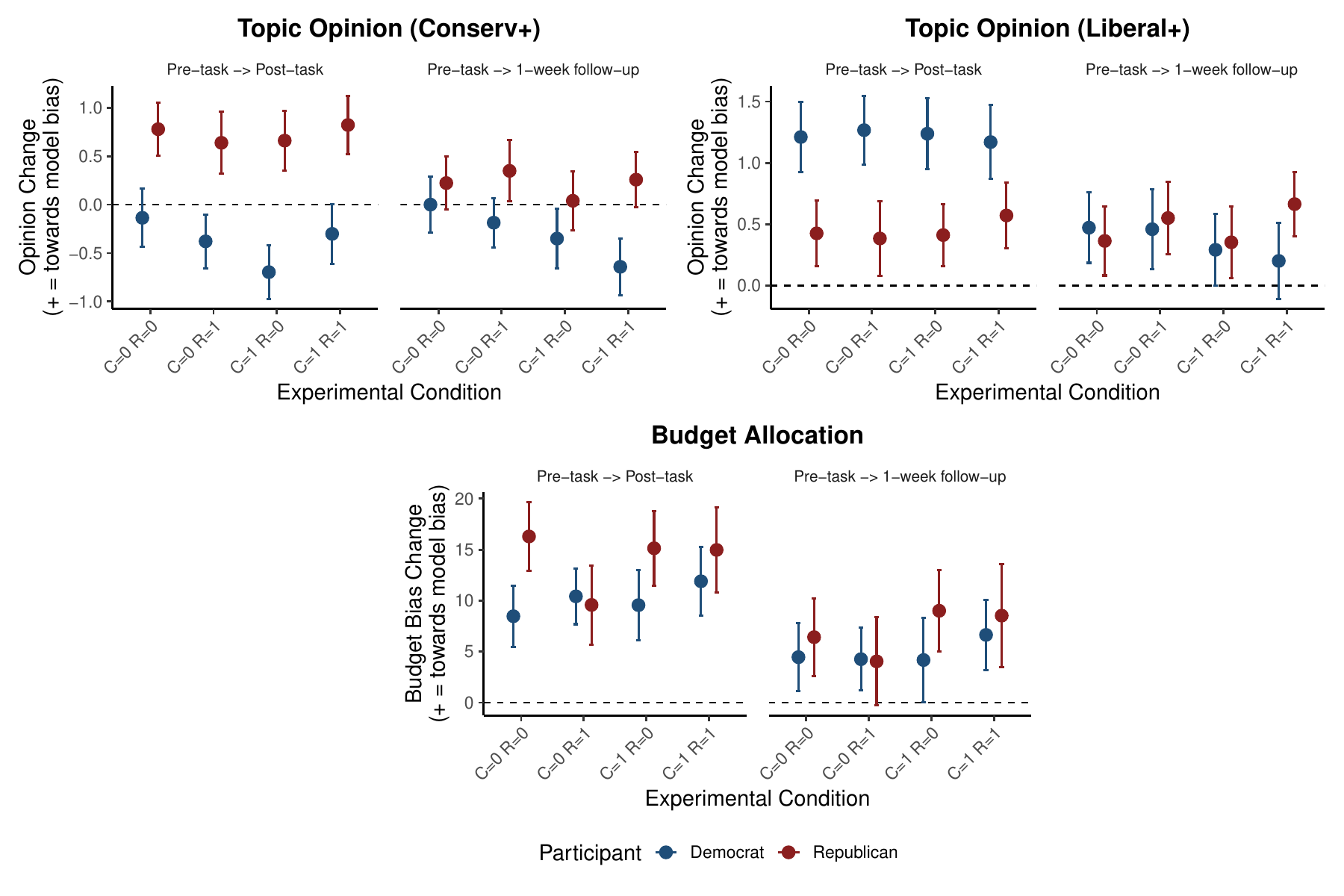}
    \caption{Mean change in outcomes from pre-task to post-task and pre-task to one-week follow-up. Points represent mean within-participant change, with 95\% confidence intervals, separately for Democrats and Republicans across experimental conditions. Experimental conditions are indicated by \texttt{Causal}, ``C'' (0=absent, 1=present) and \texttt{Risk}, ``R'' (0=inaccuracy, 1=political bias). Positive values indicate movement toward the model’s bias. The panels show changes in opinion for the Conservative+ and Liberal+ topic tasks and changes in budget allocations.}
    \Description{The figure contains three panels showing mean within-participant change in outcomes from pre-task to post-task and from pre-task to a one-week follow-up. The top-left panel shows change in opinion for the Conservative+ topic, the top-right shows change in opinion for the Liberal+ topic, and the bottom-center shows change in budget allocation. Each panel displays four experimental conditions, defined by combinations of C and R conditions, with separate blue points for Democrats and red points for Republicans. Error bars represent 95\% confidence intervals. Positive values indicate movement toward the model’s bias.}
    \label{fig:mean_time_change_all}
\end{figure}
To better understand the persuasive effects over time, we include \Cref{fig:mean_time_change_all} as a visualization of average opinion change per task between pre-task and post-task, compared to pre-task and 1-week follow up. Consistent with the patterns in our regression (\Cref{tab:followup_pre_all}, the mean change in \Cref{fig:mean_time_change_all} generally remained in the direction of the immediate effects rather than returning fully to baseline a 0.

\subsection{Sensitivity Checks}\label{appx:sensitivity_checks}
As mentioned in our pre-registration, we include a few sensitivity analyses below. All show strong robustness of results.  

\paragraph{Follow-Up Analysis with Full Model}
We present the H4 analysis using the full model, both as a minimal specification with no additional covariates and a specification that additionally controls for all demographic characteristics on which Democrats and Republicans who completed the follow-up differed significantly: age, education, Hispanic ethnicity, and race.

Interestingly, when pooling the data in the minimal model (\Cref{tab:followup_pre_minimal_all}), we observe a statistically significant decrease in persuasion from the pre-task measure to the one-week follow-up for the Causal condition in the Conserv+ task. However, this effect is reduced to marginal significance ($p < .10$) when demographic covariates are included (\Cref{tab:followup_full_pre_all}). Taken together, these results lend additional support to our hypothesis that the effects of the intervention diminish over time but do not fully reverse one week later.

\begin{table}[ht]
\centering
\begin{threeparttable}
\caption{Follow-Up Sensitivity Check: Disclaimer Effects Controlling for Partisanship (Pre-Task, Minimal Model)- All Participants}
\label{tab:followup_pre_minimal_all}
\small
\begin{tabular}{l r@{} l r@{} l r@{} l}
\toprule
& \multicolumn{2}{c}{Conserv+ Opinion} & \multicolumn{2}{c}{Liberal+ Opinion} & \multicolumn{2}{c}{Budget Allocation} \\
\cmidrule(lr){2-3} \cmidrule(lr){4-5} \cmidrule(lr){6-7}
Predictor & \multicolumn{2}{c}{$b$ (SE)} & \multicolumn{2}{c}{$b$ (SE)} & \multicolumn{2}{c}{$b$ (SE)} \\
\midrule
Risk               & $-0.11$ & $(0.14)$ & $0.06$ & $(0.15)$ & $-0.71$ & $(1.82)$ \\
Causal             & \sigfive{-0.24} & $(0.14)$ & $-0.04$ & $(0.14)$ & $1.35$ & $(1.89)$ \\
Risk$\times$Causal & $0.11$ & $(0.20)$ & $-0.02$ & $(0.20)$ & $1.88$ & $(2.72)$ \\
Republican         & \sigfive{0.52} & $(0.10)$ & \sigfive{0.55} & $(0.11)$ & \sigfive{3.57} & $(1.42)$ \\
\bottomrule
\end{tabular}

\begin{tablenotes}
\small
\item \textit{Note.} Entries are OLS regression coefficients with HC3 robust standard errors in parentheses. Models are estimated on the full sample with Republican (vs.\ Democrat) included as a covariate. Pre-treatment values of the corresponding outcome are included as covariates. Hypothesis tests for disclaimer conditions (Risk, Causal, Risk$\times$Causal) are one-tailed, directional negative; Republican is tested two-tailed. \sigfive{} = $p < .05$; \sigten{} = $p < .10$.
\end{tablenotes}

\end{threeparttable}
\end{table}

\begin{table}[ht]
\centering
\begin{threeparttable}
\caption{Follow-Up Sensitivity Check: Disclaimer Effects Controlling for Partisanship (Pre-Task, Full Model)- All Participants}
\label{tab:followup_full_pre_all}
\small
\begin{tabular}{l r@{} l r@{} l r@{} l}
\toprule
& \multicolumn{2}{c}{Conserv+ Opinion} & \multicolumn{2}{c}{Liberal+ Opinion} & \multicolumn{2}{c}{Budget Allocation} \\
\cmidrule(lr){2-3} \cmidrule(lr){4-5} \cmidrule(lr){6-7}
Predictor & \multicolumn{2}{c}{$b$ (SE)} & \multicolumn{2}{c}{$b$ (SE)} & \multicolumn{2}{c}{$b$ (SE)} \\
\midrule
Risk
    & $-0.12$ & $(0.14)$
    & $0.07$ & $(0.15)$
    & $-0.63$ & $(1.82)$ \\
Causal
    & $\sigten{-0.25}$ & $(0.14)$
    & $-0.04$ & $(0.14)$
    & $1.31$ & $(1.88)$ \\
Risk$\times$Causal
    & $0.11$ & $(0.20)$
    & $-0.03$ & $(0.20)$
    & $1.79$ & $(2.72)$ \\
Republican
    & $\sigfive{0.52}$ & $(0.10)$
    & $\sigfive{0.59}$ & $(0.12)$
    & $\sigfive{4.15}$ & $(1.41)$ \\
\bottomrule
\end{tabular}

\begin{tablenotes}
\small
\item \textit{Note.} Entries are OLS regression coefficients with HC3 robust standard errors in parentheses. Models are estimated on the full sample with Republican (vs.\ Democrat) included as a covariate, along with age, education, and race covariates. Pre-task values of the corresponding outcome are included as covariates. Hypothesis tests for disclaimer conditions (Risk, Causal, Risk$\times$Causal) are one-tailed, directional negative; Republican is tested two-tailed. \sigfive{} = $p < .05$; \sigten{} = $p < .10$.
\end{tablenotes}

\end{threeparttable}
\end{table} 

\paragraph{Analysis with Change Scores}
Next, we re-analyze H1-H4 using the change in responses ($y_{POST}$ - $y_{PRE}$) model interaction instead of regressing for $y_{POST}$ using $y_{PRE}$ as a covariate. Results are seen in \Cref{tab:sensitivity_difference_all} for the main study and \Cref{tab:followup_difference_all} for the follow-up analysis. We find no difference in significance found compared to the results in our main study.

\begin{table}[ht]
\centering
\begin{threeparttable}
\caption{Main Study Sensitivity Check: Disclaimer Effects on Opinion and Budget Change Scores ($y_{\text{post}} - y_{\text{pre}}$) -  All Participants}
\label{tab:sensitivity_difference_all}
\small
\begin{tabular}{ll r@{} l r@{} l r@{} l}
\toprule
& & \multicolumn{2}{c}{Conserv+ Opinion} & \multicolumn{2}{c}{Liberal+ Opinion} & \multicolumn{2}{c}{Budget Allocation} \\
\cmidrule(lr){3-4} \cmidrule(lr){5-6} \cmidrule(lr){7-8}
Party & Predictor & \multicolumn{2}{c}{$b$ (SE)} & \multicolumn{2}{c}{$b$ (SE)} & \multicolumn{2}{c}{$b$ (SE)} \\
\midrule
\multirow{3}{*}{Democrat}
 & Risk               & $-0.18$ & $(0.19)$ & $-0.03$ & $(0.19)$ & $-0.38$ & $(1.91)$ \\
 & Causal             & $\sigfive{-0.39}$ & $(0.19)$ & $0.03$ & $(0.19)$ & $1.49$ & $(2.20)$ \\
 & Risk$\times$Causal & $0.49$ & $(0.27)$ & $-0.10$ & $(0.27)$ & $0.56$ & $(3.01)$ \\
\multirow{3}{*}{Republican}
 & Risk               & $0.03$ & $(0.19)$ & $0.01$ & $(0.18)$ & $\sigfive{-6.74}$ & $(2.27)$ \\
 & Causal             & $-0.13$ & $(0.19)$ & $-0.01$ & $(0.16)$ & $-1.98$ & $(2.25)$ \\
 & Risk$\times$Causal & $0.15$ & $(0.26)$ & $0.16$ & $(0.24)$ & $5.03$ & $(3.28)$ \\
\bottomrule
\end{tabular}

\begin{tablenotes}
\small
\item \textit{Note.} Entries are OLS regression coefficients with HC3 robust standard errors in parentheses. The dependent variable is the post-pre difference score rather than the ANCOVA specification used in primary analyses. All hypothesis tests are one-tailed, directional negative. AI Trust is excluded as it has no pre-measure. \sigfive{} = $p < .05$; \sigten{} = $p < .10$.
\end{tablenotes}

\end{threeparttable}
\end{table} %
\begin{table}[ht]
\centering
\begin{threeparttable}
\caption{Follow-Up Sensitivity Check: Disclaimer Effects on Opinion and Budget Change Scores ($y_{\text{fu}} - y_{\text{pre}}$)- All Participants}
\label{tab:followup_difference_all}
\small
\begin{tabular}{ll r@{} l r@{} l r@{} l}
\toprule
& & \multicolumn{2}{c}{Conserv+ Opinion} & \multicolumn{2}{c}{Liberal+ Opinion} & \multicolumn{2}{c}{Budget Allocation} \\
\cmidrule(lr){3-4} \cmidrule(lr){5-6} \cmidrule(lr){7-8}
Party & Predictor & \multicolumn{2}{c}{$b$ (SE)} & \multicolumn{2}{c}{$b$ (SE)} & \multicolumn{2}{c}{$b$ (SE)} \\
\midrule
\multirow{3}{*}{Democrat}
 & Risk               & $-0.19$ & $(0.20)$ & $-0.01$ & $(0.22)$ & $-0.20$ & $(2.33)$ \\
 & Causal             & \sigten{$-0.35$} & $(0.22)$ & $-0.18$ & $(0.21)$ & $-0.28$ & $(2.73)$ \\
 & Risk$\times$Causal & $-0.10$ & $(0.29)$ & $-0.08$ & $(0.31)$ & $2.66$ & $(3.62)$ \\[4pt]

\multirow{3}{*}{Republican}
 & Risk               & $0.13$ & $(0.21)$ & $0.19$ & $(0.21)$ & $-2.38$ & $(2.96)$ \\
 & Causal             & $-0.18$ & $(0.21)$ & $-0.01$ & $(0.21)$ & $2.59$ & $(2.83)$ \\
 & Risk$\times$Causal & $0.09$ & $(0.30)$ & $0.12$ & $(0.29)$ & $1.90$ & $(4.43)$ \\
\bottomrule
\end{tabular}

\begin{tablenotes}
\small
\item \textit{Note.} Entries are OLS regression coefficients with HC3 robust standard errors in parentheses. The dependent variable is the follow-up minus pre-task difference score rather than the ANCOVA specification used in primary follow-up analyses. All hypothesis tests are one-tailed, directional negative. \sigfive{} = $p < .05$; \sigten{} = $p < .10$.
\end{tablenotes}
\end{threeparttable}
\end{table} 
\paragraph{Analysis with Logistic Regression}
Next, we re-estimated our primary models using ordinal logistic regression rather than ordinary least squares. We note, that we do not include Budget Analysis as this is a continuous number. The post-treatment outcomes and trust measures, which were originally analyzed as approximately continuous Likert-type scales, were instead treated as ordered categorical variables. Specifically, each outcome was converted to an ordered factor preserving its original response ordering, and proportional odds models were estimated using `polr`. The same model specifications, including all covariates and interaction terms, were retained to assess whether our substantive conclusions were robust to alternative assumptions about the scale properties of the dependent variables. We again see no significant change in the follow-up results (\Cref{tab:sensitivity_ordinal_all}, \Cref{tab:followup_ordinal_all}). 

\begin{table}[ht]
\centering
\begin{threeparttable}
\caption{Main Study Sensitivity Check: Ordinal Logistic Regression of Disclaimer Effects -  All Participants}
\label{tab:sensitivity_ordinal_all}
\small
\begin{tabular}{ll r@{} l r@{} l r@{} l}
\toprule
& & \multicolumn{2}{c}{Conserv+ Opinion} & \multicolumn{2}{c}{Liberal+ Opinion}  & \multicolumn{2}{c}{AI Trust} \\
\cmidrule(lr){3-4} \cmidrule(lr){5-6} \cmidrule(lr){7-8} 
Party & Predictor & \multicolumn{2}{c}{$b$ (SE)} & \multicolumn{2}{c}{$b$ (SE)} & \multicolumn{2}{c}{$b$ (SE)} \\
\midrule
\multirow{3}{*}{Democrat}
 & Risk               & $\sigfive{-0.31}$ & $(0.18)$ & $0.02$ & $(0.18)$ &   $-0.10$ & $(0.18)$ \\
 & Causal             & $\sigfive{-0.36}$ & $(0.18)$ & $0.06$ & $(0.18)$ &  $-0.15$ & $(0.17)$ \\
 & Risk$\times$Causal & $0.60$ & $(0.26)$ & $-0.14$ & $(0.25)$  & $0.34$ & $(0.25)$ \\[4pt]

\multirow{3}{*}{Republican}
 & Risk               & $-0.05$ & $(0.18)$ & $0.03$ & $(0.18)$ & $0.17$ & $(0.17)$ \\
 & Causal             & $-0.22$ & $(0.18)$ & $0.05$ & $(0.18)$ & $0.16$ & $(0.17)$ \\
 & Risk$\times$Causal & $0.31$ & $(0.25)$ & $0.05$ & $(0.26)$ & $\sigfive{-0.44}$ & $(0.25)$ \\
\bottomrule
\end{tabular}

\begin{tablenotes}
\small
\item \textit{Note.} Entries are log-odds coefficients from proportional odds (ordinal logistic) regression models estimated via \texttt{polr}, with standard errors in parentheses. Post-treatment outcomes and trust measures were treated as ordered factors. The same model specifications as the primary analyses were retained. All hypothesis tests are one-tailed, directional negative. \sigfive{} = $p < .05$; \sigten{} = $p < .10$.
\end{tablenotes}

\end{threeparttable}
\end{table}

\begin{table}[ht]
\centering
\begin{threeparttable}
\caption{Follow-Up Sensitivity Check: Ordinal Logistic Regression of Disclaimer Effects- All Participants}
\label{tab:followup_ordinal_all}
\small
\begin{tabular}{ll r@{} l r@{} l}
\toprule
& & \multicolumn{2}{c}{Conserv+ Opinion} & \multicolumn{2}{c}{Liberal+ Opinion} \\
\cmidrule(lr){3-4} \cmidrule(lr){5-6} 
Party & Predictor & \multicolumn{2}{c}{$b$ (SE)} & \multicolumn{2}{c}{$b$ (SE)} \\
\midrule
\multirow{3}{*}{Democrat}
 & Risk               & \sigten{$-0.29$} & $(0.20)$ & $0.07$ & $(0.20)$   \\
 & Causal             & \sigfive{$-0.35$} & $(0.20)$ & $-0.14$ & $(0.19)$  \\
 & Risk$\times$Causal & $-0.01$ & $(0.28)$ & $-0.16$ & $(0.28)$  \\[4pt]

\multirow{3}{*}{Republican}
 & Risk               & $0.08$ & $(0.20)$ & $0.09$ & $(0.21)$  \\
 & Causal             & $-0.23$ & $(0.20)$ & $0.10$ & $(0.20)$  \\
 & Risk$\times$Causal & $0.15$ & $(0.29)$ & $0.10$ & $(0.29)$  \\
\bottomrule
\end{tabular}

\begin{tablenotes}
\small
\item \textit{Note.} Entries are log-odds coefficients from proportional odds (ordinal logistic) regression models estimated via \texttt{polr}, with standard errors in parentheses. Follow-up outcomes were treated as ordered factors; post-treatment values of the corresponding outcome were included as covariates. All hypothesis tests are one-tailed, directional negative. \sigfive{} = $p < .05$; \sigten{} = $p < .10$.
\end{tablenotes}
\end{threeparttable}
\end{table} 
Given these results we feel confident in our analysis in the main paper.

\end{document}